\documentclass[acmtog, balance = false]{acmart}
\acmJournal{TOG}
\setcitestyle{square}
\usepackage{lipsum}
\usepackage{url}

\usepackage{wrapfig}
\usepackage{nicefrac}
\usepackage{mathtools}
\usepackage{graphicx}
\usepackage{booktabs} 
\usepackage{combelow} % for comma underneath a word
\usepackage[ruled,vlined,linesnumbered]{algorithm2e}
\usepackage{tabularx} % beautiful table
\usepackage{colortbl} % add row color to a table
\usepackage{cancel}
\usepackage{makecell} % multiple lines in a cell
\usepackage{enumitem}
\usepackage{manfnt} % cube symbol

\usepackage{amsmath}
\usepackage{amsfonts}
\usepackage{amsbsy}

\usepackage{listings} % code env
\usepackage{courier}
\definecolor{mygreen}{rgb}{0, 0.53, 0.66}
\definecolor{mygray}{rgb}{0.5,0.5,0.5}
\definecolor{mymauve}{rgb}{0.58,0,0.82}
\definecolor{superlightgray}{RGB}{240,240,240}
\usepackage[mathletters]{ucs}
\usepackage[utf8x]{inputenc}

\definecolor{darkBlue}{RGB}{0, 94, 184}
\definecolor{derekBlue}{RGB}{144,210,236}
\definecolor{derekTableBlue}{RGB}{189,235,252}
\definecolor{iglGreen}{RGB}{153,203,67}
\definecolor{coralRed}{RGB}{250,114,104}
\definecolor{gray}{RGB}{180,180,180}
\definecolor{orange}{RGB}{255,165,0}
\definecolor{Purple}{RGB}{137, 99, 198}
\definecolor{lightgray}{gray}{0.65}

\SetCommentSty{commentFont}
\SetNlSty{textbf}{}{.}

\newcommand{\refequ}[1] {Eq.~\eqref{equ:#1}}

\newcommand{\reffig}[1] {Fig.~\ref{fig:#1}}
\newcommand{\figref}[1]{\reffig{#1}}

\newcommand{\refsec}[1] {Sec.~\ref{sec:#1}}
\newcommand{\secref}[1] {\refsec{#1}}

\newcommand{\refalg}[1] {Alg.~\ref{alg:#1}}

\newcommand{\R}{\mathbb{R}}

\newcommand{\after}[1]{\widetilde{#1}}
\newcommand{\T}{\mathcal{T}}
\newcommand{\error}{C}

\newcommand{\vecFont}[1]{\mathbf{#1}}

\def\vc{{\vecFont{c}}}

\def\ve{{\vecFont{e}}}

\def\vp{{\vecFont{p}}}

\def\vt{{\vecFont{t}}}
\def\vu{{\vecFont{u}}}

\def\vw{{\vecFont{w}}}
\def\vx{{\vecFont{x}}}

\def\vz{{\vecFont{z}}}

\newcommand{\matFont}[1]{\mathbf{#1}}

\def\mD{{\matFont{D}}}

\def\mP{{\matFont{P}}}

\def\mR{{\matFont{R}}}

\newcommand{\eg}{\emph{e.g.}} % "for example"
\newcommand{\ie}{\emph{i.e.}} % "that is"
\newcommand{\etc}{\emph{etc.}} % "et cetera" (and so on)
\newcommand{\resp}{\emph{resp.}} % "respectively"
\newcommand{\ala}{\emph{\`{a} la}} % "in the style of"
\newcommand{\highlight}[1]{{#1}}

\renewcommand*{\ij}{{i\hspace{-0.1em}j}}
\newcommand{\ji}{{j\hspace{-0.05em}i}}
\newcommand{\jk}{{jk}}

\newcommand{\ki}{{ki}}

\newcommand{\ijk}{{i\hspace{-0.05em}j\hspace{-0.01em}k}}

\renewcommand{\deg}[1]{\text{degree}(#1)} % face degree of a vertex

\newcommand{\lComment}[1]{\text{\color{mygreen} $\triangleright$\ #1}}
\newcommand{\rComment}[1]{\hfill \text{\color{mygreen} $\triangleright$\ #1}}

\AtBeginDocument{%
  \providecommand\BibTeX{{%
    \normalfont B\kern-0.5em{\scshape i\kern-0.25em b}\kern-0.8em\TeX}}}

\begin{document} 

\title{Surface Simplification using Intrinsic Error Metrics}
% \title{Intrinsic Coarsening {\small Surface Simplification with Intrinsic Curvature Error, Intrinsic Coarsening with Curvature Transport Error}}

\author{Hsueh-Ti Derek Liu}
\affiliation{%
  \institution{Roblox \& University of Toronto}
  \country{Canada}
  }
\email{hsuehtil@gmail.com}

\author{Mark Gillespie}
\affiliation{%
  \institution{Carnegie Mellon University}
  \country{USA}
  }
\email{mgillesp@cs.cmu.edu}

\author{Benjamin Chislett}
\affiliation{%
  \institution{University of Toronto}
  \country{Canada}
  }
\email{chislett.ben@gmail.com}

\author{Nicholas Sharp}
\affiliation{%
  \institution{University of Toronto \& NVIDIA}
  \country{USA}
  }
\email{nmwsharp@gmail.com}

\author{Alec Jacobson}
\affiliation{%
  \institution{University of Toronto \& Adobe Research}
  \country{Canada}
  }
\email{jacobson@cs.toronto.edu}

\author{Keenan Crane}
\affiliation{%
  \institution{Carnegie Mellon University}
  \country{USA}
  }
\email{kmcrane@cs.cmu.edu}

\begin{teaserfigure}
  \includegraphics[width=\linewidth]{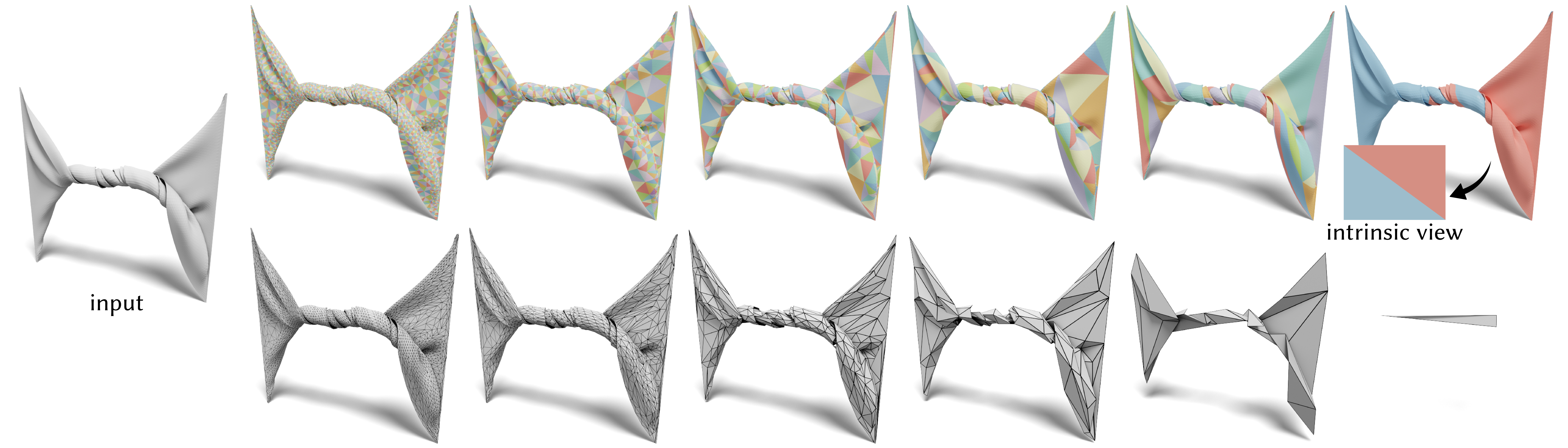}
   \caption{Whereas traditional extrinsic simplification \emph{(bottom row)} must simultaneously juggle element quality and approximation error, triangles produced by our intrinsic scheme \emph{(top row)} can wrap around the original surface---nicely approximating the underlying function space without changing the geometry.  Coarse meshes or hierarchies produced by this scheme can be used in ``black box'' fashion to accelerate solvers without changing user inputs/outputs.\label{fig:teaser}}
\end{teaserfigure} 


\begin{abstract}
   This paper describes a method for fast simplification of surface meshes.  Whereas past methods focus on visual appearance, our goal is to solve equations on the surface.  Hence, rather than approximate the extrinsic geometry, we construct a coarse \emph{intrinsic triangulation} of the input domain.  In the spirit of the \emph{quadric error metric (QEM)}, we perform greedy decimation while agglomerating global information about approximation error.  In lieu of extrinsic quadrics, however, we store intrinsic tangent vectors that track how far curvature ``drifts'' during simplification.  This process also yields a bijective map between the fine and coarse mesh, and prolongation operators for both scalar- and vector-valued data.  Moreover, we obtain hard guarantees on element quality via intrinsic retriangulation---a feature unique to the intrinsic setting.  The overall payoff is a ``black box'' approach to geometry processing, which decouples mesh resolution from the size of matrices used to solve equations.  We show how our method benefits several fundamental tasks, including geometric multigrid, all-pairs geodesic distance, mean curvature flow, geodesic Voronoi diagrams, and the discrete exponential map.
\end{abstract}



\setcopyright{acmlicensed}
\acmJournal{TOG}
\acmYear{2023} \acmVolume{42} \acmNumber{4} \acmArticle{} \acmMonth{8} \acmPrice{15.00}\acmDOI{10.1145/3592403}

\maketitle

% wrapfig appearance
\setlength{\intextsep}{0pt} % no space above wrapfig
\setlength{\columnsep}{.5em} % better margin size between text and wrapfig

\section{Introduction}
\label{sec:Introduction}

% TODO Does it make sense to say something about Delaunay refinement and the quality guarantees you get?

% TODO A good high-level takeaway from our application section is that it's overkill to compute a lot of the quantities we compute on high-res meshes---but we've been lacking a fast "bridge" between extrinsic and intrinsic meshes that makes it easy---and cheap!---to simplify -> compute -> transfer

%
% \subsection{Motivation}
% \label{sec:Motivation}
%
Algorithms for mesh simplification were originally motivated by the need to maintain visual fidelity for rendering and display~\cite{Hoppe96}.  Since image generation fundamentally depends on \emph{extrinsic} quantities like vertex positions and surface normals, extrinsic metrics (like the distance to the original surface) were a natural choice~\cite{GarlandH97}.  However, in a wide variety of problems throughout computer graphics, geometry processing, and scientific computing, the goal is to solve equations on surface meshes, rather than display them on screen.  Since functions on surfaces have derivatives only in tangential directions, differential operators appearing in these equations (such as the Laplacian) are almost always \emph{intrinsic}---even in cases where one solves for extrinsic quantities~\cite{Finnendahl2023ARAP}.  Hence, to develop fast, accurate solvers for equations on surfaces, it is natural to seek error metrics that focus on intrinsic geometry, \ie, quantities that depend only on distances along the surface, rather than coordinates in space.

\begin{figure}
    \begin{center}
    \includegraphics[width=1\linewidth]{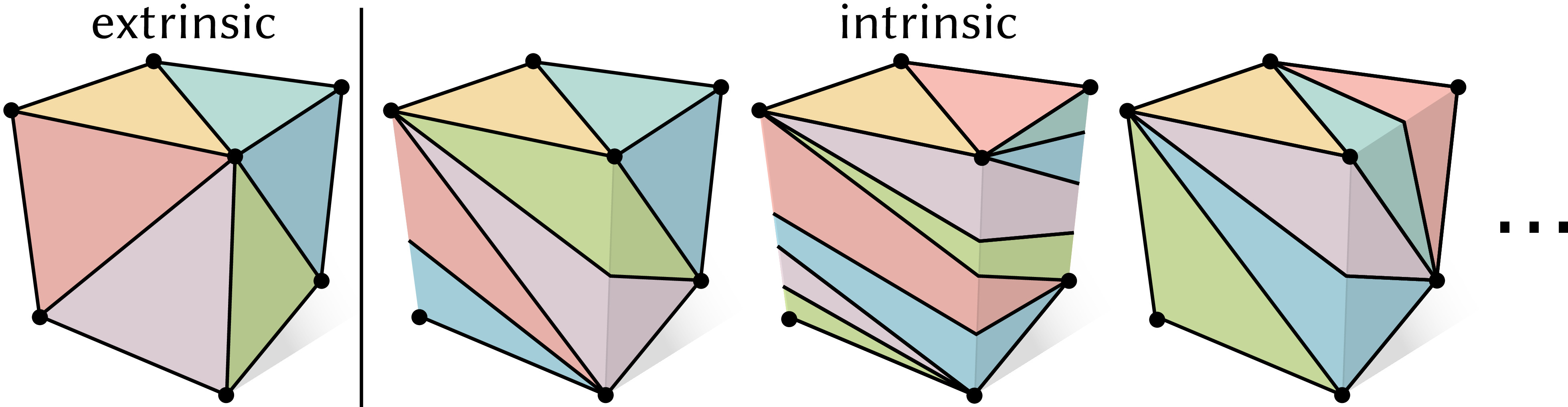}
    \end{center}
    \caption{In contrast to standard extrinsic meshes \emph{(left)}, where edges are straight line segments in \(\mathbb{R}^n\), intrinsic triangulations connect vertices along arbitrary geodesic paths between vertices \emph{(right)}.}
    \label{fig:many_intrinsic_triangulations}
\end{figure}

\newpage

A principled approach to this problem is enabled by recent work on flexible data structures for \emph{intrinsic triangulations}~\cite{FisherSBS06,SharpSC19a,GillespieSC21a}.  Roughly speaking, an intrinsic triangulation is one where edges need not be straight line segments in space, but can instead be any \emph{geodesic arc} along the surface (\figref{many_intrinsic_triangulations}).  More generally, an intrinsic triangulation is an assignment of positive lengths to edges---with no requirement that there be vertex positions that realize these lengths.  Not surprisingly, this construction provides a vastly larger space of possibilities for mesh processing, by de-coupling the elements used to approximate the geometry from those used to approximate functions on the surface.  Moreover, standard objects like the Laplacian can still be built directly from edge lengths, allowing many existing algorithms to be run as-is.  However, past work has not provided any mechanism for \emph{coarsening} intrinsic triangulations---as proposed in this paper.

\subsection{Method Overview}
\label{sec:MethodOverview}

\begin{figure}
   \includegraphics[width=1\linewidth]{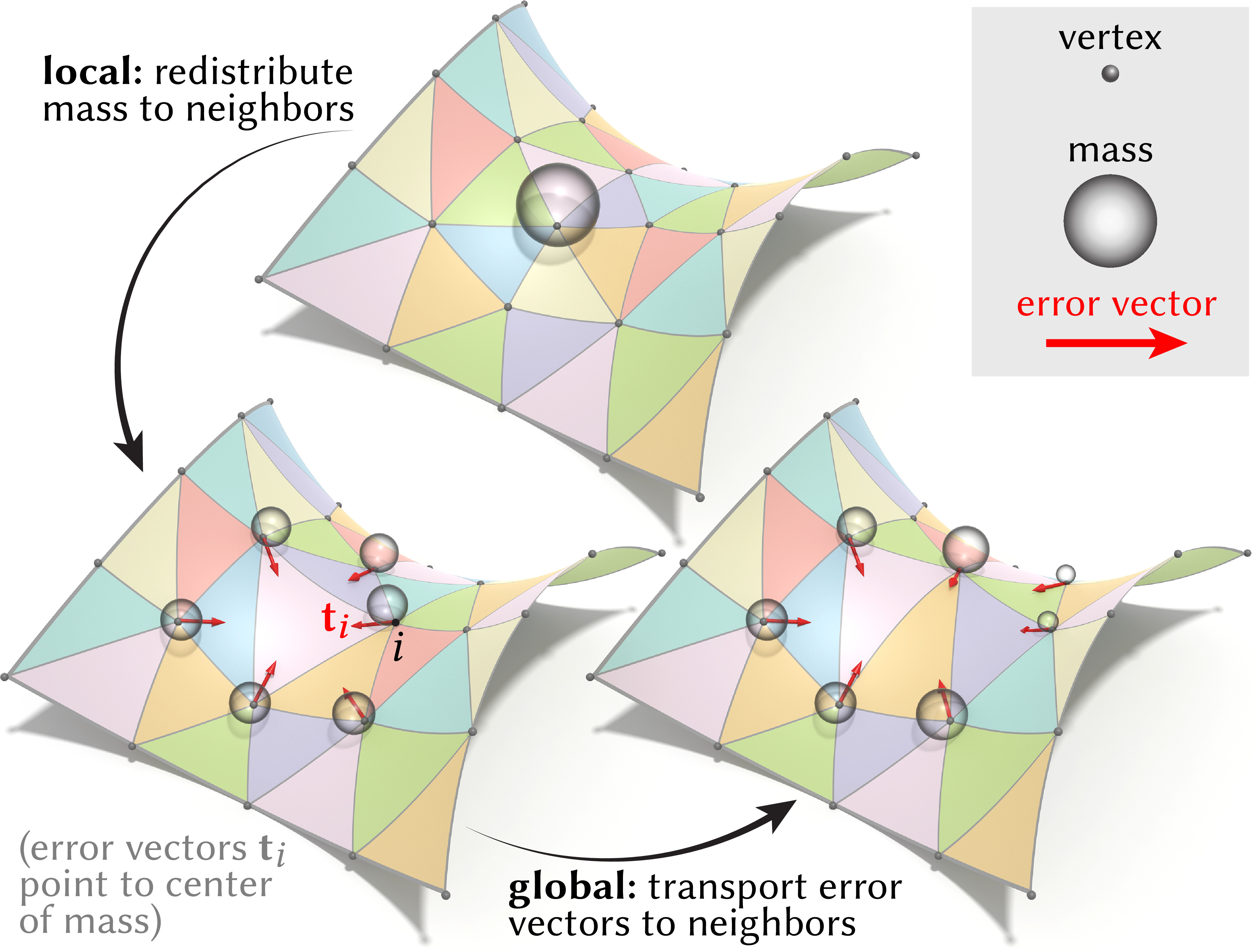}
   \caption{Our method constructs a coarse triangulation over a fixed geometric domain.  In each \textbf{local} step we redistribute curvature or other quantities from a removed vertex to its neighbors.  From step to step we also agglomerate \textbf{global} information about error via tangent vectors that point toward the approximate center of mass of the decimated vertices. \label{fig:MethodOverview}}
\end{figure}

In the extrinsic case, greedy iterative decimation has proven remarkably effective, with a notable example being the \emph{quadric error metric (QEM)} of \citet{GarlandH97}.  Despite being over a quarter-century old, QEM remains the method of choice in many modern systems~\cite{nanite}.  The key insight of QEM is that one obtains useful information about global approximation error by aggregating the distortion induced by each local operation into a constant-size record at each vertex.  Our method, which we dub the \emph{intrinsic curvature error (ICE) metric}, adopts the same basic strategy, but tracks intrinsic rather than extrinsic data.  There are two geometric perspectives on this metric, developed further in \secref{ErrorMetric}:

% @keenan: I'm putting these two paragraphs in an "itemzie," because otherwise Stephen Spencer will complain that we have used the "bold paragraph hack"! ;-P
\begin{itemize}[leftmargin=*]
   \item \textbf{Local Picture.}  Whereas extrinsic methods penalize deviation of positions in space, we penalize changes in the intrinsic metric.  The local cost of removing a vertex is determined by the \emph{optimal transport cost} of redistributing its curvature to neighboring vertices.  More precisely, we compute an approximate \emph{1-Wasserstein distance} between Gaussian curvature distributions before and after flattening the removed vertex (\figref{MethodOverview}, \emph{top left}).
   \item \textbf{Global Picture.} One can also aggregate information about where error comes from in order to make better greedy decisions.  In QEM, aggregation is achieved by summing \emph{quadrics}, which approximate the squared distance to the ancestors of each vertex.  In contrast, the ICE metric tracks an approximate curvature-weighted center of mass.  More precisely, at each vertex \(i\) we track (i) the total accumulated curvature and (ii) a tangent vector \(\mathbf{t}_i\) such that the exponential map \(\exp_i(\mathbf{t}_i)\) approximates the curvature-weighted \emph{Karcher mean} of all ancestors (\figref{MethodOverview}, \emph{bottom}).
\end{itemize}
By greedily applying decimation operations that keep these tangent vectors small, we maintain a good approximation of the original intrinsic geometry---for instance, curvature does not ``drift'' from one place to another.  Moreover, unlike the extrinsic setting, we can freely change mesh connectivity without incurring any additional geometric error---making it trivial to improve element quality via tools like \emph{intrinsic Delaunay triangulation} (\secref{IntrinsicDelaunayTriangulation}).

\section{Related Work}
\label{sec:RelatedWork}

Numerous tasks in geometry processing and simulation use coarsened or hierarchical representations of geometry \cite{Garland99, GuskovSS99}, including compression (\eg{} using wavelets \cite{schroder1996wavelets, peyre2005surface} or Laplacian bases \cite{KarniG00}), surface modeling \cite{ZorinSS97,KobbeltCVS98, BotschK04}, physical simulation \cite{GrinspunKS02,progressiveCloth2022}, parameterization \cite{RayL03}, and eigendecomposition \cite{NasikunBH18, NasikunH22}.  Though the core computation in many of these tasks is inherently intrinsic, the coarsening process itself has so far been performed in the extrinsic domain.  We briefly review work on surface simplification and remeshing relevant to our approach; see \citet[Chapter 7]{BotschKPAL10} for a more detailed discussion.

\subsection{Intrinsic Triangulations}
\label{sec:RelatedWorkIntrinsicTriangulations}

As noted in \secref{Introduction}, the machinery of intrinsic triangulations is central to our approach \citet{SharpGC21}.
Early work on intrinsic triangulations explored the basic formulation~\cite{regge1961GRC} and its deep connection to \emph{Delaunay triangulations}~\cite{rivin1994ESS,indermitte2001VDP,bobenko2007discrete}.
Subsequent work on \emph{discrete uniformization}~\cite{luo2004CYF,SpringbornSP08,gu2018DUT} and an intrinsic \emph{Laplace operator} \cite{bobenko2007discrete} has recently stimulated broader applications in geometry processing~\cite{SharpSC19a,Sharp:2019:YCF,SharpC20,GillespieSC21,GillespieSC21a,Takayama22,Finnendahl2023ARAP}.  Several general-purpose data structures have been developed for intrinsic triangulations~\cite{FisherSBS06}, including those that support refinement operations~\cite{SharpSC19a,GillespieSC21a}.  However none support operations needed for intrinsic coarsening, as proposed here.

\subsection{Surface Simplification}
\label{sec:mesh_simplification}

\subsubsection{Local Decimation}
\label{sec:LocalDecimation}

A large class of coarsening methods apply incremental local decimation via vertex removal \cite{SchroederZL92}, vertex redistribution \cite{Turk92}, vertex clustering \cite{RossignacB93, LowT97, AlexaK15}, face collapse \cite{GiengHJST97}, and edge collapse strategies \cite{Hoppe96}, including QEM \cite{GarlandH97}.  Although methods based on global energy minimization can produce impressive results \cite{hoppe93,CohenSteinerAD04}, local decimation schemes remain popular since they are easy to implement, typically exhibit near-linear scaling (each decimation operation is essentially \(O(1)\)), and can easily meet an exact target vertex budget (by stopping when the budget is reached).  QEM-based simplification in particular has endured because its cheap \emph{local} heuristic yields excellent \emph{global} approximation when aggregated over many decimation operations.  Moreover, QEM is easily adapted to other attributes such as color or texture \cite{GarlandH98,Hoppe99}.

In the intrinsic setting, we likewise favor a local decimation approach because it yields fast execution times and near-linear scaling (\secref{Benchmark}), and easily incorporates rich geometric criteria (\secref{AdaptiveCoarsening}); as in QEM, aggregating information across local operations leads to high-quality global approximation (\secref{ErrorMetric}).  Unlike extrinsic, visualization-focused methods, the intrinsic setting also furnishes \emph{quality guarantees} valuable for simulation (\secref{IntrinsicRetriangulation}).

\subsubsection{Global Remeshing}
\label{sec:GlobalRemeshing}

In contrast to local decimation, which incrementally mutates the given triangulation, global remeshing methods seek only to approximate the given geometry, using an entirely new triangulation (possibly a coarser one).  Global remeshing can be performed via, \eg{}, streamline tracing \cite{Alliez:2003:IPR} or global parameterization \cite{Alliez:2002:IGR, Alliez:2005:CVD}, with significant emphasis in recent years on field-aligned methods \cite{bommes2013quad}. However a high-quality global parameterization is notoriously difficult to compute---especially if a seamless grid is needed for mesh generation \cite{Bommes:2013:IGM}.  Unlike our ICE metric (which is based on transport cost), error metrics based on local parametric distortion \cite{schmidt2019distortion} or pointwise changes in curvature \cite{EbkeSCK16} can fail to account for global tangential ``drift'' across the surface.  Moreover, parameterization-based methods do not take advantage of the flexibility of intrinsic triangulations, requiring at all times an explicit embedding into the 2D plane.  A key exception are methods based on modification of edge lengths~\cite{capouellez2022metric}, most notably the \emph{conformal equivalence of triangle meshes (CETM)} algorithm of \citet{SpringbornSP08}, which we apply locally to define our vertex removal operation (\secref{VertexFlattening}).  A more global intrinsic coarsening strategy might be to conformally map the whole surface to a high-quality \emph{cone metric}~\cite{Soliman:2018:OCS,fang2021computing,li2022computing} then remove all flat vertices; however, even just computing a good cone configuration is already more expensive than our entire algorithm.

\begin{figure}
    \begin{center}
    \includegraphics[width=1\linewidth]{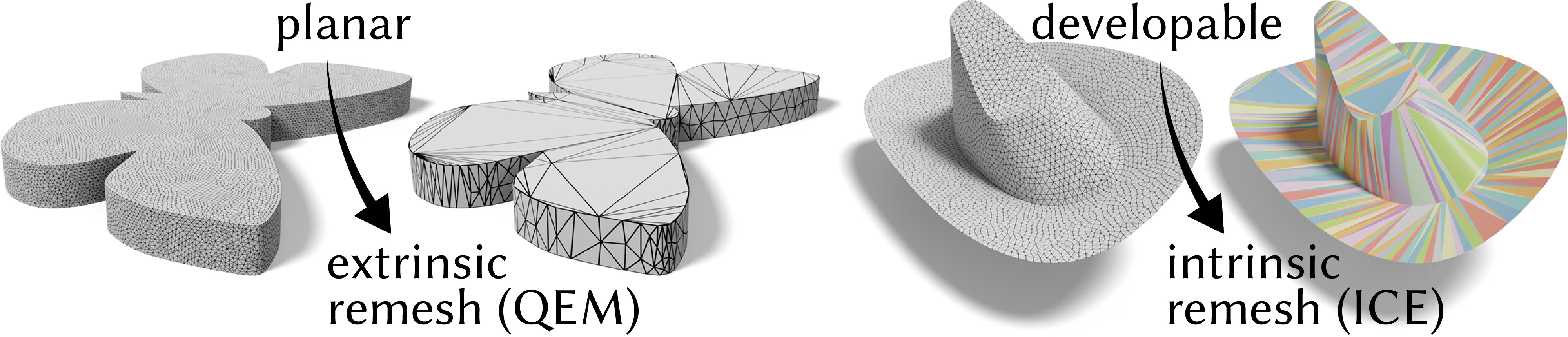}
    \end{center}
    \caption{Just as extrinsic simplification pays no cost for removing vertices from an initially flat region \emph{(left)}, our intrinsic method pays no cost for removing vertices from an initially developable region \emph{(right)}.}
    \label{fig:dev_remesh}
\end{figure}

\subsubsection{Tracking Correspondence}
\label{sec:TrackingCorrespondence}

\begin{wrapfigure}[12]{r}{80pt}
	\includegraphics[width=1\linewidth]{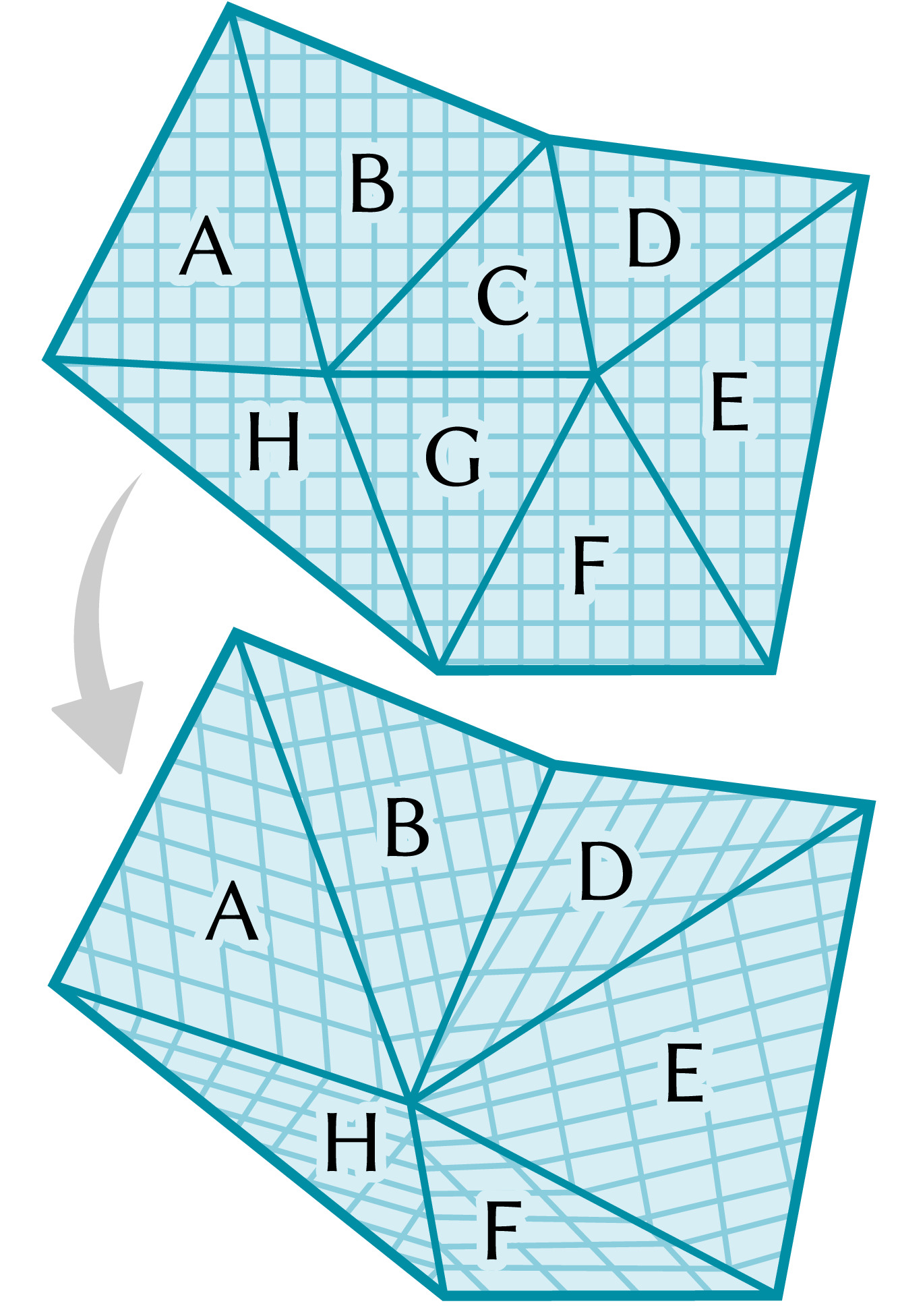}
	\label{fig:edge_onering_map}
\end{wrapfigure}

\begin{figure}
    \begin{center}
    \includegraphics[width=1\linewidth]{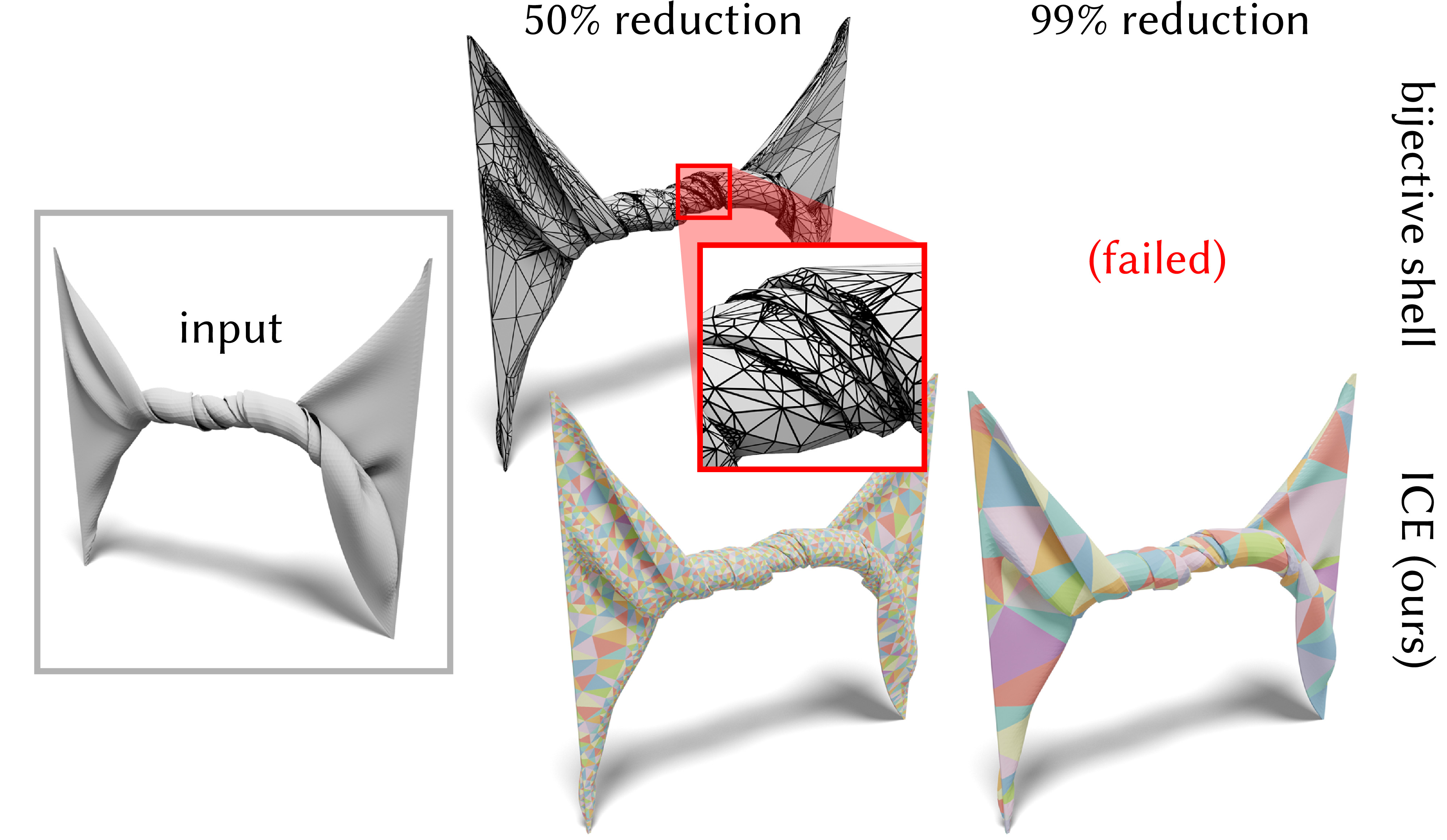}
    \end{center}
    \caption{Methods that rely on extrinsic information to construct a mapping during coarsening can fail in the presence of self-intersections.  Here, the \emph{bijective shell} method, which relies on extrinsic ray casting \cite[\S3.2]{JiangSZP20}, fails to coarsen below 50\% of the input size.  In contrast, our intrinsic approach easily obtains an extremely coarse decimation.}
    \label{fig:bijective_shell}
\end{figure}

Many applications require not only a coarse mesh, but also a way of mapping various attributes between coarse and fine meshes \cite{KharevychMOD09,LiFZ15,LiuJO19}.  Early methods compose local 2D mappings (inset) to obtain a so-called \emph{successive mapping} \cite{CohenMO97,LeeSSCD98,KhodakovskyLS03}, as more recently discussed by \citet{LiuKCAJ20, LiuZBJ21}.  Alternatively, correspondence can be defined via normal offsets \cite{GuskovVSS00}, texture domain chart boundaries \cite{SanderSGH01}, or bijective projection \cite{JiangSZP20, JiangZHSZP21}, though methods based on extrinsic correspondence can struggle in the presence of, \eg{}, mesh self-intersections (\figref{bijective_shell}).  Since our method performs both simplification \emph{and} mapping via intrinsic triangulations, we can achieve lower-distortion mappings than methods restricted to the smaller space of extrinsic mesh sequences (Figs. \ref{fig:jacobian_distortion} and \ref{fig:small_K}), in turn improving the numerical behavior of many algorithms (\secref{EvaluationAndResults}).

\subsection{Mesh Hierarchies}
\label{sec:mesh_hierarchies}

Coarsening and correspondence tracking are also key components of multi-resolution methods.  Whereas extrinsic coarsening is essential for, \eg{}, adaptive visualization~\cite{Hoppe96}, intrinsic coarsening is well-suited to multiresolution solvers such as geometric multigrid~\cite{LiuZBJ21}.  Here, \emph{coarse-to-fine} schemes, \eg{} based on subdivision surfaces~\cite{zorin2000subdivision}, yield regular connectivity and principled prolongation operators based on subdivision basis functions~\cite{GoesDMD16, ShohamVB19}. However, without careful preprocessing \cite{EckDDHLS95, HuLGCHMM22} subdivision behaves poorly on coarse, low-quality meshes encountered in the wild \cite{Zhou:2016:TDT}.  We instead adopt a robust \emph{fine-to-coarse} strategy, via repeated decimation (\secref{SurfaceMultigrid}).  Though extensively studied for extrinsic meshes, both for adaptive rendering~\cite{Hoppe96,Hoppe97,PopovicH97} and modeling/simulation~\cite{AksoyluKS05,MansonS11a,LiuZBJ21}, a fine-to-coarse hierarchy based on intrinsic triangulations enables geometric multigrid to succeed on extremely low-quality meshes where past methods fail (\secref{SurfaceMultigrid}). 

\section{Background}\label{sec:background}

An intrinsic mesh encodes geometry via edge lengths rather than vertex positions; this data is in turn sufficient to support a wide variety of surface processing and simulation tasks.  Here we give a brief review---see \citet{SharpGC21} for an in-depth introduction.

\subsection{Connectivity}\label{sec:connectivity}

\begin{wrapfigure}[5]{r}{65pt}
   \centering
   \includegraphics[width=\linewidth, trim={-1mm 4mm 0mm 0mm}]{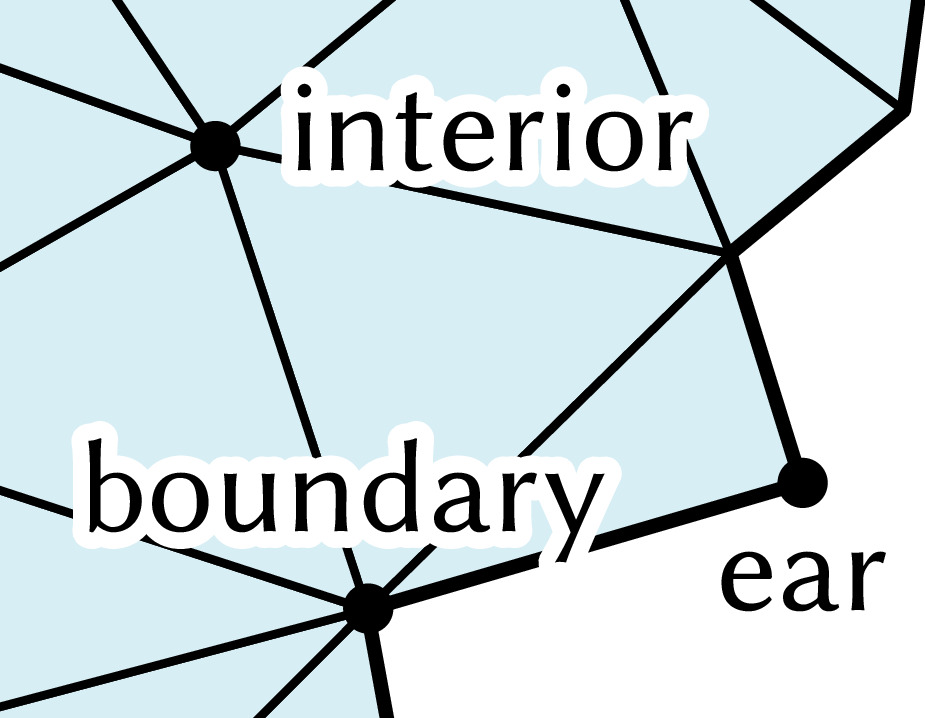}
   \label{fig:vertex_types}
\end{wrapfigure}

We consider a manifold, orientable triangle mesh \(M=(V\!, E, F)\) with vertices \(i \in V\), edges \(\ij \in E\), and faces \(\ijk \in F\).  Likewise, we write \(\after{M} = (\after{V}\!, \after{E}, \after{F})\) for any modification of \(M\).  We use \(\deg{i}\) to denote the \emph{(face) degree}, \ie, the number of faces containing \(i\).  A degree-1 boundary vertex is called an \emph{ear}; otherwise it is a \emph{regular} boundary vertex.  We use \(u_i\), \(u_{ij}\), \(u_{ijk}\) to denote a value at a vertex, edge, and face, respectively, and \(\smash{u_\jk^i}\) for a quantity at corner \(i\) of triangle \(ijk\).  Sometimes we also consider values \(u_\ij\) on \emph{oriented} edges, where \(u_\ij \neq u_\ji\).

\begin{wrapfigure}[10]{l}{53pt}
	\includegraphics[width=1\linewidth]{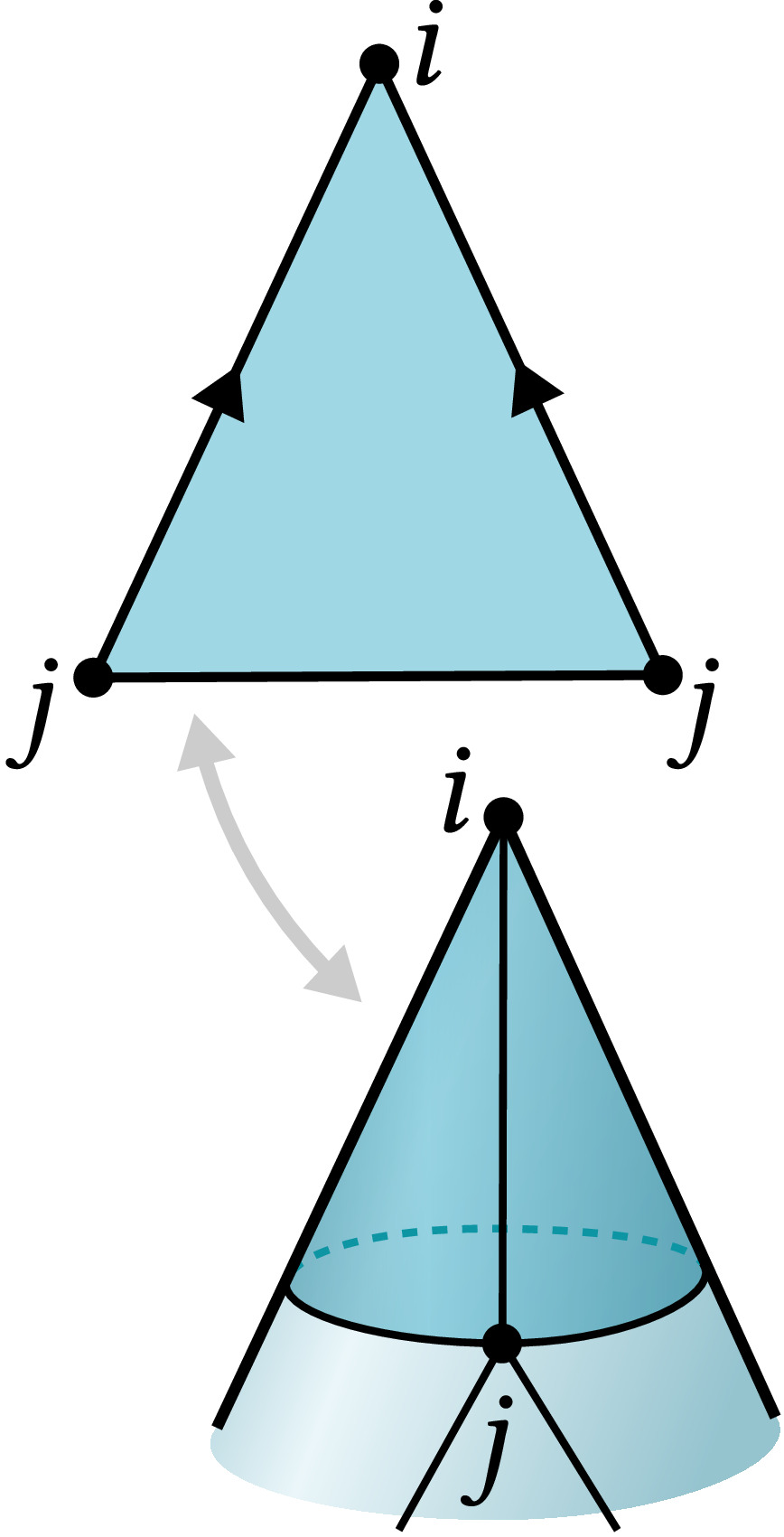}
\end{wrapfigure}
To represent the full space of intrinsic triangulations (which furnishes important algorithmic guarantees~\cite{bobenko2007discrete}), we assume that \(M\) is a \emph{\(\Delta\)-complex}~\cite[\S2.1]{Hatcher02}.  Unlike a simplicial complex, elements in a \(\Delta\)-complex are not uniquely determined by their vertices.  For instance, two edges of the same face may be glued together to form a cone (see inset).  Hence, we write \(ijk \in F\) we refer to only one of possibly several faces with vertices \(i\), \(j\), and \(k\)---which themselves need not be distinct.  Likewise, an edge \(\ij\) may connect a vertex to itself (\(i=j\)); we refer to such edges as \emph{self-edges}.  Throughout we let \(\mathcal{N}_i := \{ j \in V | \ij \in E\}\) be the neighbors of vertex \(i\), excluding \(i\) itself in the case of a self-edge \(ii\).  The connectivity of a \(\Delta\)-complex can be encoded via an \emph{edge gluing matrix} \cite[\S4.1]{SharpC20} or a \emph{halfedge mesh} \cite[Chapter 2]{BotschKPAL10}.

\subsection{Geometry}\label{sec:geometry}

\begin{figure}
   \includegraphics[width=1\linewidth]{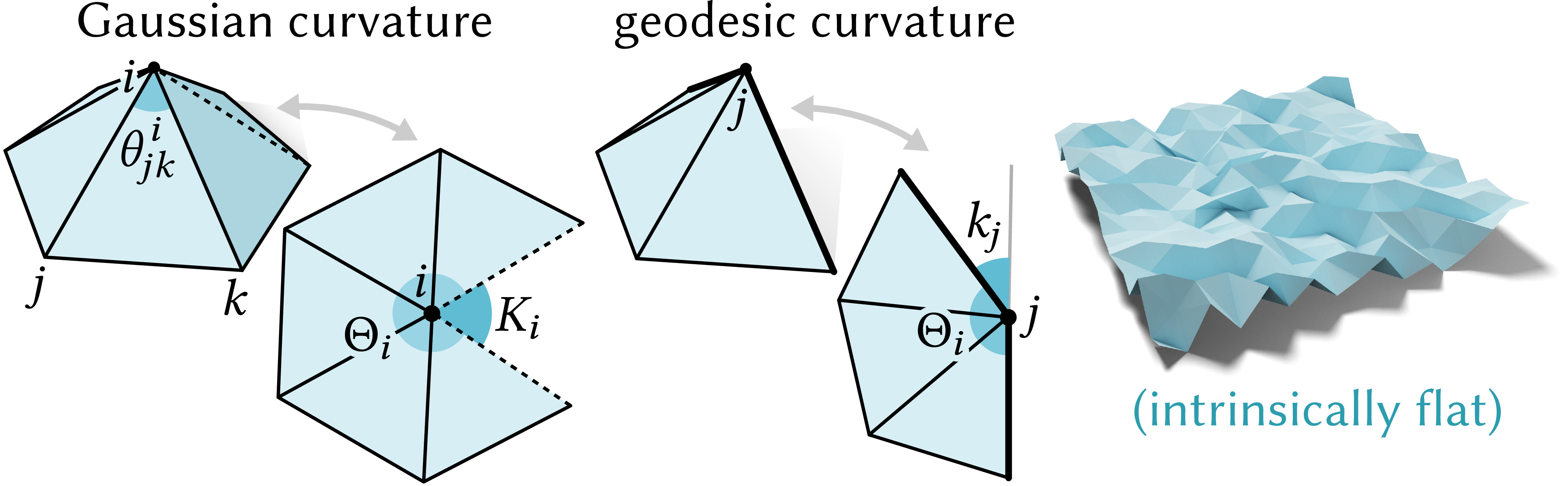}
   \caption{Gaussian curvature \(K\) and geodesic curvature \(\kappa\) depend only on corner angles \(\theta\), which in turn depend only on edge lengths \(\ell\).  Hence, surfaces that appear bent in \(\mathbb{R}^3\) may in fact have zero intrinsic curvature \emph{(right)}.\label{fig:DiscreteCurvature}}
\end{figure}

Any set of positive edge lengths \(\ell: E \to \R_{>0}\) that satisfy the triangle inequality \(\ell_{ij} + \ell_{jk} > \ell_{ki}\) at each triangle corner determines a valid intrinsic metric, i.e., a Euclidean geometry for each triangle.  We typically obtain initial edge lengths \(\ell_{ij} = \|p_i - p_j\|\) from input vertex positions \(p:V\to\smash{\R^3}\), but in principle could start with any abstract metric (e.g., coming from a \emph{cone flattening}~\cite{Soliman:2018:OCS}).  Interior angles \(\theta^i_{jk} \in (0,\pi)\) at corners can be determined from the edge lengths, via the law of cosines.  We use \(\|\mathbf{v}\|\) to denote the Euclidean norm of any vector \(\mathbf{v} \in \mathbb{R}^n\).

\subsubsection{Curvature} Intrinsically, the curvature of a surface is completely described by its Gaussian and geodesic curvature---though an intrinsically flat surface can still be extrinsically bent like a crumpled sheet of paper (\figref{DiscreteCurvature}).  On a triangle mesh, let \(\smash{\Theta_i = \sum_{\ijk \in \mathcal{N}_i} \theta_{jk}^i}\) be the angle sum around vertex \(i\).  The \emph{discrete Gaussian curvature} at interior vertex \(i\) is then given by the angle defect
\begin{equation}
  K_i := 2\pi - \Theta_i,
\end{equation}
measuring deviation from the angle sum \(2\pi\) of the flat plane.  Likewise, the \emph{discrete geodesic curvature} at boundary vertex \(i\) is
\begin{equation}
  \kappa_j := \pi - \Theta_j,
\end{equation}
measuring deviation from a straight line.

\begin{figure}
  \includegraphics[width=1\linewidth]{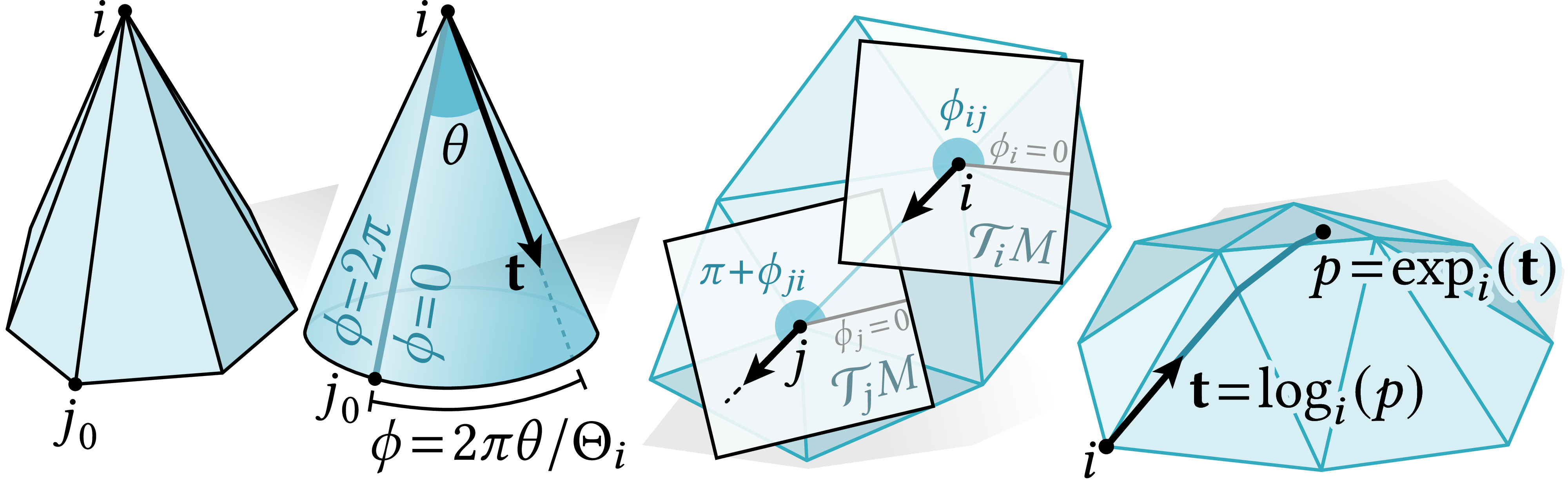}
   \caption{Notation and conventions for tangent vectors \emph{(left)}, parallel transport \emph{(center)}, and the logarithmic/exponential maps \emph{(right)}.\label{fig:vertex_tangent_space}}
\end{figure}

\subsubsection{Tangent Vectors}
\label{sec:TangentVectors}

% TODO Maybe use a letter different from t here, since t will have a special meaning later on.

In a small neighborhood of any vertex \(i\), the intrinsic metric looks like a cone of total angle \(\Theta_i\); we use \(\T_i M\) to denote the set of tangent vectors at \(i\) (\figref{vertex_tangent_space}, \emph{left}).  Following \citet[\S6]{KnoppelCPS13}, we express the direction of any tangent vector \(\vt \in \T_i M\) as a normalized angle \(\phi := 2\pi\theta/\Theta \in [0,2\pi)\), where \(\theta\) is the angle of \(\vt\) relative to an arbitrary but fixed reference edge \(ij_0\).  The vector itself is then encoded as a complex number \(\|\vt\|e^{\imath\phi} \in \mathbb{C}\), where \(\imath\) is the imaginary unit.  In particular, the \emph{angular coordinate} \(\phi_\ij \in [0, 2\pi)\) encodes the outgoing direction of an oriented edge \(\ij\); we use \(\ve_\ij \in \T_i M\) to denote the vector with direction \(\phi_{ij}\) and magnitude \(\ell_{ij}\).  The corresponding direction at vertex \(j\) is given by \(\phi_\ji + \pi\).  Hence, we can \emph{parallel transport} vectors from \(\T_i M\) to \(\T_j M\) (\figref{vertex_tangent_space}, \emph{center}) via a rotation by \(\mR_{ij} := e^{\imath((\phi_\ji + \pi) - \phi_\ij)}\) (encoded by a unit complex number).  See \citet[\S3.3 and \S5.2]{SharpSC19} for further discussion.

\subsubsection{Exponential and Logarithmic Map}
\label{sec:ExponentialAndLogarithmicMap}

The \emph{exponential map} \(\exp_i(\vt)\) of a tangent vector \(\vt\) at vertex \(i\) computes the point \(p\) reached by starting at vertex \(i\) and walking straight (\ie, along a geodesic) with initial direction \(\vt\) for a distance \(\|\vt\|\) (\figref{vertex_tangent_space}, \emph{right}).  In particular, for any oriented edge \(ij\) we have \(\exp_i(\ve_{ij}) = j\).  For a given point \(p \in M\), the \emph{logarithmic map} \(\log_i(p)\) gives the smallest tangent vector \(\vt\) at \(i\) such that \(\exp_i(\vt) = p\).  Note that, in general, \(\log_i(j)\) may not yield the edge vector \(\ve_{ij}\), since there may be a shorter path from \(i\) to \(j\). 

\subsection{Retriangulation}\label{sec:retriangulation}

\subsubsection{Intrinsic Edge Flip}
\label{sec:IntrinsicEdgeFlip}

\begin{figure}
   \includegraphics[width=1\linewidth]{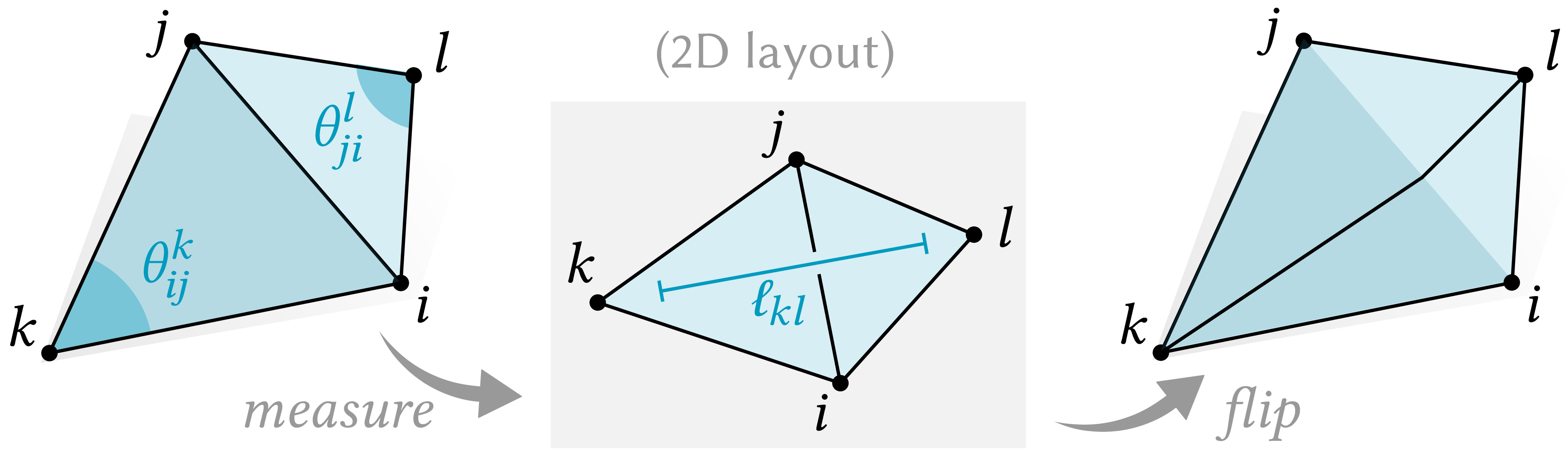}
%    \caption{\emph{Left:} an intrinsic flip of edge \(\ij\) measures the length \(\ell_{kl}\) of the opposite diagonal in a 2D layout, then updates connectivity. \emph{Right:} greedily flipping until \(\theta^k_{\ij} + \theta^l_{\ji} \leq \pi\) at all edges \(\ij\) yields an \emph{intrinsic Delaunay triangulation}.\label{fig:edge_flip}}
    \caption{An intrinsic flip of edge \(\ij\) measures the length \(\ell_{kl}\) of the opposite diagonal in a 2D layout, then updates connectivity. Greedily flipping until \(\theta^k_{\ij} + \theta^l_{\ji} \leq \pi\) at all edges \(\ij\) yields an \emph{intrinsic Delaunay triangulation}.\label{fig:edge_flip}}
\end{figure}

Consider an edge \(ij\) contained in triangles \(ijk,jil\).  An \emph{intrinsic edge flip} replaces \(ij\) with a geodesic arc along the opposite diagonal \(kl\), where the length of the new edge \(kl\) is determined via a planar unfolding (\figref{edge_flip}).  Hence, an intrinsic flip exactly preserves the original geometry---in particular, the discrete curvature is unchanged.  An edge \(ij\)  is \emph{flippable} if and only if (i) \(\deg{i},\deg{j} \geq 2\) and (ii) triangles \(ijk,jil\) form a convex quadrilateral, \ie, if both \(\theta^i_{jk} + \theta^i_{lj}\) and \(\theta^j_{ki} + \theta^j_{il}\) are less than \(\pi\) \cite[\S3.1.3]{Sharp:2019:YCF}.  Note that these conditions are considerably easier to check than in the extrinsic case \cite[App. C]{LiuKCAJ20}.

\subsubsection{Intrinsic Delaunay Triangulation}
\label{sec:IntrinsicDelaunayTriangulation}

A triangulation is \emph{intrinsic Delaunay} if it satisfies the angle sum condition \(\smash{\theta_\ij^k + \theta_\ji^l < \pi}\) at all interior edges \(\ij \in E\) (\figref{edge_flip}).  Such triangulations extend many useful properties of 2D Delaunay triangulations to surface meshes---\cite[\S4.1.1]{SharpGC21} gives a detailed list.  A triangulation can be made intrinsic Delaunay via a simple greedy algorithm: flip non-Delaunay edges until none remain~\cite{bobenko2007discrete}.

\section{Vertex Removal}
\label{sec:VertexRemoval} 

\begin{wrapfigure}{r}{37pt}
	\includegraphics[width=1\linewidth]{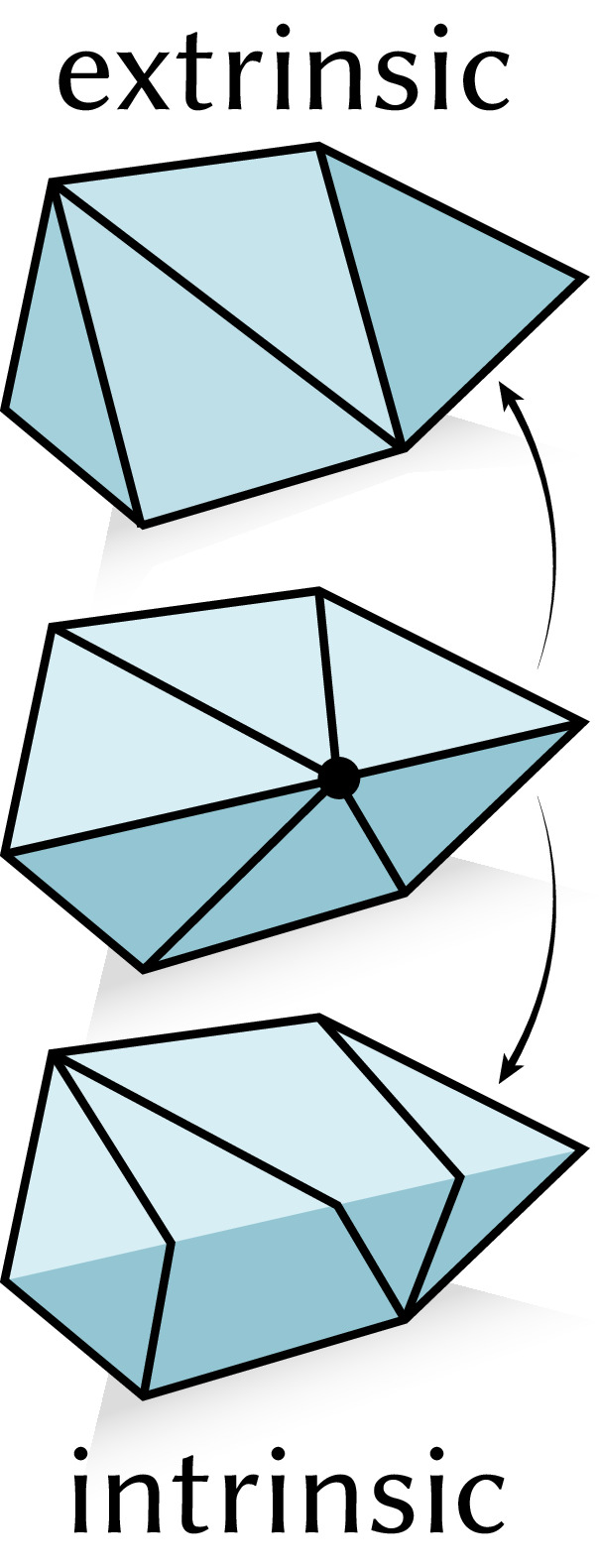}
\end{wrapfigure}
Extrinsic simplification methods reduce vertex count by making local changes to connectivity~\cite{SchroederZL92,Hoppe96,GarlandH97}.  We extend local simplification to the setting of intrinsic triangulations, using \emph{vertex removal} as our atomic operation.  Intrinsic simplification provides strictly more possibilities than its extrinsic counterpart (see inset), since any extrinsic operation can be represented intrinsically.

\begin{figure}
    \begin{center}
    \includegraphics[width=1\linewidth]{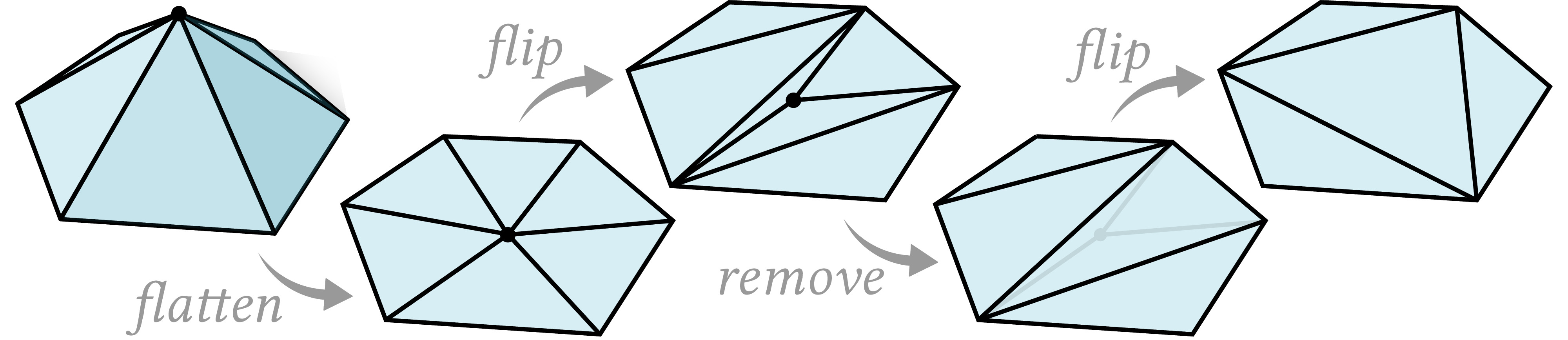}
    \end{center}
    \caption{We decimate an interior vertex by intrinsically flattening it, flipping to degree 3, removing it from the mesh, then flipping back to an intrinsic Delaunay triangulation.  (For boundary vertices, we instead flip to degree 2.)} 
    \label{fig:cetm_vertex_removal}
\end{figure}

Our method removes a vertex $i$ in three steps, illustrated in \reffig{cetm_vertex_removal}:
\begin{enumerate}
\item Intrinsically flatten \(i\) (\refsec{VertexFlattening}).
\item Remove \(i\) from the triangulation (\refsec{interior_vertex_removal}).
\item Flip to an intrinsic Delaunay triangulation (\secref{IntrinsicDelaunayTriangulation}).
\end{enumerate}
The vertex removal step extends the scheme from \cite[\S3.5]{GillespieSC21a} to boundary vertices.  Note that all changes to the geometry occur in the first step, redistributing the curvature at \(i\) to neighboring vertices \(j \in \mathcal{N}_i\). The second step merely retriangulates a flat region, and the third step performs only intrinsic edge flips.  Hence, our error metric in \secref{ErrorMetric} will need only consider the first (flattening) step to prioritize vertex removals.  Maintaining a Delaunay triangulation at each step helps ensure numerical robustness throughout simplification.

\begin{wrapfigure}[4]{r}{81pt}
	\includegraphics[width=1\linewidth]{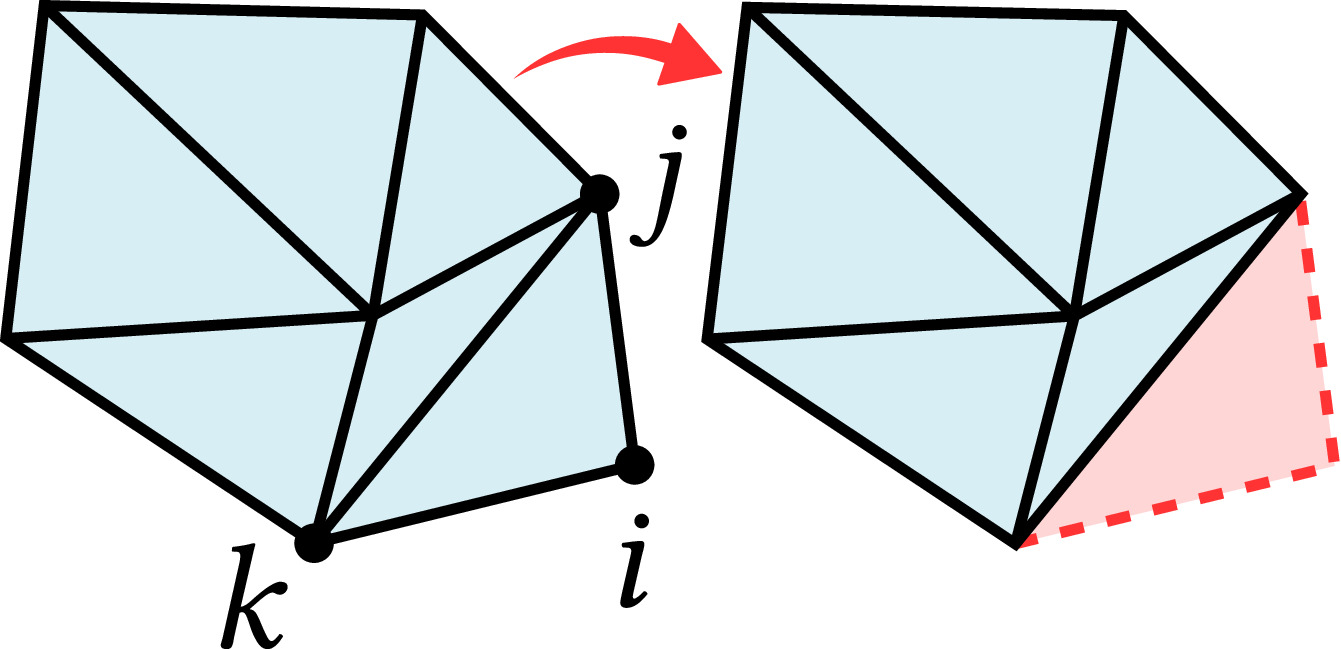}
	\label{fig:remove_ear_non_bijective}
\end{wrapfigure}
\paragraph{Special cases} To remove an ear vertex \(i\), it is tempting to simply remove the triangle \(ijk\) containing \(i\).  However, doing so leaves points on the fine mesh that do not map to any point on the coarse mesh.  Instead, we transform any ear into a regular boundary vertex by first flipping the opposite edge \(jk\).

\begin{wrapfigure}{r}{99pt}
  \includegraphics[width=1\linewidth]{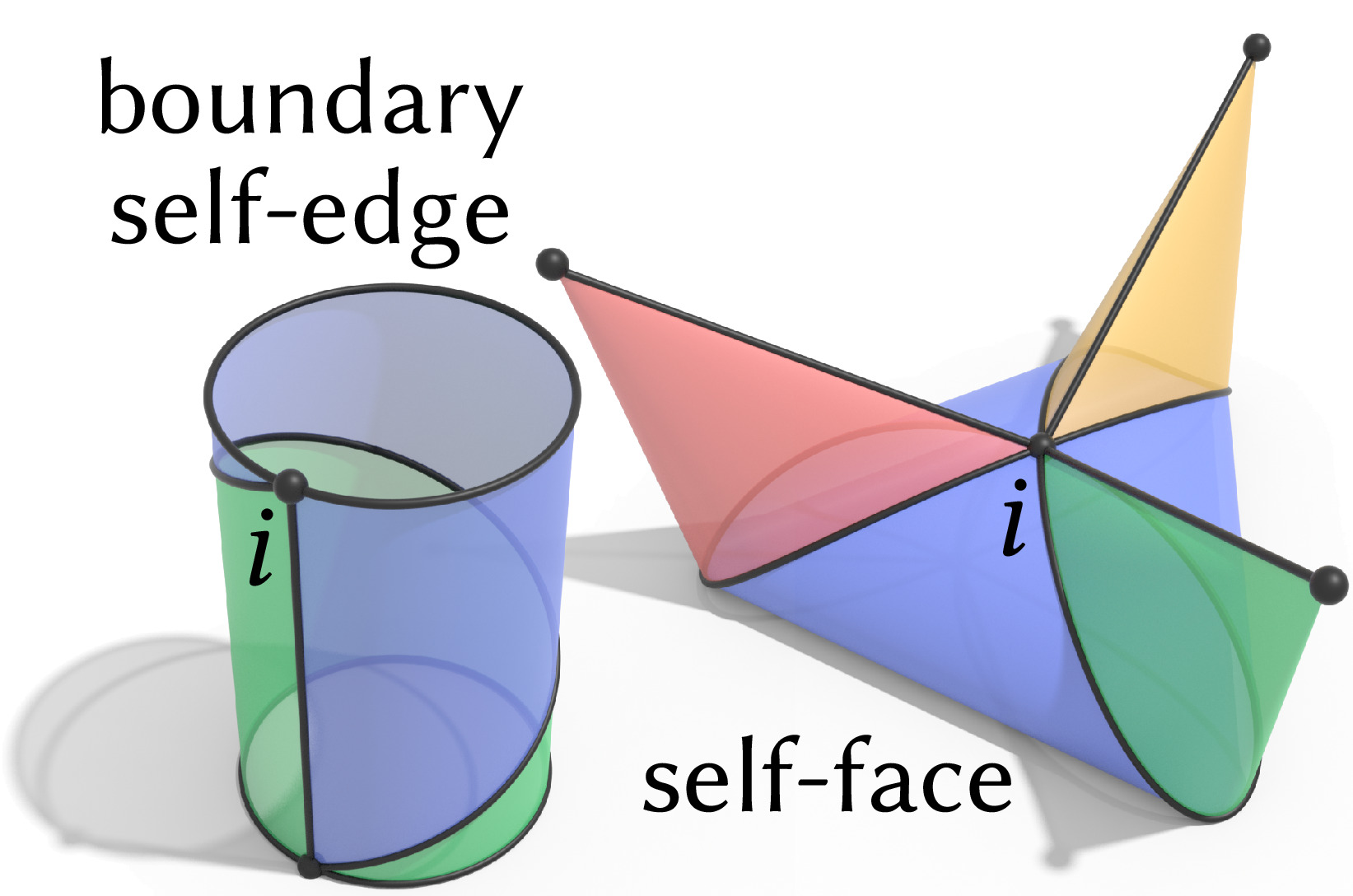}
  \label{fig:edge_cases}
\end{wrapfigure}

We cannot remove vertices \(i\) incident on a boundary self-edge, since every boundary loop must contain at least one vertex.  Likewise, vertices \(i\) of self-faces (\ie{}, triangles with only a single distinct vertex) can cause trouble for flipping, and are skipped.

\subsection{Vertex Flattening}
\label{sec:VertexFlattening}

We first eliminate all curvature at vertex \(i\).  For this operation to remain local and valid we must bijectively flatten the neighborhood \(\mathcal{N}_i\), while keeping edge lengths along the boundary of this region fixed.  \highlight{Extrinsic flattening schemes can fix boundary vertices~\cite{floater1997parametrization,Weber:2014:LIP}, but it is unclear how to construct the least-distorting boundary polygon with prescribed lengths.  In contrast, the CETM algorithm of \citet{SpringbornSP08} supports edge length constraints, and operates directly on an intrinsic triangulation.}  Moreover, using a conformal map will simplify the tangent vector prolongation scheme in \refsec{vector_field_prolongation}.

Following \citet{Luo:2004:CYF}, two sets of edge lengths \(\ell\) and \(\after \ell\) are \emph{conformally equivalent} if there exist values \(u:V \to \R\) such that
\begin{equation}
   \label{equ:conformal_equivalence}
   \after{\ell}_{\ij} = e^{(u_i + u_j)/2} \ell_{\ij}, \quad \forall ij \in E.
\end{equation}
Given initial lengths \(\ell\), CETM finds conformally equivalent edge lengths \(\tilde{\ell}\) with prescribed angle sums \(\widehat{\Theta}_i\) by minimizing a convex energy \(\mathcal{E}(u)\).

In our case, we need only determine a single scale factor \(u_i\) at the removed vertex \(i\).  We let \(\widehat{\Theta}_i = 2\pi\) for interior vertices (zero Gaussian curvature), and \(\widehat{\Theta}_i = \pi\) at regular boundary vertices (zero geodesic curvature).  Setting \(u_j = 0\) for all other vertices \(j \in \mathcal{N}_i\) ensures that the boundary lengths are unchanged---in fact, \citet[Appendix E]{SpringbornSP08} show that these boundary conditions also induce minimal area distortion.  Using the expressions for the gradient and Hessian of \(\mathcal{E}\) from \cite[Equations 9 and 10]{SpringbornSP08}, we solve the 1D root finding problem \(\nabla\mathcal{E}(u_i) = 0\) via Newton's method:  
\begin{alignat}{1}\label{equ:CETM_vertex_flattening}
   u_i \gets u_i - \frac{\widehat{\Theta}_i - \sum_{\ijk \in \mathcal{N}_i} \theta^i_{\jk}}{ \frac{1}{2}\sum_{\ijk \in \mathcal{N}_i} \cot \theta^k_{\ij} + \cot \theta^j_{\ki}}.
\end{alignat}
Notice that we express this formula as a sum over faces, so that it applies to both interior and boundary vertices.  In practice, this scheme converges in about five iterations.  Occasionally, the new edge lengths \(\after \ell\) (computed via \refequ{conformal_equivalence}) fail to satisfy the triangle inequality.  In this case we try performing edge flips, \ala{} \citet[\S3.2]{SpringbornSP08}. If these flips still fail to resolve the issue, we skip this vertex and revisit it in future coarsening iterations. 

\subsection{Flat Vertex Removal}
\label{sec:interior_vertex_removal}

\begin{wrapfigure}[15]{r}{108pt}
    \includegraphics[width=1\linewidth]{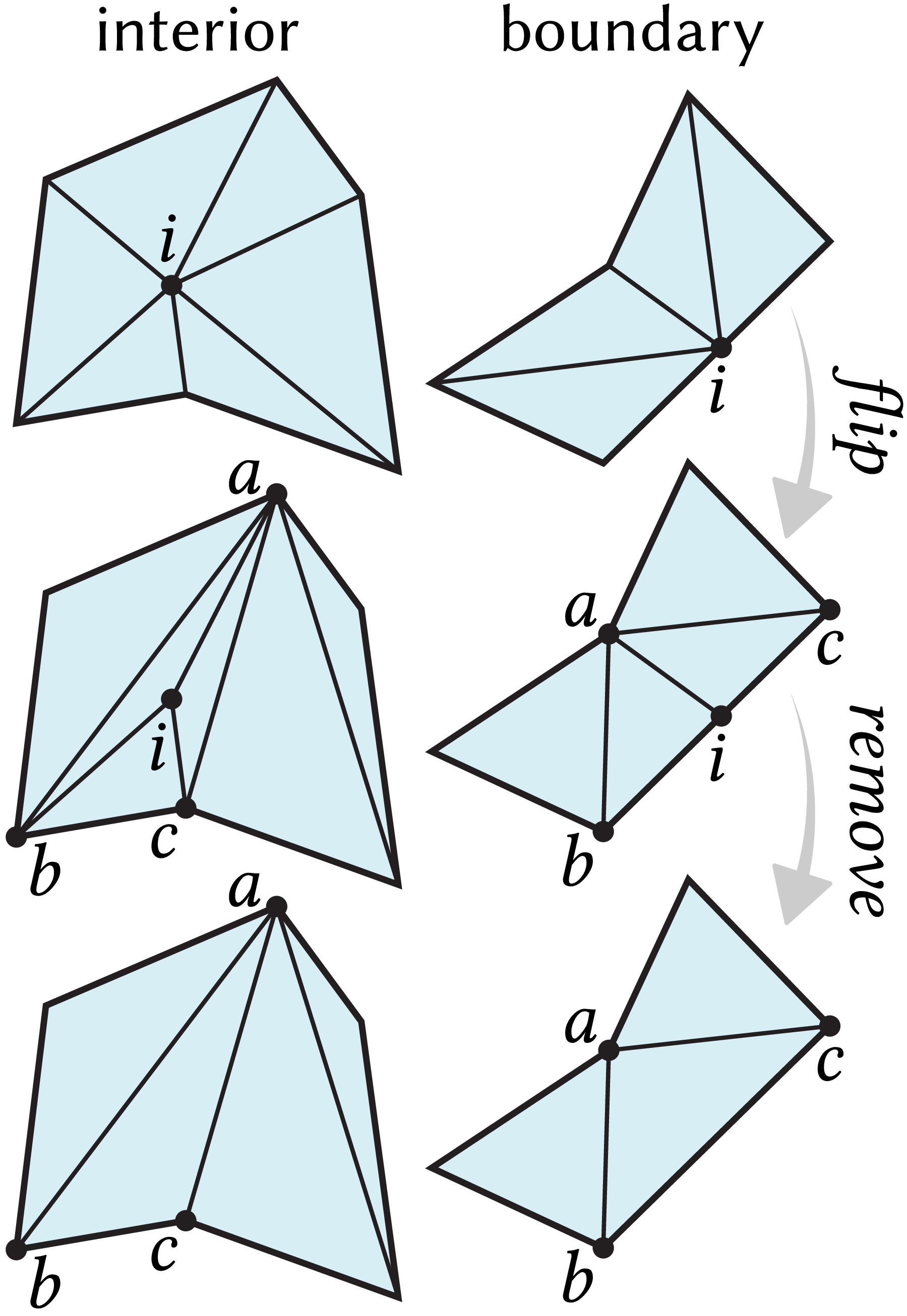}
\end{wrapfigure}
To remove a flattened vertex \(i\), we flip it to a degree-3 vertex and replace the three triangles \(iab,ibc,ica\) incident on \(i\) with the single triangle \(abc\) (inset, left).  Since the vertex neighborhood is already flat, these operations preserve the geometry.  \citet[Appendix D.1]{GillespieSC21a} show that iteratively flipping any remaining flippable edge \(ij\) incident on \(i\) will yield a degree-3 vertex, so long as the neighborhood \(\mathcal{N}_i\) remains simplicial.  Hence, at each step we first flip any self-edges (\(i=j\)); if there are none, we flip the edge \(ij\) with largest angle sum \(\theta^k_{ij} + \theta^l_{ji}\) (since only convex triangle pairs can be flipped).  In the rare case where \(\deg{i} > 3\) and no flippable edges remain, we skip this vertex removal and revert the mesh to its previous state.  If \(i\) is a boundary vertex, we perform edge flips until \(\deg{i} = 2\) and replace the two resulting triangles \(iab,ica\) with the single triangle \(abc\) (inset, right).  Here again the geometry is unchanged, since \(i\) has no geodesic curvature.  When \(i\) is an ear vertex we need only flip the opposite edge to give \(i\) degree-2, while for regular boundary vertices we use the same procedure as for interior vertices.

After removal, we must also update the angles \(\phi_{jk}\) and corresponding edge vectors \(\ve_{jk}\) for each edge \(jk\) with endpoints in \(\mathcal{N}_i\)  (\secref{TangentVectors}).  We then flip the mesh to an intrinsic Delaunay triangulation, \ala{} \secref{IntrinsicDelaunayTriangulation}.

\section{Error Metric}
\label{sec:ErrorMetric}

To prioritize vertex removals, we must define a notion of cost.  Standard extrinsic metrics, such as QEM, are not appropriate: even if they could somehow be evaluated using intrinsic data, they would attempt to preserve aspects of the geometry that are not relevant for intrinsic problems (as discussed in \secref{Introduction}).  Our method is however inspired by the remarkable effectiveness of greedy local error accumulation in QEM.  Likewise, metrics that focus on finite element equality (\ala{} \cite{Shewchuk2002WhatIA}) are not appropriate, since at intermediate steps of coarsening the triangulation used to encode the intrinsic geometry is transient and subject to change.  Standard considerations from finite element theory do however provide good justification for flipping the final triangulation to Delaunay.

Our ICE metric is instead based on two intrinsic and triangulation-independent concepts: \emph{optimal transport}~\cite{peyre2019computational}, and the \emph{Karcher mean}~\cite{karcher2014riemannian}.  Optimal transport helps quantify the effort of redistributing mass, providing the local cost for our ``memoryless'' metric (\secref{Memoryless2D}).  Karcher means encode the center of mass of all fine vertices contributing to a coarse vertex \(i\), providing the basis for our ``memory-based'' metric (\secref{memory_error_metric}).  These two pieces fit together in a natural way: after a single vertex removal, the mass-weighted norm of all error vectors \(\vt_i\) encoding Karcher means is exactly equal to the optimal transport cost.  Hence, after many vertex removals this norm approximates the cost of transporting the initial fine mass distribution to the coarsened vertices.  Vertex removals that keep cost small should hence be prioritized, since they better preserve the initial mass distribution.  Just as in QEM, this information is captured by a fixed-size representation (masses and tangent vectors at each vertex) that is easily agglomerated during coarsening.  To get a good \emph{geometric} approximation, we use curvature as our basic notion of ``mass'', but can also use other attributes such as area (\secref{IntrinsicCurvatureErrorMetric}).

\begin{wrapfigure}{r}{110pt}
   \includegraphics[width=1\linewidth]{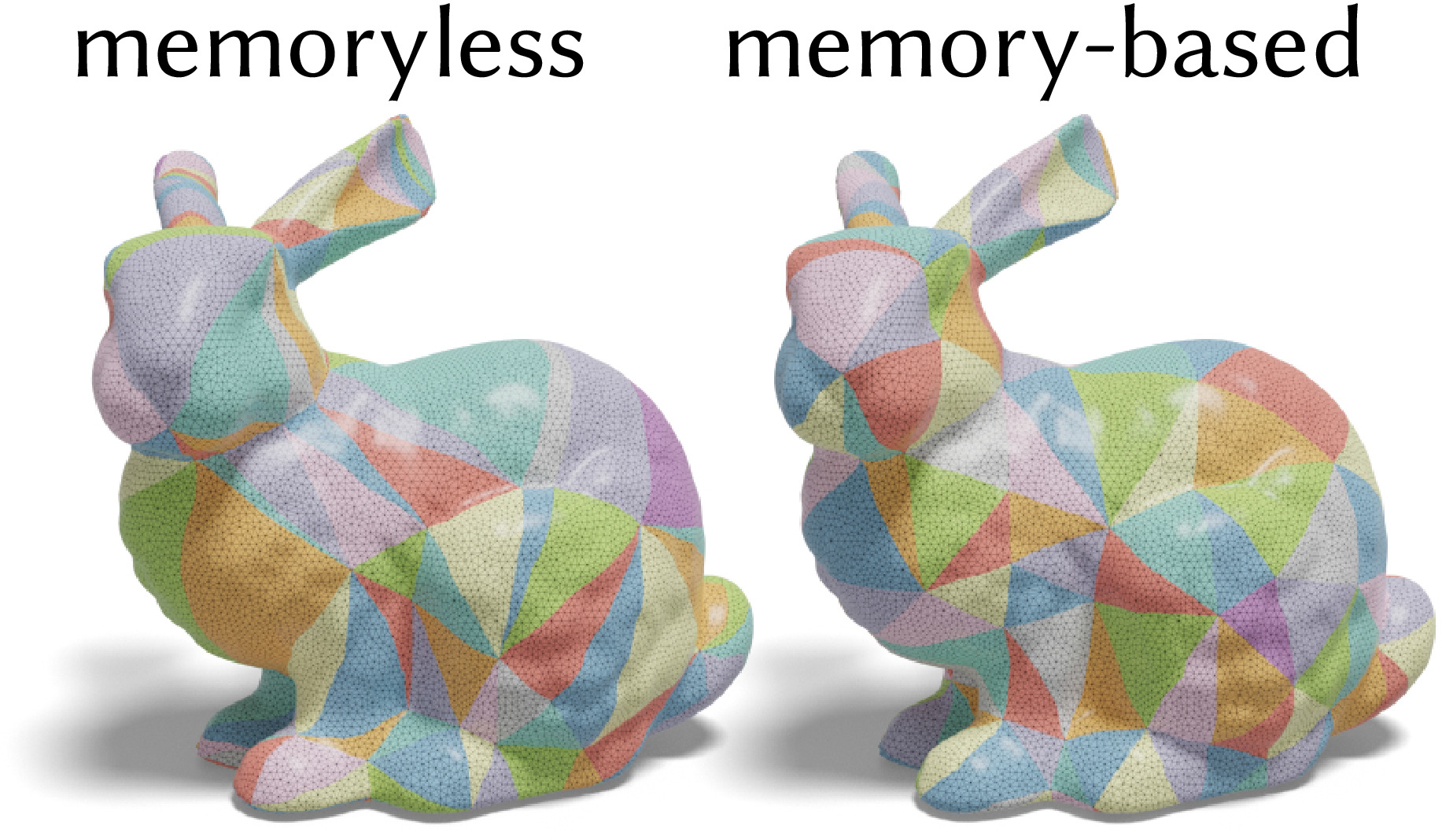}
\end{wrapfigure}
For clarity of exposition we first define error metrics in 2D, before generalizing to surfaces (\secref{memory_error_metric_3D}), and incorporating data like curvature or other attributes (\secref{IntrinsicCurvatureErrorMetric}).  Note that we view the memoryless error metric only as an intermediate step to explain the memory-based version, and use the memory-based metric for all results and experiments.  As suggested by the inset example, the memoryless metric tends to keep vertices at highly-curved points, whereas the memory-based version better distributes vertices proportional to nearby curvature.  However, there may be application contexts where the memoryless version is preferable (\eg{}, \cite{Hoppe99}).

%%%%%%%%%%%%
% ==========
%%%%%%%%%%%%
\subsection{2D Error Metric (Memoryless)}
\label{sec:Memoryless2D}

\begin{figure}[b]
    \begin{center}
    \includegraphics[width=1\linewidth]{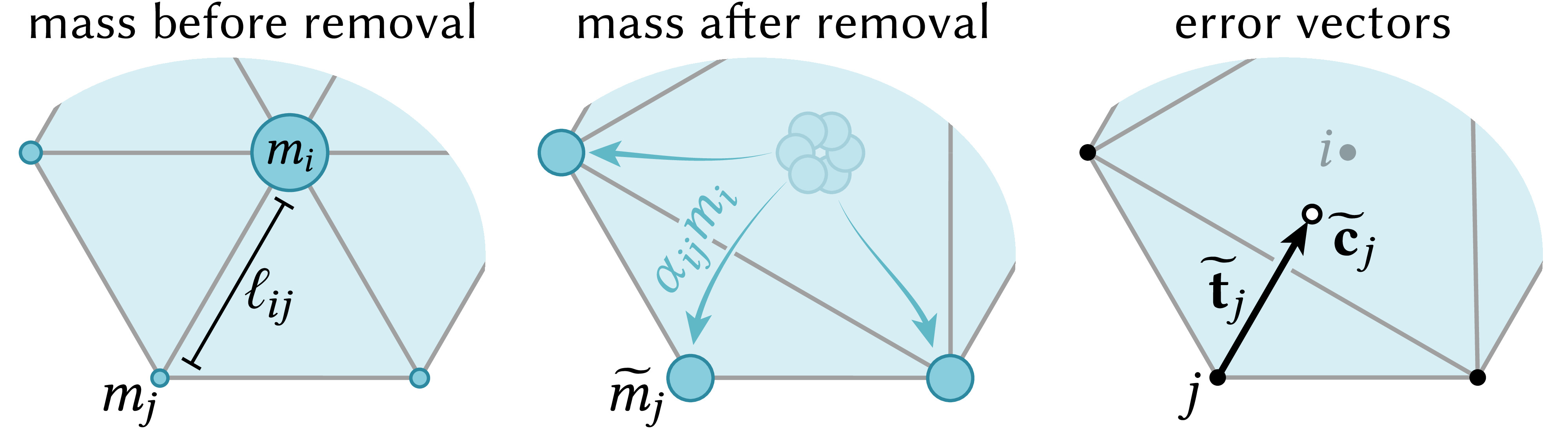}
    \end{center}
    \caption{The local cost of removing any vertex \(i\) is the \emph{optimal transport cost} of transporting its mass \(m_i\) to its neighbors \(j \in \mathcal{N}_i\).  We can also calculate this cost as the sum of new masses \(\after{m}_j\) times the length of \emph{error vectors} \(\after{t}_j\), which point to the new centers of mass \(\after{\vc}_j\).\label{fig:memoryless_transport_cost}}
\end{figure}

Consider a mass distribution \(m: V \to \mathbb{R}_{\geq 0}\) at mesh vertices, representing any nonnegative user-defined quantity (signed quantities will be addressed in \secref{IntrinsicCurvatureErrorMetric}).  Suppose we remove vertex \(i\), redistributing its mass \(m_i\) to its immediate neighbors \(j \in \mathcal{N}_i\).  In particular, let \(\alpha_{ij} \in [0,1]\) be the fraction of \(m_i\) sent to vertex \(j\) (hence \(\sum_{j \in \mathcal{N}_i} \alpha_{ij} = 1\)), so that the new mass at \(j\) is
\begin{equation}
   \label{equ:NewMass}
   \after{m}_j = m_j + \alpha_{ij}m_i.
\end{equation}

\subsubsection{Error Vectors} Suppose we want to track not only the mass distribution, but also where mass came from.  Then at each vertex \(i\) we can store an \emph{error vector} \(\vt_i\) (initially set to zero) pointing to the center of mass \(\vc_i\) of all vertices that contributed to the current value of \(m_i\).  Explicitly, after removing \(i\), the center of mass at vertex \(j\) is
\[
   \after{\vc}_j = \frac{\alpha_{ij} m_i \vx_i + m_j \vx_j}{\alpha_{ij} m_i + m_j},
\]
where \(\vx_i \in \mathbb{R}^2\) denotes the location of vertex \(i\).  Hence, the vector pointing from \(\vx_j\) to \(\after{\vc}_j\) is
\[
   \after{\vt}_j = \after{\vc}_j - \vx_j = \frac{\alpha_{ij}m_i \ve_{ji}}{\alpha_{ij}m_i + m_j},
\]
where \(\ve_{\ji} = \vx_i - \vx_j\) is the vector along edge \(ji\).  The total cost of removing \(i\) can then be measured by summing up the mass-weighted norms of these vectors.  Noting that \(\|\ve_{\ji}\| = \ell_{ij}\), we get a cost
\begin{equation}
   \label{equ:one_wasserstein_cost}
   \error_i = \sum_{j \in \mathcal{N}_i} \after{m}_j \|\after{\vt}_j\| = \sum_{j \in \mathcal{N}_i} \alpha_\ij m_i \ell_\ij.
\end{equation}
This cost also coincides with the so-called \emph{1-Wasserstein distance} between the old and new mass distribution~\cite[Chapter 2]{peyre2019computational}.  Intuitively, this distance measures the total ``effort'' of moving mass from \(i\) to neighbors \(j\), penalizing not only the amount of mass moved, but also the distance traveled.

\subsection{2D Error Metric (Memory-Based)}
\label{sec:memory_error_metric}

Rather than assign a cost to each vertex removal in isolation, we can accumulate information about how mass has been redistributed across all prior removals.  At each step, we still update the mass distribution via \refequ{NewMass}, but now update vectors encoding the centers of mass via
\begin{alignat}{1}\label{equ:memory_transport_vector}
    \after{\vt}_j = \frac{\alpha_{\ij} m_i (\vt_i + \ve_{\ji}) + m_j \vt_j}{\alpha_{\ij} m_i + m_j}.
\end{alignat}
\begin{wrapfigure}{r}{100pt}
	\includegraphics[width=1\linewidth]{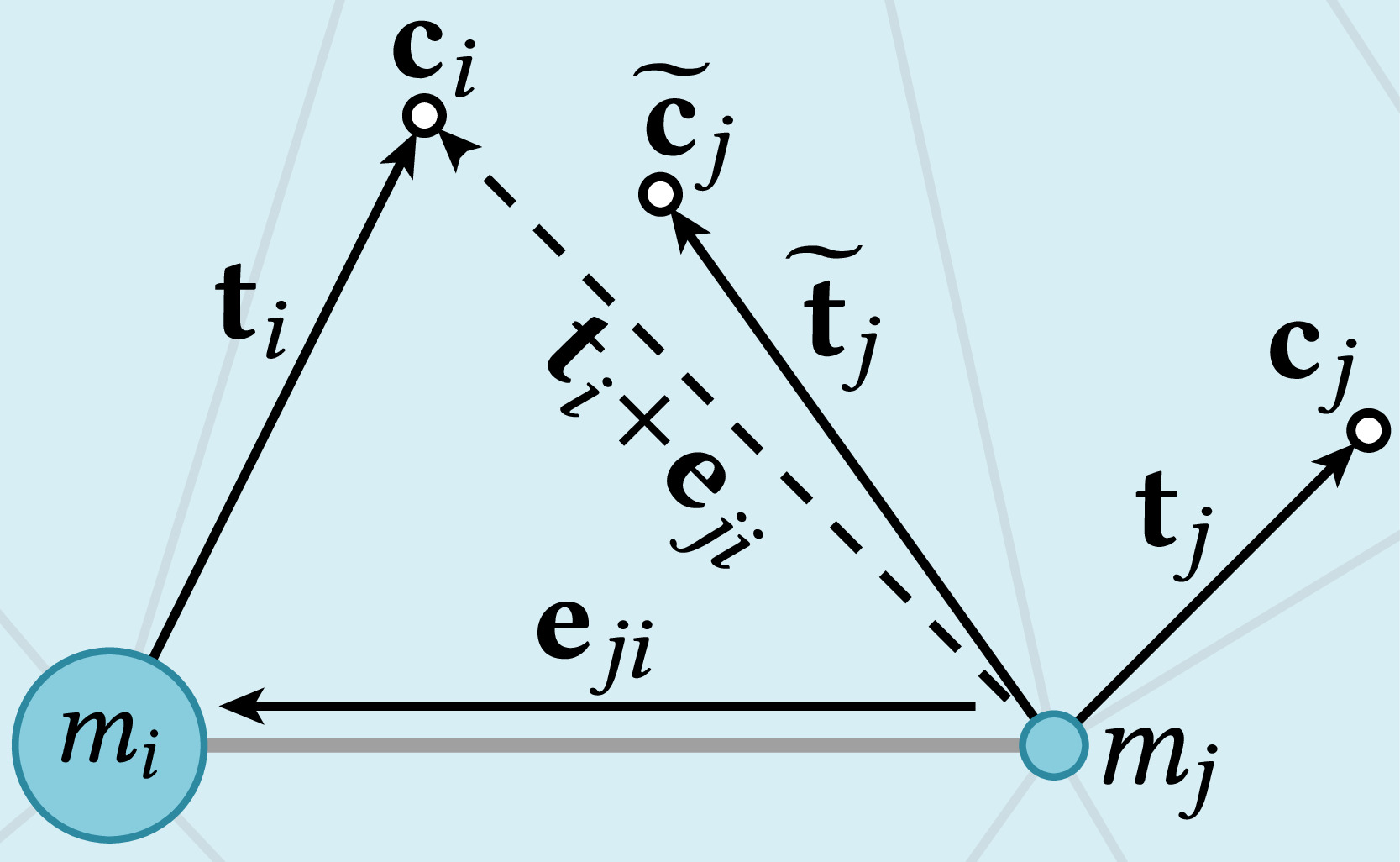}
\end{wrapfigure}
In other words, we re-express \(\vt_i\) relative to \(\vx_j\) by adding the edge vector \(\ve_\ji\), then take the mass-weighted average of the old error vector \(\vt_j\) with this new vector.  The overall cost is still evaluated via \refequ{one_wasserstein_cost}, but now approximates the effort of moving the initial mass distribution to the current one---rather than just penalizing the most recent change.  This cost is only approximate since the 1-Wasserstein distance to the center of mass is not in general equal to the distance to the original fine distribution---but it is usually quite close.  Thus, our error metric favors decimation sequences which keep each coarse vertex close to the center of all fine vertices that contribute to its mass.

\subsection{Surface Error Metric (Memory-Based)}
\label{sec:memory_error_metric_3D}

\begin{wrapfigure}{r}{111pt}
	\includegraphics[width=1\linewidth]{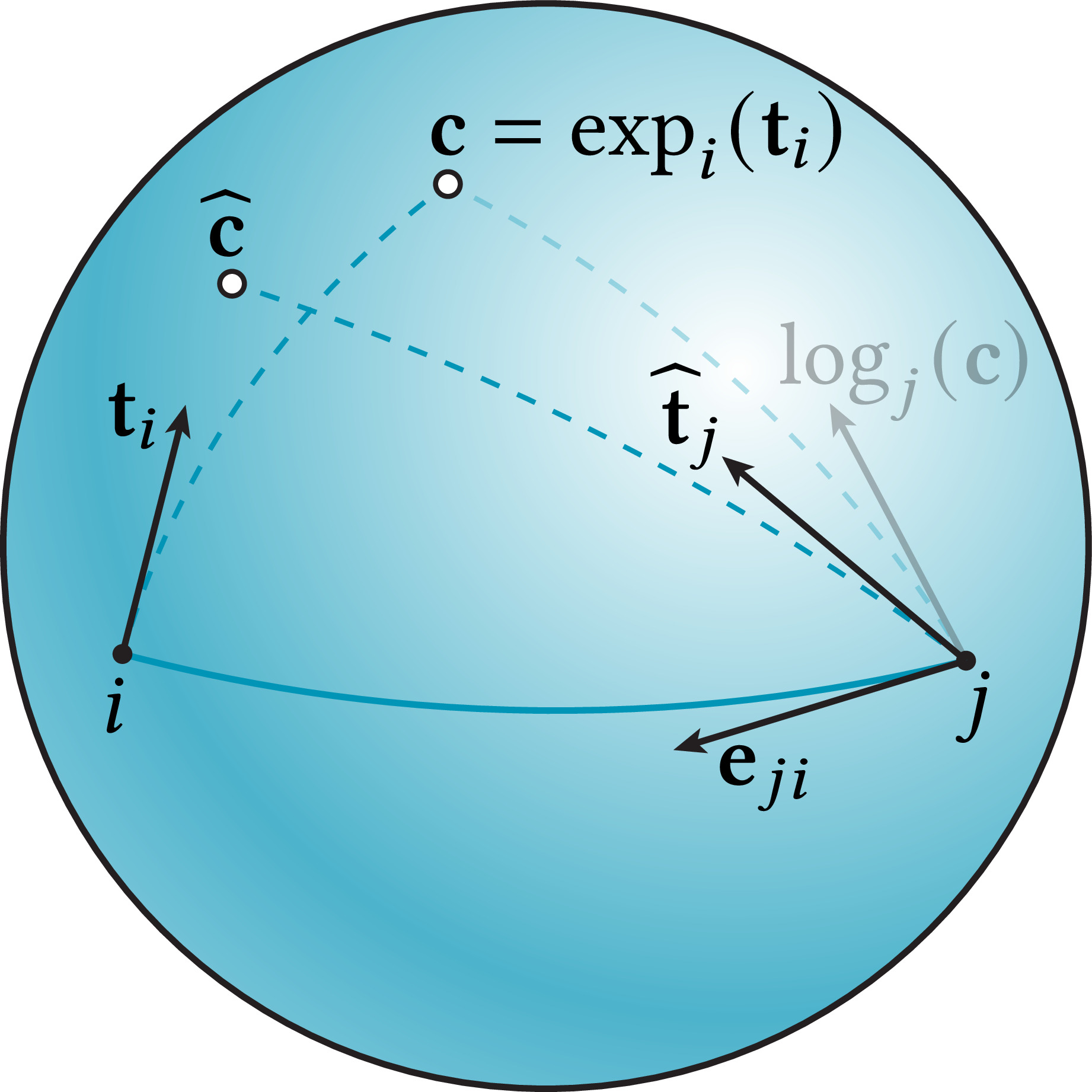}
\end{wrapfigure}
To extend this scheme to surfaces, we must address the fact that tangent vectors from different tangent spaces cannot be added directly.  In particular, we cannot re-express the error vector \(\vt_i\) at a neighboring vertex \(j\) by simply adding the edge vector \(\ve_{ji}\).  If \(\vc := \exp_i(\vt_i)\) is the center of mass encoded by \(\vt_i\), then ideally we would just compute the vector \(\log_j(\mathbf{c})\) pointing from \(j\) to \(\vc\) (see inset).  Although there are algorithms for computing the log map (\eg{}, \cite[\S8.2]{SharpSC19} and \cite[\S6.5]{Sharp:2019:YCF}), they are far too expensive to apply for each vertex removal.  Instead, we approximate this vector by parallel transporting \(\vt_i\) from \(i\) to \(j\) (\ala{} \secref{TangentVectors}) and offsetting by \(\ve_{ji}\) as in 2D, yielding a new vector \(\widehat{\vt}_j := \mR_{\ij}\vt_i + \ve_{ji}\).  The final error vector \(\widetilde{\vt}_j\) stored at \(j\) is then given by a weighted average, just as in \secref{memory_error_metric}:
\begin{alignat}{1}
   \label{equ:surface_transport_vector}
    \after{\vt}_j = \frac{\alpha_{\ij} m_i (\mR_{\ij}\vt_i + \ve_{\ji}) + m_j \vt_j}{\alpha_{\ij} m_i + m_j}.
\end{alignat}
The mass is updated as in \refequ{NewMass}, and the overall cost is again given by \refequ{one_wasserstein_cost}.  Note, then, that on curved surfaces the vector \(\after{\vt}_i\) merely approximates the center of mass of the fine mass distribution corresponding to coarse vertex \(i\).  Yet since this approximation is reasonably accurate and cheap to compute, it provides an efficient error metric akin to QEM.

\subsection{Intrinsic Curvature Error Metric}
\label{sec:IntrinsicCurvatureErrorMetric}

\begin{figure}
    \begin{center}
    \includegraphics[width=1\linewidth]{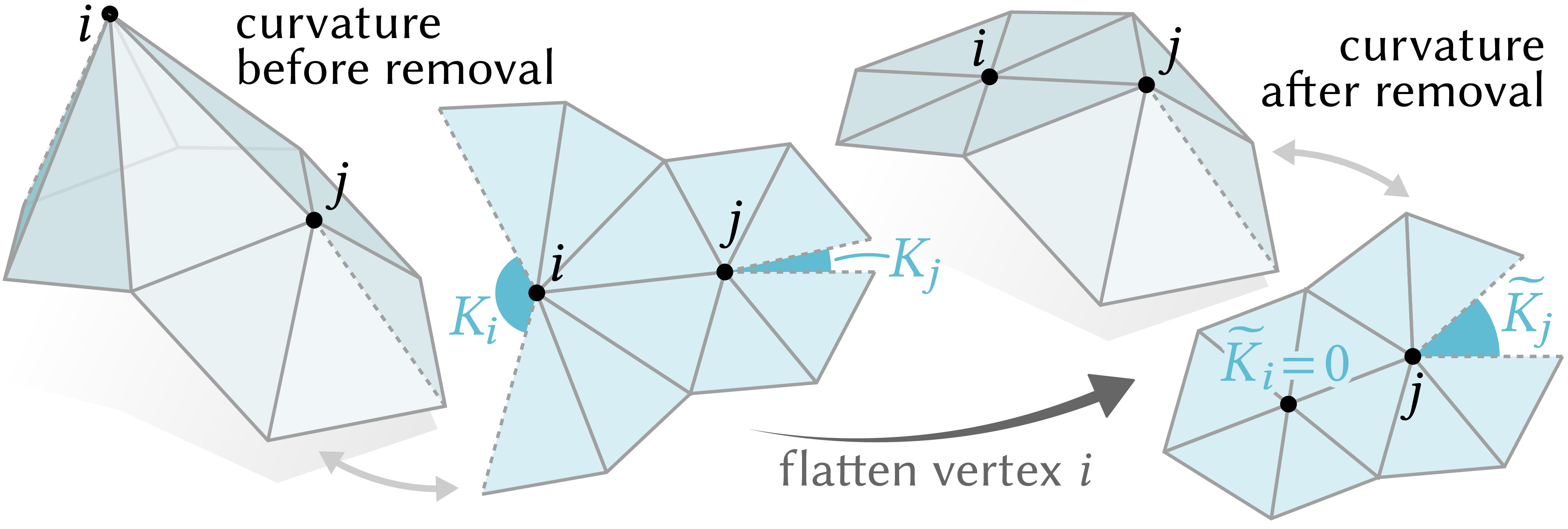}
    \end{center}
    \caption{Flattening a vertex \(i\) changes the angle sums \(\Theta\) at neighboring vertices \(j\), effectively redistributing the discrete curvature \(K = 2\pi - \Theta\).  We use the change in curvature from \(K\) to \(\after{K}\) to guide simplification.}
    \label{fig:curvature_distribution}
\end{figure} 

Due to the Gauss-Bonnet theorem, flattening a vertex \(i\) (\ala{} \refsec{VertexFlattening}) conservatively redistributes curvature to neighboring vertices \(j\), making curvature a natural ``mass'' distribution to guide simplification.  A challenge here is that the old and new curvatures \(K\) and \(\after{K}\) are not in general positive quantities.  One possibility might be to use a transport cost for signed measures such as~\citep{mainini2012description}, but doing so would require us to solve a small optimal transport problem for each vertex removal.  We instead adopt a cheap alternative.  In particular, we define convex weights
\begin{alignat}{1}
   \label{equ:transfer_proportion}
   \alpha_{\ij} := \frac{|\after{K}_j - K_j|}{\sum_{l \in \mathcal{N}_i} |\after{K}_l - K_l|}.
\end{alignat}
For boundary vertices we use the same formula, but replace Gaussian curvature \(K\) with geodesic curvature \(\kappa\).  If vertex \(i\) is already flat prior to removal, then there is no change in curvature and we simply distribute mass equally to all neighbors.  We then split the initial fine curvature function \(K\) (or \(\kappa\)) into two positive mass functions \(K_i^+ := \max(K_i,0)\) and \(K_i^- := -\min(K_i,0)\).  Each of these quantities is tracked throughout simplification exactly like \(m_i\) in \secref{memory_error_metric_3D}, using two separate vectors \(\vt_i^+\) and \(\vt_i^-\) (\resp{}), and weights \(\alpha\) from \refequ{transfer_proportion}.  The overall error, which defines the ICE metric, is then the sum of the errors in the two curvature functions (\ala{} \refequ{one_wasserstein_cost}).  Note that if a vertex \(i\) cannot be flattened or removed, we assign it an infinite cost (which may later get updated to a finite value when its neighbors are removed---see \refalg{coarsening}).

\subsubsection{Auxiliary Data}
\label{sec:AuxiliaryData}

\begin{figure*}
    \begin{center}
    \includegraphics[width=1\linewidth]{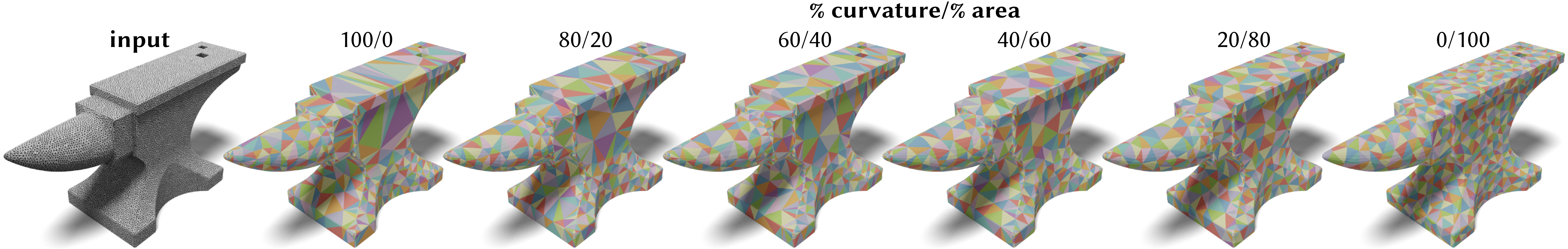}
    \end{center}
    \caption{We can mix and match different quantities to guide coarsening. Here, for instance, strongly weighting Gaussian curvature emphasizes preservation of intrinsic geometry, whereas strongly weighting area prioritizes uniform triangle size.}
    \label{fig:different_mixture_weights}
\end{figure*}

Similar to \cite{GarlandH98}, other quantities at vertices (areas, colors, \etc{}) can be used to drive simplification in an analogous fashion: each signed quantity is split into two positive mass functions, and a list of all ``channels'' \(m^0, \ldots, m^k\) is tracked along with associated tangent vectors \(\vt^0, \ldots, \vt^k\).  The cost is then given by
\[
   C_i = \sum_{j \in \mathcal{N}_i} \sum_{p=0}^k w^p m_i^p \|\vt^p_j\|,
\]
where a choice of weights \(w^1, \ldots, w^k \in \mathbb{R}_{\geq 0}\) puts an emphasis on different features.  For instance, \figref{different_mixture_weights} shows the impact of different weightings on curvature versus area.

\section{Simplification}
\label{sec:Simplification}

We now have all the ingredients to perform intrinsic simplification.  Just as in QEM, we first initialize a priority queue by evaluating the ICE metric at all vertices (\secref{Initialization}), then greedily remove the lowest-cost vertex from this queue until we reach a target vertex count \(n\) (\secref{Coarsening}), or until no more vertices can be removed.  \refalg{coarsening} provides pseudocode; a full reference implementation can be found at \url{https://github.com/HTDerekLiu/intrinsic-simplification}.

\subsection{Initialization}
\label{sec:Initialization}

For each vertex \(i \in V\) we compute the initial masses \(m_i\) (\eg{}, the curvature functions \(K^+\) and \(K^-\) from \secref{IntrinsicCurvatureErrorMetric}) and an initial error vector \(\vt_i = 0\).  To compute the cost of removing \(i\) we perform a tentative flattening \ala{} \refsec{VertexFlattening} and use the resulting weights \(\alpha_{ij}\) from \refequ{transfer_proportion} to evaluate the cost \(C_i\) via the second sum in \refequ{one_wasserstein_cost}.  If flattening yields invalid edge lengths, or vertex \(i\) cannot be flipped to degree 3 (or 2 on the boundary), we let \(C_i = \infty\).  After evaluating the cost function, we ``undo'' the tentative removal, \ie, we restore the previous connectivity and revert any changes to edge lengths.

\subsection{Coarsening}
\label{sec:Coarsening}

At each iteration, we pick the vertex \(i\) with the minimum cost \(C_i\) from our priority queue.  If \(C_i = \infty\), then no more vertices can be removed and we terminate.  Otherwise, we apply the removal procedure from \refsec{VertexRemoval}.  The resulting weights \(\alpha_{\ij}\) (\refequ{transfer_proportion}) are used to compute new masses \(\after{m}_j\) (\refequ{NewMass}) and updated transport vectors \(\after{\vt}_j\) (\refequ{surface_transport_vector}) for each neighbor \(j \in \mathcal{N}\).  We then flip the mesh back to intrinsic Delaunay \ala{} \secref{IntrinsicDelaunayTriangulation}---note that to initialize the greedy flipping algorithm we need only enqueue edges in \(\mathcal{N}_i\), since the mesh was already Delaunay prior to removing vertex \(i\).  Finally, we must also update the priority queue with new costs \(C_j\) by tentatively flattening each neighbor \(j\), and evaluating the first sum in \refequ{one_wasserstein_cost} (this time over neighbors \(k \in \mathcal{N}_j\)).  Here, finite costs may become infinite (or vice versa), since vertices that were previously removable may no longer be removable.

\IncMargin{1em}
\begin{algorithm}[t]
  \SetKwInOut{Input}{Input}
  \SetKwInOut{Output}{Output}
  \SetKwInOut{Parameter}{Param.}
    \caption{Intrinsic Coarsening}  
    \label{alg:coarsening}
    \Indentp{-1em}
    \Input{$\ M, \ell, n$ \rComment{mesh, edge lengths, target vertex count}}
    \Output{$\ \widetilde{M}, \tilde{\ell}$ \rComment{coarsened mesh \& edge lengths}}
    \Indentp{1.3 em}
    \lComment{initialization}\\
    $Q \gets \textsc{EmptyPriorityQueue}()$ \\
    $M, \ell \gets \textsc{FlipToDelaunay}(M, \ell)$ \rComment{\refsec{retriangulation}}\\
    \ForEach {vertex $i \in \widetilde{V}$}
    {
        $(m_i,\vt_i) \gets ((K^-_i,K^+_i),0)$ \rComment{initial mass \& error}
        $c \gets \textsc{IntrinsicCurvatureError}(M, \ell, i)$ \rComment{\refsec{IntrinsicCurvatureErrorMetric}} \\
        $\textsc{Enqueue}(Q, i, c)$ \\
    }
    \lComment{coarsening}\\
    \While {\normalfont{\textsc{VertexCount}($M$) > $n$ \textbf{and}} $!\textsc{Empty}(Q)$}
    {
        $i \gets \textsc{Pop}(Q)$ \rComment{extract minimum-cost vertex} \\
        $\ell \gets \textsc{Flatten}(M, \ell, i)$ \rComment{\refsec{VertexFlattening}} \\
        $M, \ell, m, \vt \gets \textsc{RemoveVertex}(M, \ell, i)$ \rComment{Sec.\ \ref{sec:interior_vertex_removal}, \ref{sec:memory_error_metric_3D}} \\
        $M, \ell \gets \textsc{FlipToDelaunay}(M, \ell, i)$ \\
        \ForEach{vertex $j \in \mathcal{N}_i$} 
        {
           $c \gets \textsc{IntrinsicCurvatureError}(M, \ell, j)$\\
           $Q \gets \textsc{UpdatePriority}(Q, j, c)$ \\
        }
    }
    \Return $(M,\ell)$
\end{algorithm}
\DecMargin{1em}

\subsection{Intrinsic Retriangulation}
\label{sec:IntrinsicRetriangulation}

\begin{figure}
    \begin{center}
    \includegraphics[width=1\linewidth]{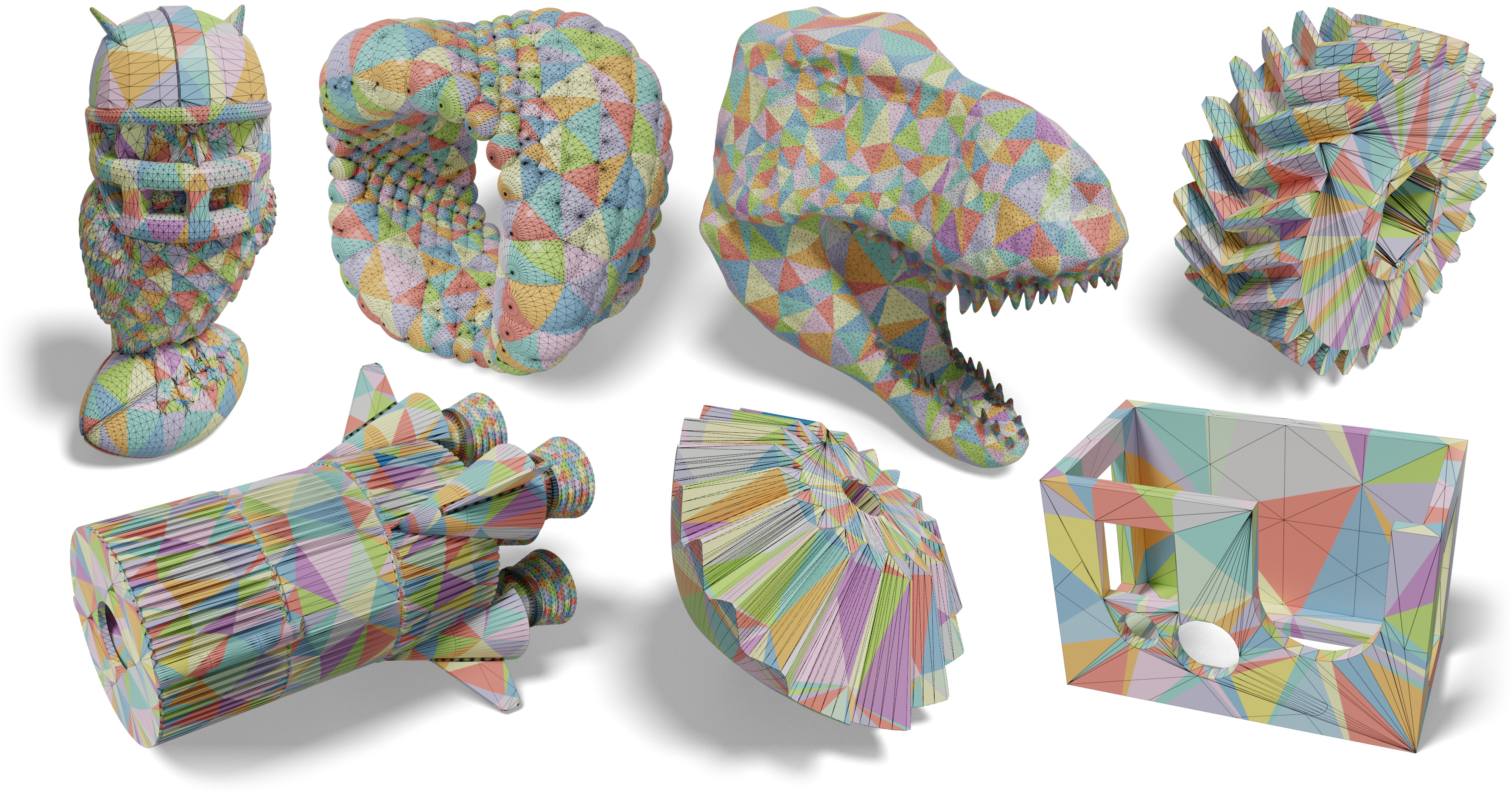}
    \end{center}
    \caption{Adding vertices to the mesh prior to coarsening provides more options for simplification and remeshing, since coarse vertices \(\after V\) no longer need be a subset of fine vertices \(V\).}
    \label{fig:remeshing}
\end{figure}

Though our overall goal is to coarsen the input, carefully \emph{adding} vertices to the mesh via \emph{intrinsic Delaunay refinement}~\citep[Section 4.2]{SharpSC19a} can be quite valuable in two distinct ways.  On typical models, this pre- or post-processing adds only a fraction of a second to overall execution time~\citep[Section 6]{SharpSC19a}.

\begin{figure}
    \begin{center}
    \includegraphics[width=1\linewidth]{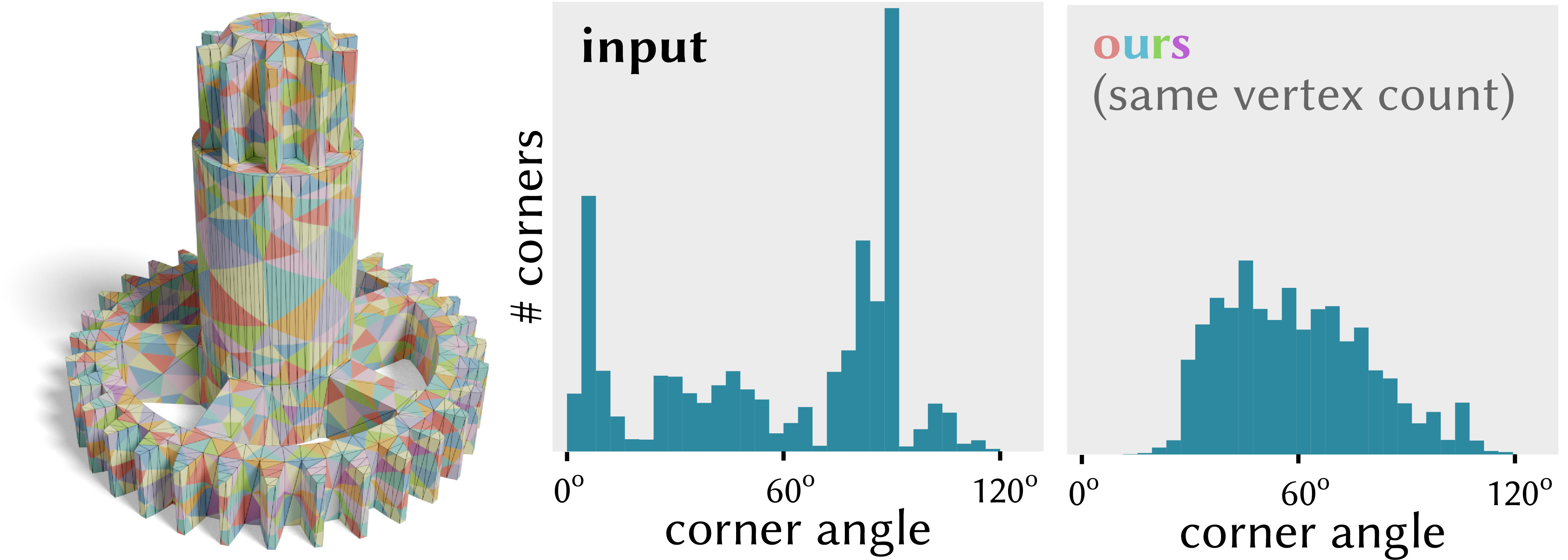}
    \end{center}
    \caption{Even for an identical vertex count, refining then coarsening yields higher quality elements, as quantified by histograms of corner angles \(\theta^i_{jk}\).}
    \label{fig:remesh_better_quality}
\end{figure}

\paragraph{Pre-refinement} As stated, our coarsening algorithm (\refalg{coarsening}) can produce only meshes whose vertices \(V\) are a subset of the input vertices \(\after{V}\).  However, we can often obtain a smaller or more accurate representation of the original metric by first inserting a larger set of candidate vertices, prior to simplification.  For instance, in \figref{remeshing} we run Delaunay refinement on the input until all corner angles \(\smash{\theta^i_{jk}}\) are no smaller than \(25^\circ\).  As a result, it becomes much easier to construct high-quality coarse triangles in regions with few input vertices (\eg{}, on the rocket model).  \figref{remesh_better_quality} shows how this strategy can even improve quality on coarse input models, where we do not seek to reduce the vertex count.

\paragraph{Post-refinement} Since coarsening maintains an intrinsic Delaunay triangulation, the final mesh is already guaranteed to exhibit properties valuable for applications---such as positive edge weights for the discrete Laplacian~\citep{bobenko2007discrete}.  As an optional post-process, we can also provide hard guarantees on element \emph{quality}: as recently proven by \citet{GillespieSC21a}, intrinsic Delaunay refinement yields a triangulation where the smallest corner angle \(\theta_i^{jk}\) is no smaller than \(30^\circ\) (hence no greater than \(120^\circ\)), so long as (i) \(M\) is closed and (ii) \(\Theta_i \geq 60^\circ\) for all input vertices \(i \in V\).  In practical terms, this guarantee ensures that even very low-quality input meshes can be used successfully in numerical algorithms such as those explored in \secref{EvaluationAndResults}.  This feature is unique to the intrinsic setting: extrinsic surface meshing algorithms like \emph{restricted Delaunay refinement} provide guarantees on global geometry and topology, but not on element quality~\citep[Chapter 13]{cheng2012delaunay}, nor even positive weights for the Laplacian.  The only danger is that the final triangulation is no longer guaranteed to be coarser than the input mesh, though in practice this situation is unlikely to occur unless the input is already quite coarse.

\subsection{Numerics}
\label{sec:Numerics}

As in any mesh processing algorithm, near-degenerate triangles (whether in the input or constructed during simplification) can cause numerical issues due to floating point error.  We hence follow best practices wherever possible~\cite{shewchuk1999lecture}, \eg{}, to determine if a point \(p\) is inside a triangle during pointwise mapping (\secref{PointwiseMapping}) we compute the sign of \(\det(p-x_i, p-x_j)\) for each triangle edge \(ij\), rather than directly computing barycentric coordinates (thereby avoiding division by area).

\section{Mapping and Prolongation}
\label{sec:MappingandProlongation}

For some tasks (say, computing Laplacian eigenvalues) the coarsened triangulation can be used directly; more broadly we need some way of evaluating correspondence between the coarse and fine mesh.  Here we consider two basic viewpoints: correspondence of \emph{points} (\secref{PointwiseMapping}) and correspondence of \emph{functions} (\secref{Prolongation}).

\subsection{Pointwise Mapping}
\label{sec:PointwiseMapping}

To map any point \(p\) on the fine mesh to a point \(\after{\prime}\) on the coarse mesh, we track its barycentric coordinates through local coarsening operations (namely: edge flips, vertex flattenings, and vertex removals).  This map is trivially bijective, since at each step we simply re-write the given barycentric coordinates with respect to a different triangulation of the same planar region.  The only way to violate bijectivity would be to perform a non-bijective vertex flattening---which we explicitly forbid (see \secref{VertexFlattening}).  In the applications we consider (\secref{EvaluationAndResults}), all points \(p\) that must be tracked are known ahead of time, and can be tracked during simplification.  To evaluate this map on demand, one could record the list of local operations, and ``re-play'' these operations for each new query point, as in \cite{LiuKCAJ20}.

\begin{wrapfigure}[6]{r}{80pt}
	\includegraphics[width=1\linewidth]{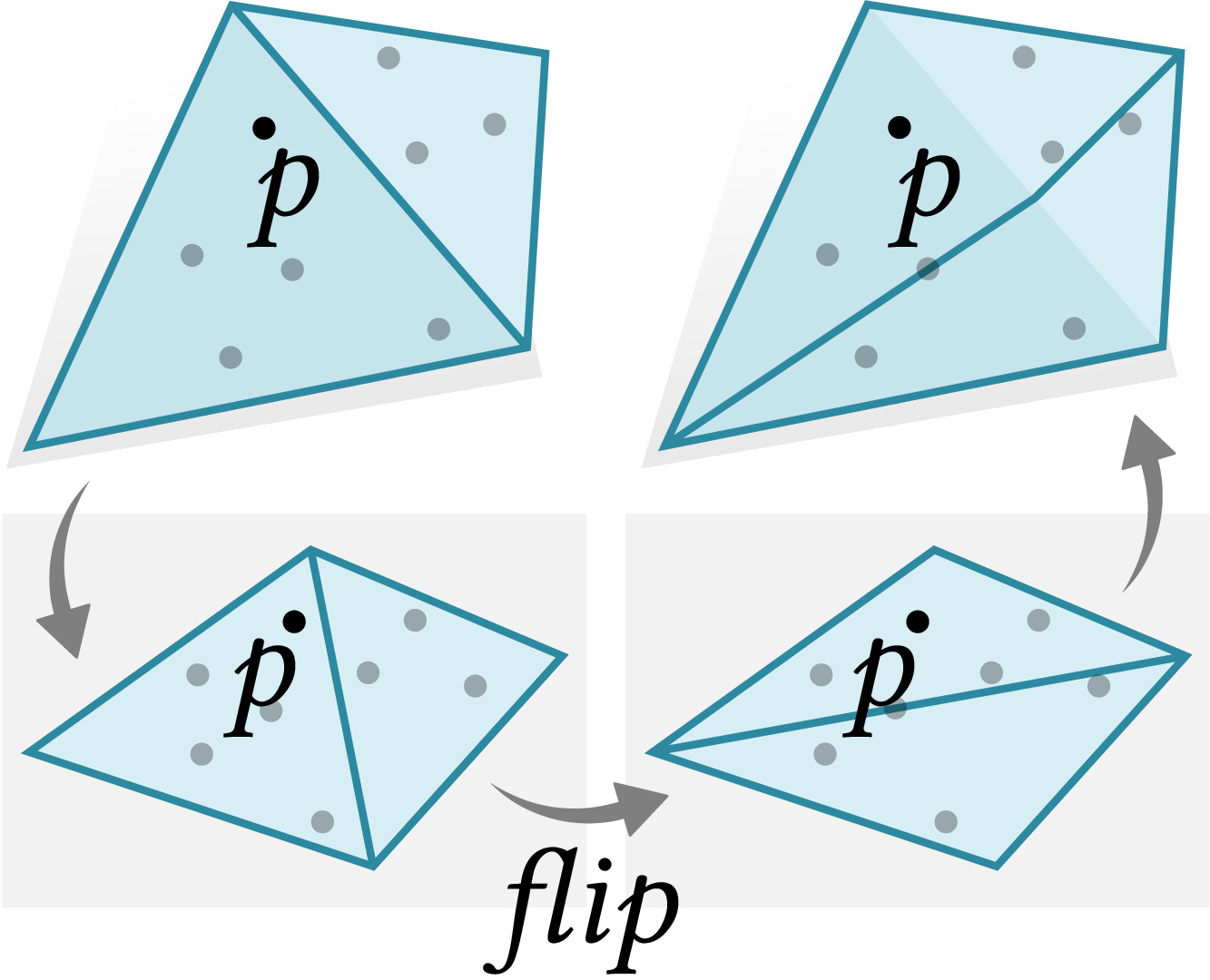}
\end{wrapfigure}

\paragraph{Edge flips.} To track a point \(p\) through an intrinsic flip of edge \(ij\), we unfold the two triangles \(ijk,jil\) into the plane (\eg{}, using formulas from \citet[Section 2.3.7]{SharpGC21}), and compute the barycentric coordinates of \(p\) in the new triangle (see inset).

\paragraph{Vertex flattening.} We must also compute new barycentric coordinates \(\after{b}\) after each vertex flattening (\refsec{VertexFlattening}). Here we use the projective interpolation scheme of \citet[\S3.4]{SpringbornSP08}.  Since edge lengths satisfy \refequ{conformal_equivalence}, this scheme defines a continuous (\(C^0\)) bijective map.  Let $b_i, b_j, b_k$ be barycentric coordinates for a point in face $\ijk$, and let $u_i$ be the scale factor at \(i\).  Then
\begin{alignat}{1}
   (\after{b}_i, \after{b}_j, \after{b}_k) = \frac{(e^{u_i} b_i, b_j, b_k)}{e^{u_i} b_i + b_j + b_k},
\end{alignat}
where the denominator ensures our updated values still sum to 1.

\begin{wrapfigure}[8]{r}{70pt}
	\includegraphics[width=1\linewidth]{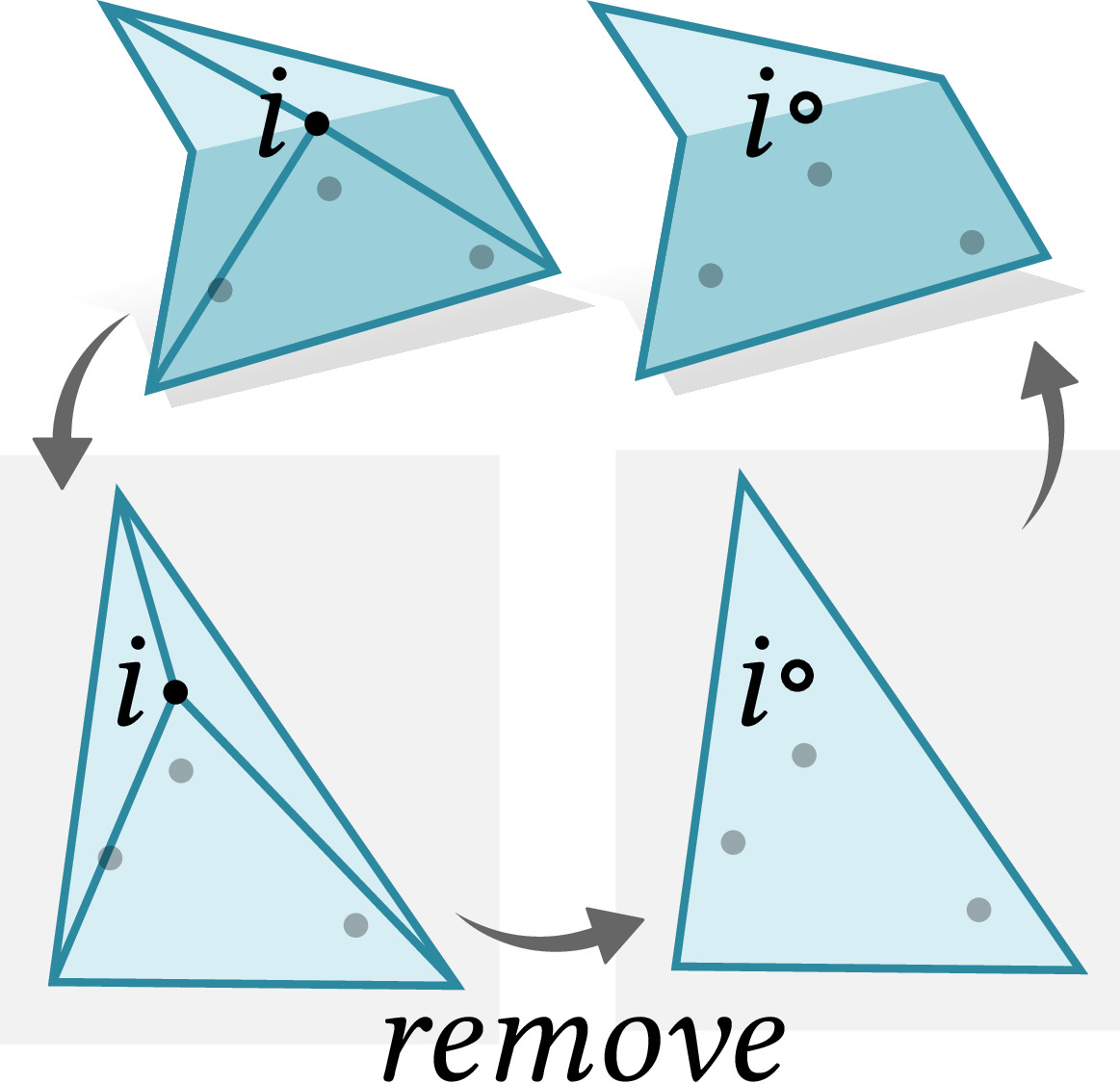}
	\label{fig:vertex_removal_barycentric}
\end{wrapfigure}
\paragraph{Vertex removal.} Once vertex \(i\) is flattened and flipped to degree three, its neighborhood can be laid out in the plane without distortion (see inset). Here we apply standard formulas to compute barycentric coordinates for vertex \(i\) in the new triangle, along with coordinates for any points located in the three removed triangles.
% After flattening vertex $i$ to have zero Gaussian curvature and doing projective updates on the barycentric coordinates, we can then remove the vertex $i$ and insert a new barycentric point in a straightforward way. This is because the one-ring of vertex $i$ has zero Gaussian curvautre. In other words, the one-ring patch can be flattened to 2D without any distortion. Therefore, we can simply remove vertex $i$ and insert a new barycentric point the same way as operating on 2D Euclidean domain (see the inset).

\subsection{Prolongation}
\label{sec:Prolongation}

Algorithms such as multigrid (\secref{SurfaceMultigrid}) often require not only pointwise correspondences, but also \emph{prolongation operators} which transfer functions from a coarse mesh to a finer one.  We define a prolongation operator via an approach similar to \citet{LeeSSCD98} and \citet{LiuZBJ21}.  In particular, we track the barycentric coordinates of all fine vertices \ala{}\ \secref{PointwiseMapping}.  A function on the coarse mesh is then mapped to the fine mesh via barycentric linear interpolation.  Explicitly, suppose each fine vertex \(i \in V\) has barycentric coordinates \(b_1, b_2, b_3\) in coarse triangle \(n_1 n_2 n_3 \in \after{F}\).  We then build a sparse matrix \(\mP \in \mathbb{R}^{|V| \times |\after{V}|}\) where row \(i\) has three nonzeros \(\mP_{i,n_j} = b_{n_j}\), for \(j=1,2,3\).  Prolongation then amounts to a matrix-vector product \(f = \mP \after{f}\), where \(\after{f} \in \mathbb{R}^{|\after{V}|}\) is a function on the coarse mesh.

\subsection{Vector Field Prolongation}
\label{sec:vector_field_prolongation}

\begin{figure}
    \begin{center}
    \includegraphics[width=1\linewidth]{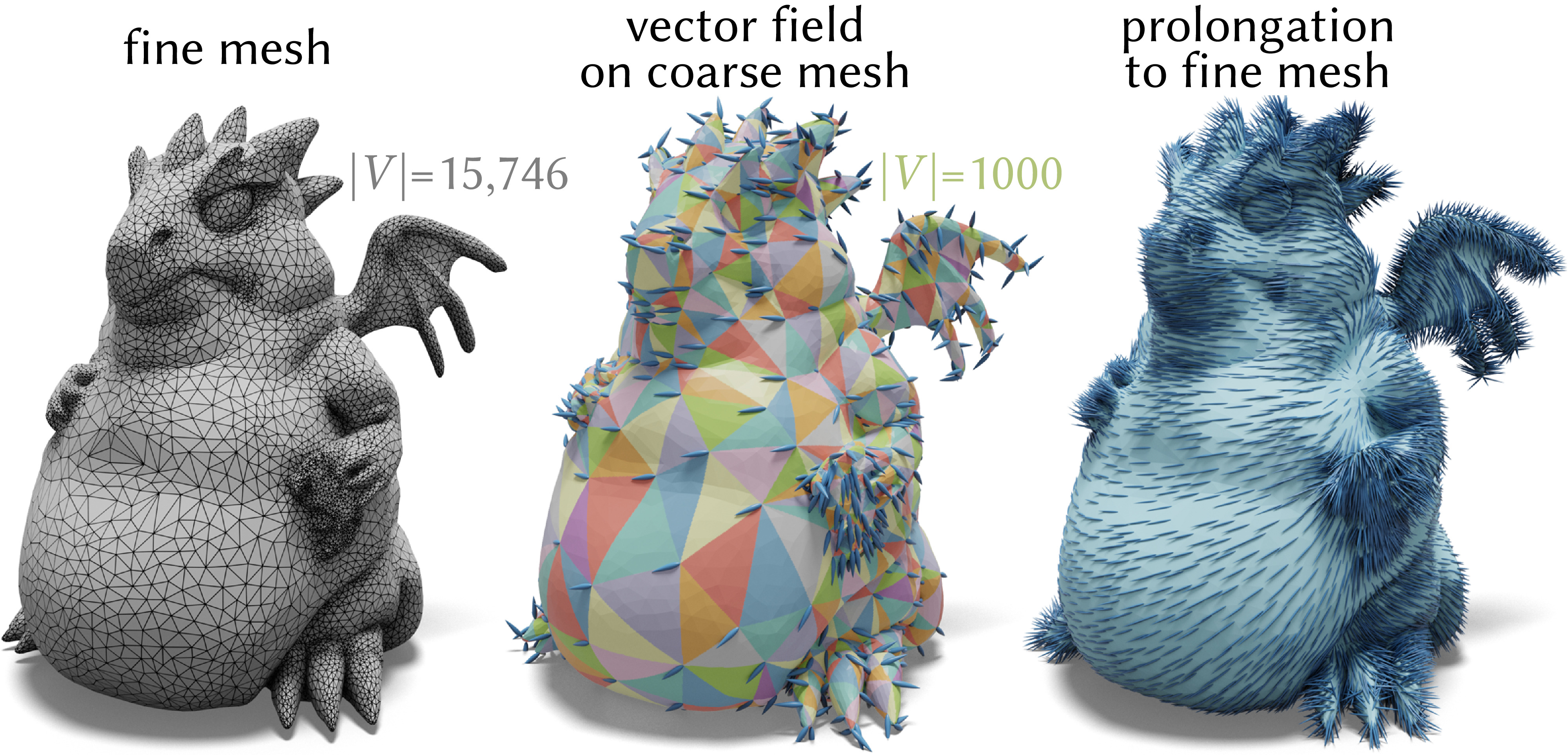}
    \end{center}
    \caption{We significantly reduce the cost of computing smooth vector fields on a fine mesh \emph{(left)} by solving on a coarse mesh \emph{(center)} and applying vector prolongation \emph{(right)}.\label{fig:vector_field_prolongation}}
  \end{figure}

We can also transfer vector fields from coarse to fine (\reffig{vector_field_prolongation}). This process has two steps (detailed below): first interpolate vectors over the coarse mesh, then sample the interpolated vector field onto the vertices of the fine mesh. Both steps are again represented by a single prolongation matrix \(\mP^{\text{vec}} \in \mathbb{C}^{|V| \times |\after{V}|}\), this time with complex entries.  However, since we store tangent vectors in a different normalized coordinate system at each vertex (\secref{TangentVectors}), we must compute unnormalized vectors before performing interpolation.  In particular, let \(\vu^0_i \in \mathbb{C}\) be the angle-normalized vectors stored at vertices, and let \(\phi_i := \arg(\vu^0_i)\) be the corresponding angles.  Then \(\vu_i := \|\vu^0_i\| e^{\imath \Theta_i \phi_i/2\pi}\) are the corresponding unnormalized vectors, and prolongation amounts to a matrix-vector multiply \(\mP^{\text{vec}} \vu\).  Note that this same prolongation scheme can be applied as-is to symmetric direction fields (line fields, cross fields, \etc{}), via the complex encoding introduced by \citet{KnoppelCPS13}.

\begin{wrapfigure}{r}{91pt}
   \includegraphics[width=1\linewidth]{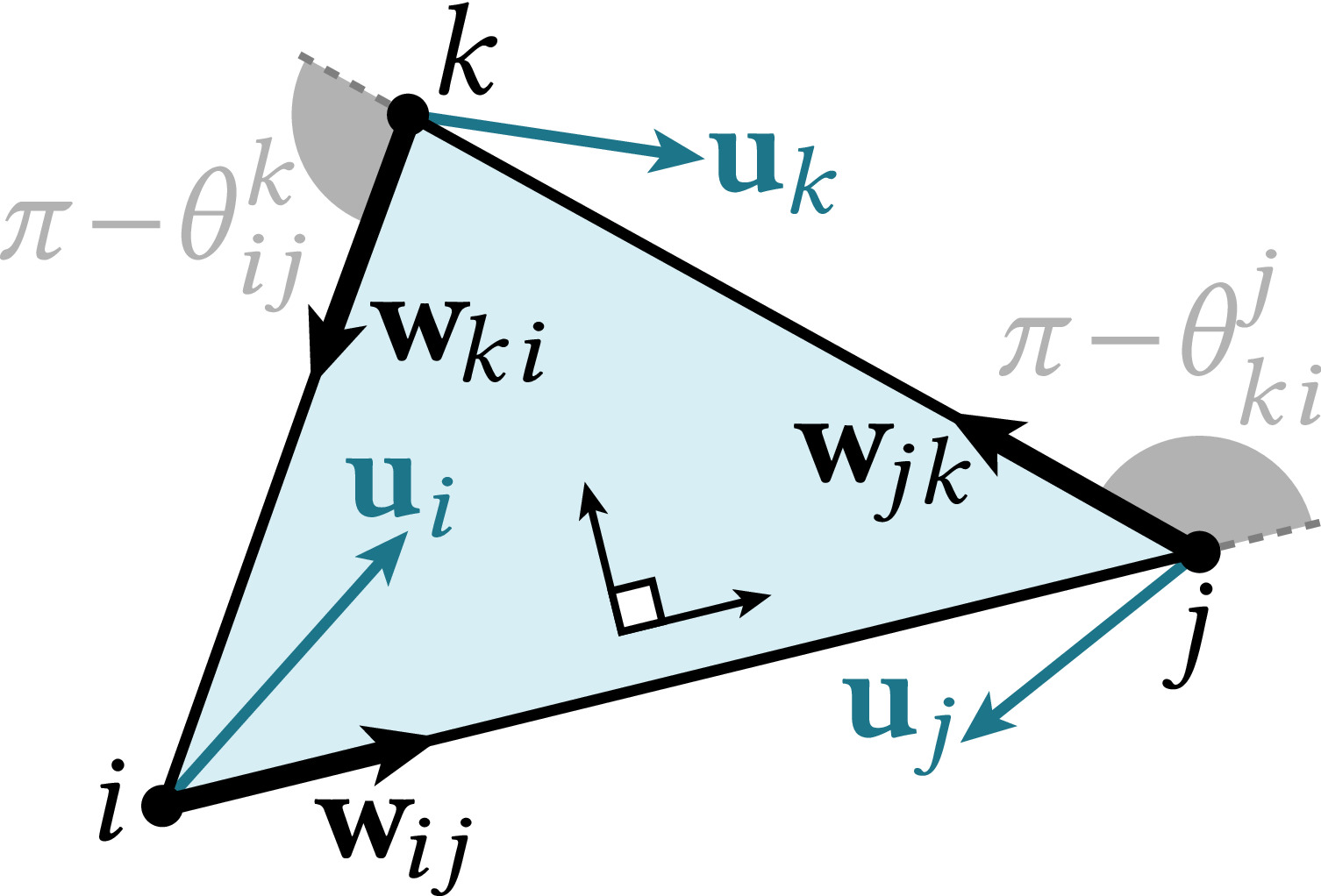}
\end{wrapfigure}

\subsubsection{Coarse Interpolation}

In each triangle \(ijk\), we adopt a coordinate system where oriented edge \(\ij\) points in the \(\phi = 0\) direction.  Let \(\vw_{\ij} := (\ve_{\ij}/\ell_{\ij})^{\Theta_i/2\pi}\) be the unit vectors along each oriented edge \(ij\), in unnormalized coordinates at \(i\), and let
\[
   \beta_{\ij} := 0,\ \ \beta_{\jk} := (\pi - \theta^j_{ki}),\ \  \text{and}\ \  \beta_{\ki} := (\pi - \theta^j_{ki}) + (\pi - \theta^k_{ij})
\]
be the angles of these unit vectors with respect to the coordinate system of the triangle.  Then \(\vz_i := e^{\imath\beta_{\ij}}/\vw_{\ij}\) is a rotation taking \(\vu_i\) to the corresponding vector in the triangle's coordinate system, and we can express the interpolated vector at any point \(\vp\) with barycentric coordinates \(b(\vp)\) as
\[
   \vu(\vp) := b_i(\vp)\vz_i\vu_i + b_j(\vp)\vz_j\vu_j + b_k(\vp)\vz_k\vu_k.
\]
Note that the resulting field is not continuous across edges, but is sufficiently regular for prolongation; if desired, better continuity can be achieved via the scheme of \citet[Section 4.3]{liu2016discrete}.

\subsubsection{Fine Sampling}
\label{sec:FineSampling}

\begin{wrapfigure}[8]{r}{88pt}
   \includegraphics[width=1\linewidth]{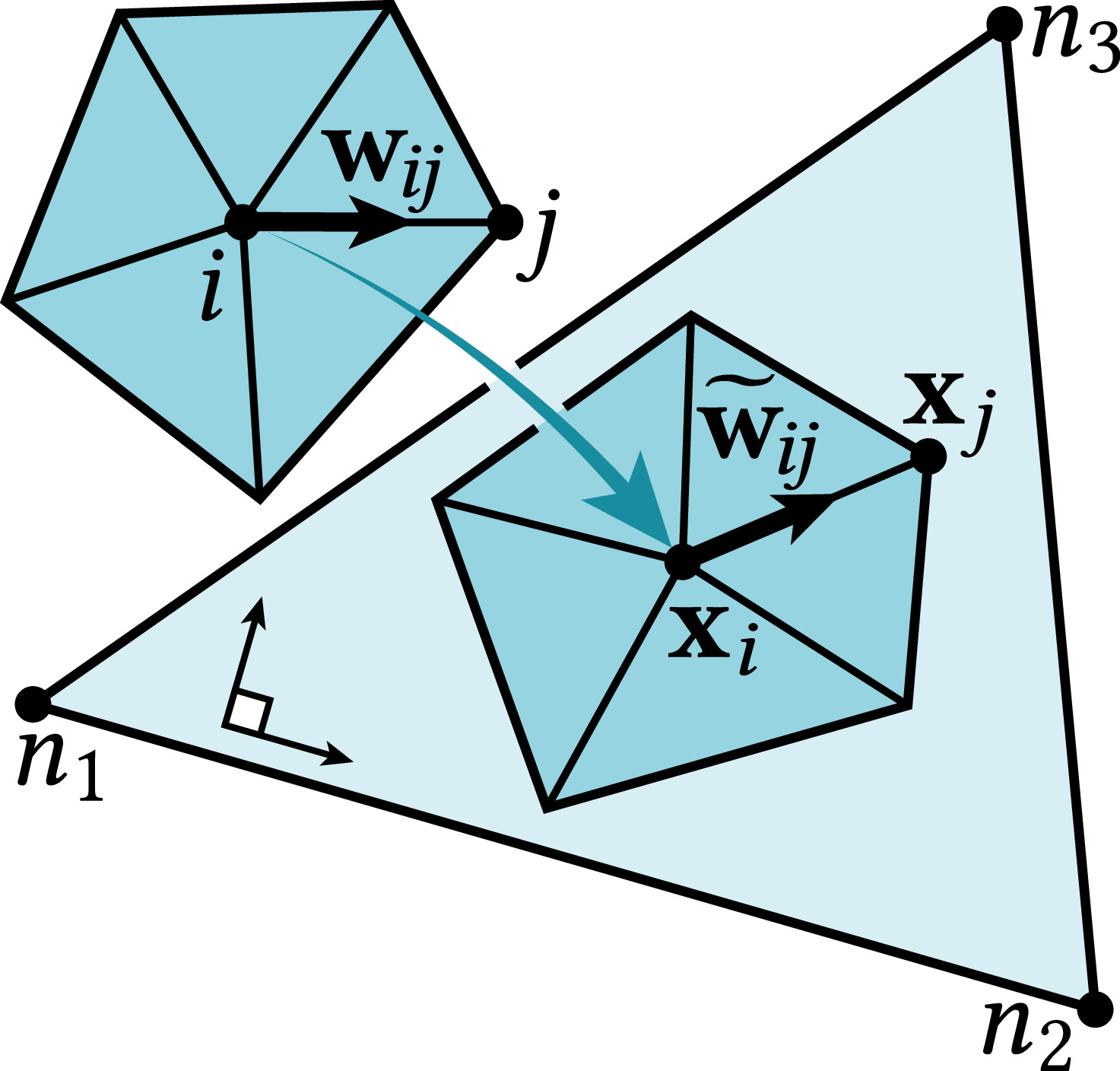}
\end{wrapfigure}
Consider a fine vertex \(i \in V\) mapped to a point \(\vx_i\) in coarse triangle \(n_1 n_2 n_3 \in \after{F}\) (expressed in a 2D layout, via the tracked barycentric coordinates).  To map the interpolated vector \(\vu(\vx_i)\) back to the vertex coordinate system, we must then compute a change of coordinates from triangle \(n_1 n_2 n_3\) to the tangent space \(\T_i M\).  Since we change the geometry via conformal flattening (\secref{VertexFlattening}), this change of coordinates is well-described by measuring the rotation and scaling of a single tangent vector.  In particular, for any \(j \in \mathcal{N}_i\) such that \(\vx_j\) is contained in the same triangle, we let \(\vw_{ij} := \ve_{ij}/\|\ve_{\ij}\|\) be the unit vector along the fine edge, and compute the corresponding unit vector \(\after{\vw}_{ij} := (\vp_j-\vp_i)/\|\vp_j-\vp_i\|\) on the coarse triangle (approximating the tangent map).  The rotation and scaling between coordinate systems is then captured by the complex number \(\psi_i := \vw_{\ij}/\after{\vw}_{\ij}\), and the final interpolated value in the normalized coordinate system at \(i\) is given by \(\psi_i \vu_i(\vx_i)\). In the rare case where no neighboring \(\vx_j\) sits in the same triangle, we simply take the average of known interpolated values.  Overall, then, row \(i\) of the vector prolongation matrix \(\mP^{\text{vec}}\) has three nonzero entries \(b_j \psi_i \vz_j\), corresponding to the three columns \(n_j\), \(j=1,2,3\).

\section{Evaluation \& Results}
\label{sec:EvaluationAndResults}

Here we evaluate the performance, robustness, and quality of our method (Sec. \ref{sec:Benchmark} and \ref{sec:MeasuringDistortion}), compare it to extrinsic alternatives (\secref{ComparisonWithExtrinsicMethods}), and explore its effectiveness in the context of several fundamental algorithms (Sec. \ref{sec:GeometricAlgorithms}, \ref{sec:AdaptiveCoarsening}, and \ref{sec:SurfaceMultigrid}).  We also describe the strategy used to visualize results throughout the paper (\secref{Visualization}).  Note that all experiments were run on a 4.1GHz Intel i7-8750H with 16GB RAM.

\subsection{Visualization}
\label{sec:Visualization}

\begin{figure}
    \begin{center}
    \includegraphics[width=1\linewidth]{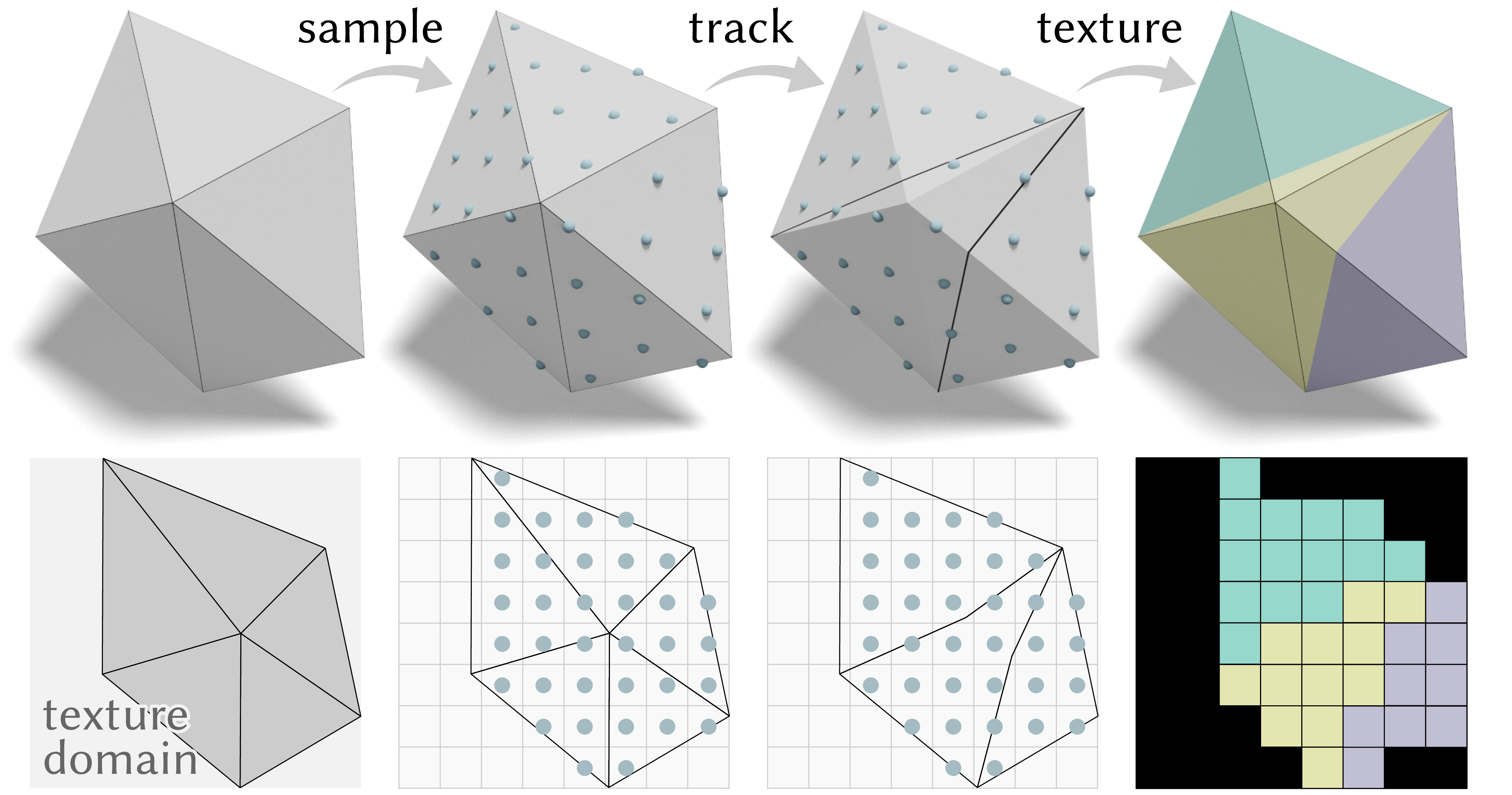}
    \end{center}
    \caption{To visualize a coarse intrinsic triangulation on a mesh with texture coordinates \emph{(left)}, we track the barycentric coordinates of each texel \emph{(center left)} through the simplification process \emph{(center right)}, and use coarse triangle IDs to color the final texels \emph{(right)}.\label{fig:bake_texture}}
    \label{fig:bake_texture} 
\end{figure}

Traditionally, intrinsic triangulations are visualized via a \emph{common subdivision}~\cite{FisherSBS06,SharpSC19a,GillespieSC21a}, \ie, the input mesh is split along geodesic arcs corresponding to intrinsic edges.  However, since coarsening does not exactly preserve intrinsic geometry, edges may no longer be geodesics.  We instead use a texture mapping approach (\figref{bake_texture}).  Given initial texture coordinates (computed via \citep{SawhneyC18}), we compute barycentric coordinates for each texel covered by a fine triangle.  These coordinates are then tracked through the coarsening process, \ala{} \secref{PointwiseMapping}.  Following \citet{SharpSC19a} we assign a greedy coloring to the coarse triangles; texels adopt the color of their associated triangle.  This visualization may not exactly depict coarse lengths or areas, but faithfully represents the bijective map.  A possible alternative is to construct an explicit \emph{topological} subdivision, where edges need not be geodesics \cite{SchmidtCBK20,Takayama22}, but which may enable more sophisticated attribute transfer~\cite[\S4.3]{GillespieSC21a}.

\subsection{Benchmark}
\label{sec:Benchmark}

We evaluated our method on about 6k manifold meshes from the \emph{Thingi10k} dataset \citep{Zhou:2016:TDT}.  For robustness experiments we preprocess meshes via intrinsic Delaunay refinement \citep[\S4.2]{SharpSC19a} with a lower angle bound of \(25^\circ\), providing more candidate locations for coarse vertices.  Since some meshes contain thousands of connected components, we normalize input/output vertex counts by the number of components.

\begin{figure}
    \begin{center}
    \includegraphics[width=1\linewidth]{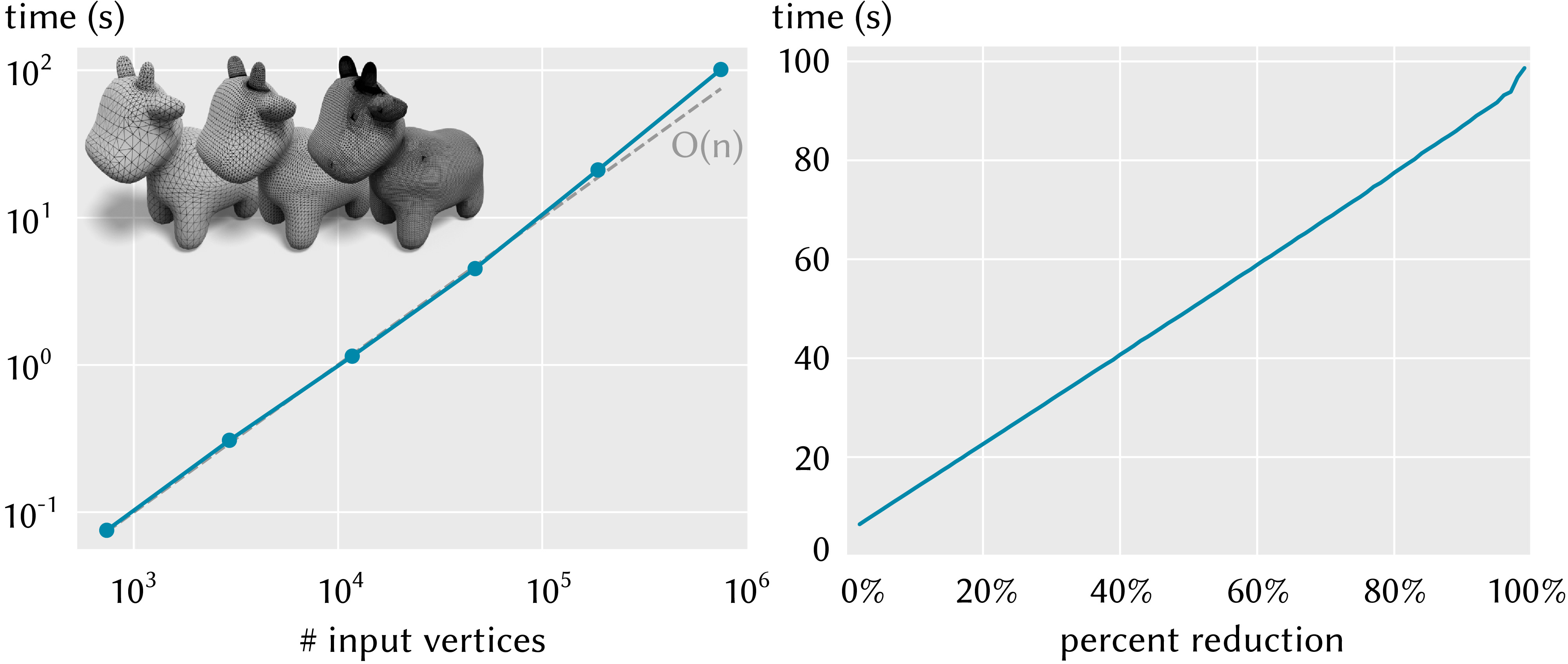}
    \end{center}
    \caption{Since we use only greedy local operations, total cost is roughly linear in both mesh size and percent reduction (including the cost to build \(\mP\)).  \emph{Left:} increasingly fine subdivisions are coarsened to 1\% of their initial size. \emph{Right:} a subdivision with 750k vertices is coarsened to various resolutions.}
    \label{fig:runtime}
\end{figure}

\paragraph{Performance} \reffig{runtime} plots the total cost of our method, including decimation, maintaining a bijective map, and constructing the prolongation operator \(\mP\).  Since each vertex removal is an \(O(1)\) operation, we achieve near-linear scaling with respect to input size, though for very large meshes the \(O(n\log n)\) cost of maintaining a priority queue will ultimately dominate.  In absolute terms, our method decimates about 10,000 vertices per second.

\begin{figure}
    \begin{center}
    \includegraphics[width=1\linewidth]{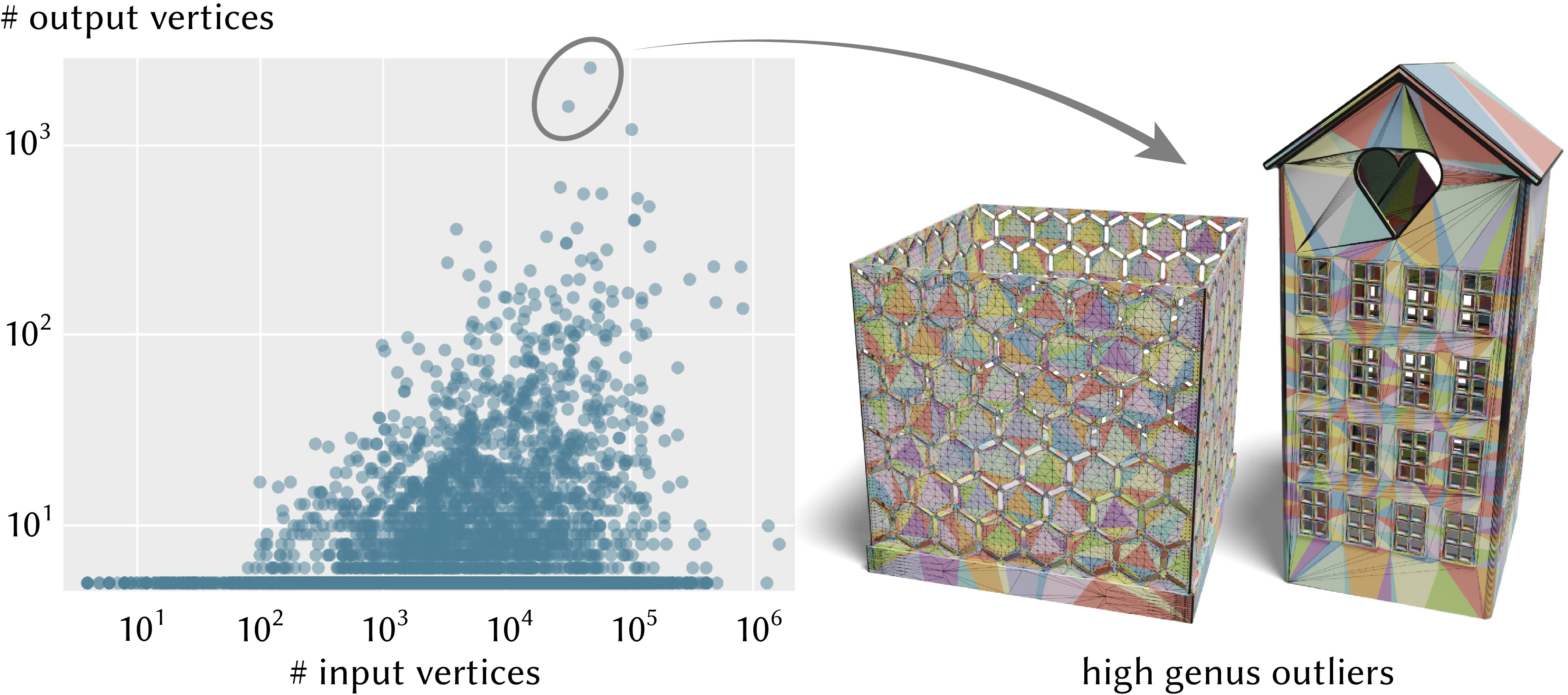}
    \end{center}
    \caption{Our method can coarsen even difficult meshes down to a small number of vertices.  Here, 98\% of \emph{Thingi10k} meshes are coarsened to less than 10\% of their input size \emph{(left)}. The exception are very high-genus meshes, which cannot be coarsened without changing the global topology \emph{(right)}.}
    \label{fig:thingi}
\end{figure}

\paragraph{Robustness} Our method fails only if no remaining vertex can be removed, \ie, if (i) flattening would violate the triangle inequality or (ii) flipping to degree-3 is not possible (\secref{VertexRemoval}).  We hence quantify robustness by coarsening as much as possible, then measuring the ratio \(|\after{V}|/|V|\). In \reffig{thingi} we successfully reduce 98\% and 84\% of models down to 10\% and 1\% (\resp{}) of the input resolution, prior to Delaunay refinement.  Large ratios occur only for high-genus models that cannot be significantly coarsened without modifying global topology.  On about 0.1\% of meshes vertex removal failed due to floating-point error; \emph{integer coordinates}~\cite{GillespieSC21a} may help to further improve robustness.

\subsection{Measuring Distortion}
\label{sec:MeasuringDistortion}

\begin{wrapfigure}{r}{72pt}
   \includegraphics[width=1\linewidth]{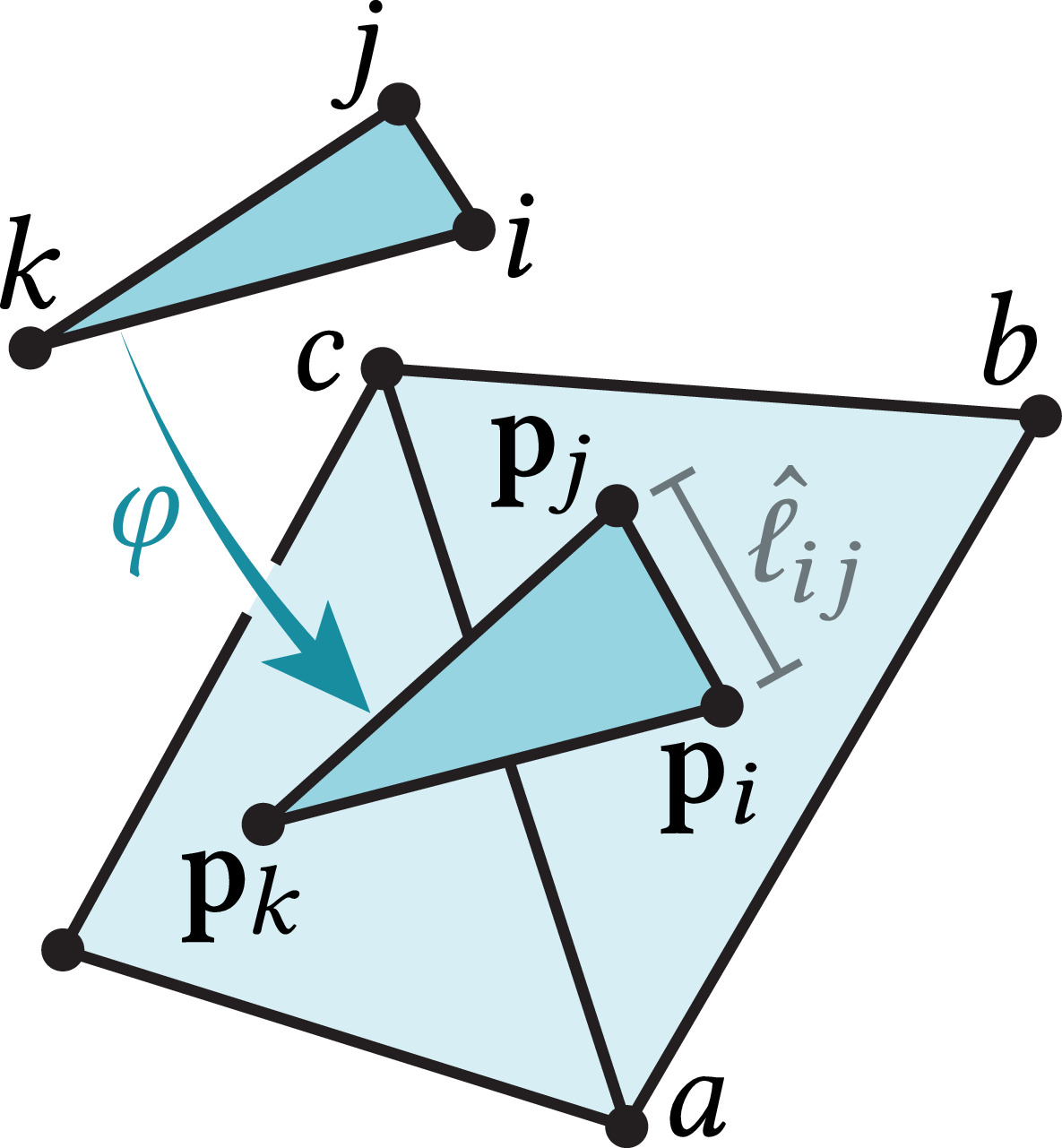}
\end{wrapfigure}
To evaluate our method, we measure the parametric distortion of the bijective map \(\varphi: M \to \after{M}\) between the fine and coarse mesh---which is distinct from the ICE metric used for coarsening.  For each fine edge \(\ij\) we first find an approximation \(\hat\ell_{\ij}\) of its length under \(\varphi\). Since all fine edges are minimal geodesics in the extrinsic mesh, we assume this property is preserved under \(\varphi\) and compute the minimal geodesic distance between \(\vp_i\) and \(\vp_j\)  If both endpoints sit in the same coarse triangle \(abc\), we can simply measure the distance between image points \(\vp_i := \varphi(i)\) and \(\vp_j := \varphi(j)\) in a local layout of \(abc\), computed via barycentric coordinates.  Otherwise, we compute geodesic distance using the method of \citet{mitchell1987discrete}.  Finally, for each fine triangle \(\ijk\) we use lengths \(\hat\ell_{\ij},\hat\ell_{\jk},\hat\ell_{\ki}\) to construct representative vertex positions \(\vp_i,\vp_j,\vp_k \in \mathbb{R}^2\), via \citet[\S 2.3.7]{SharpGC21}.  Distortion relative to input vertices \(\vp^0_i, \vp^0_j, \vp^0_k\) can then be quantified using any standard per-triangle measure---we use the \emph{anisotropic distortion} and \emph{area distortion} as defined by \citet[\S 2]{KhodakovskyLS03}.

\subsection{Comparison with Extrinsic Methods}
\label{sec:ComparisonWithExtrinsicMethods}

\begin{figure}
    \begin{center}
    \includegraphics[width=1\linewidth]{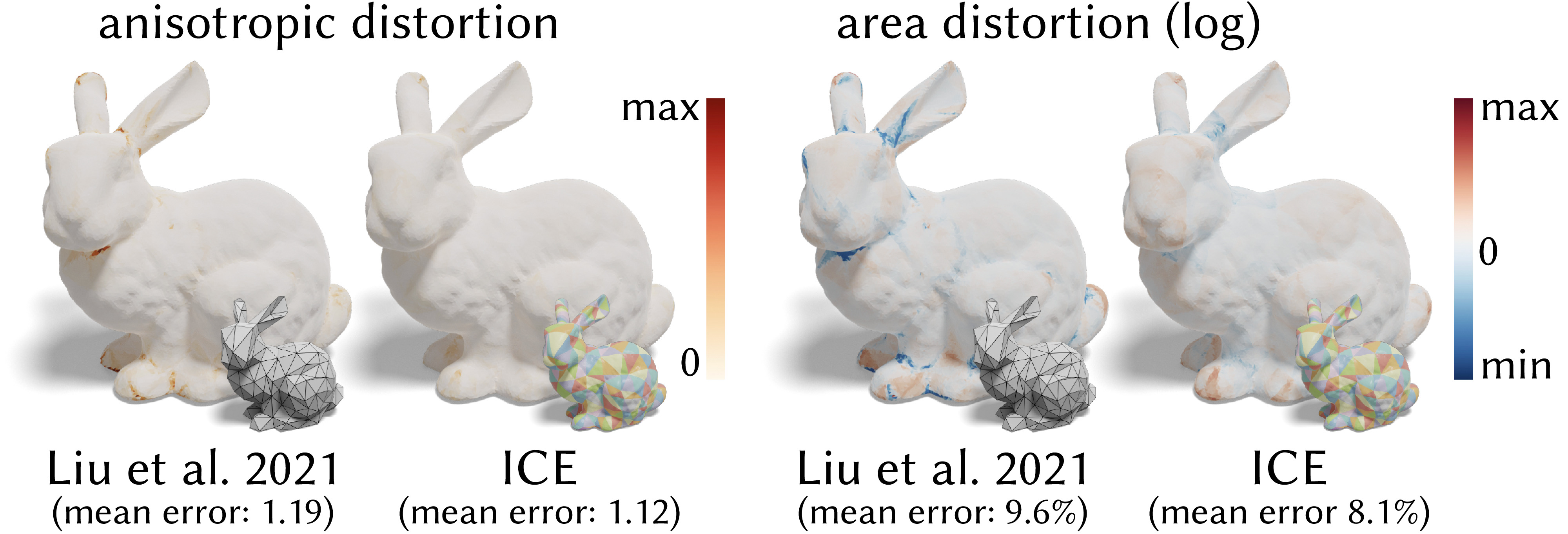}
    \end{center}
    \caption{Even on an extremely nice triangulation of a highly regular surface we see a reduction in distortion relative to past methods---owing to the much larger space of intrinsic triangulations.\label{fig:jacobian_distortion}}
    
\end{figure}

\begin{figure}
    \begin{center}
    \includegraphics[width=1\linewidth]{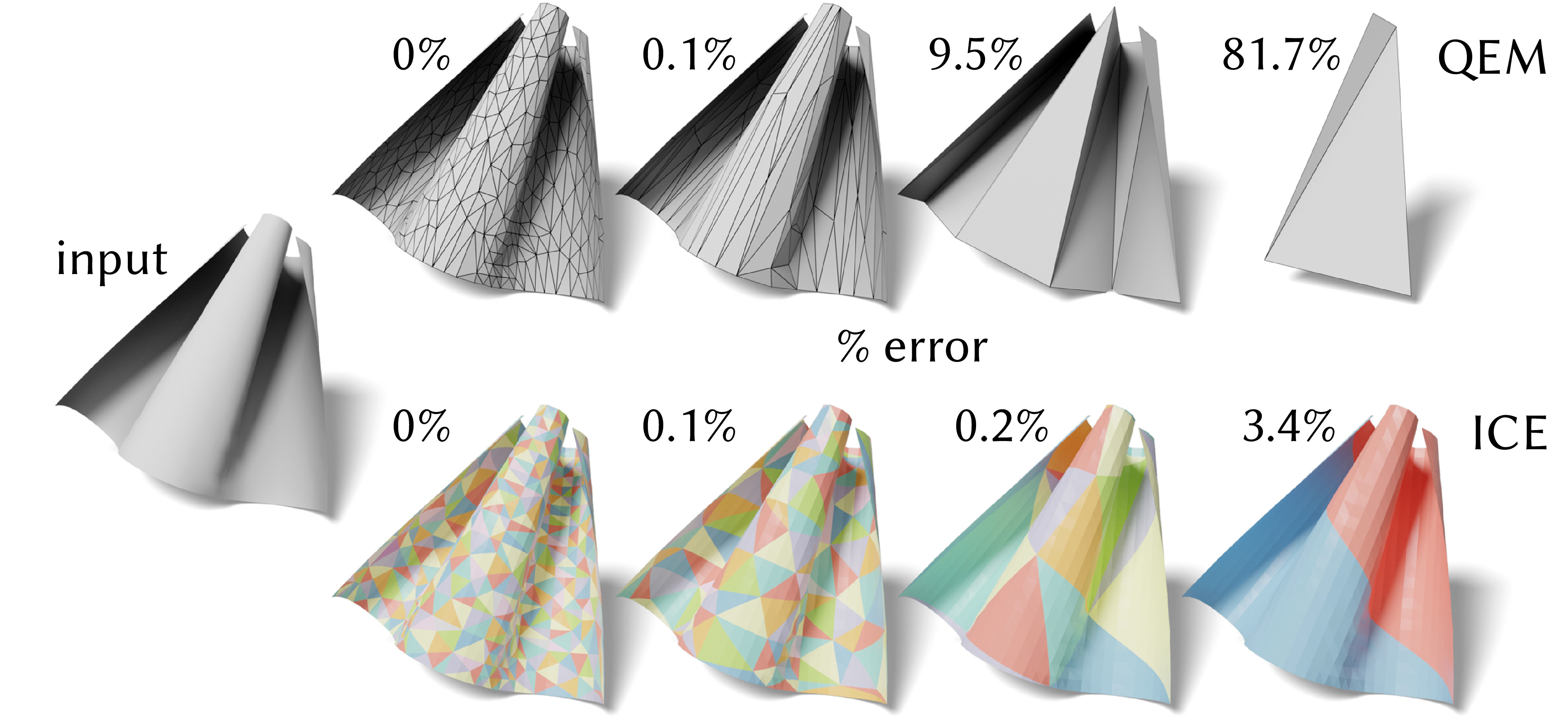}
    \end{center}
    \caption{On surfaces with small intrinsic curvature, such as a developable surface obtained from \cite{dog}, we achieve dramatically lower error in surface area compared to extrinsic methods like QEM.}
    \label{fig:small_K}
\end{figure}

Relative to past methods, the flexibility gained by working in the larger space of intrinsic triangulations leads to smaller geometric distortion on meshes of equivalent size.  For instance, in \reffig{jacobian_distortion} we coarsen a 28k bunny mesh down to 200 vertices with both the method of \citet{LiuZBJ21} and our method.  Even on this highly regular geometry we observe a modest reduction of both area distortion and anisotropic distortion.  For more difficult triangulations, or surfaces with lower intrinsic curvature (\eg{}, \figref{teaser}), we observe more significant gains.  As an extreme case, \figref{small_K} coarsens a developable surface from \cite{dog, VerhoevenVHS22} via both QEM and ICE.  Since coarse extrinsic edges are shortest paths in \(\mathbb{R}^n\), they underestimate intrinsic distances (hence areas); in contrast, intrinsic edges are essentially embedded in the original surface, providing better approximation of the original geometry.

\newpage

\subsection{Geometric Algorithms}
\label{sec:GeometricAlgorithms}

\subsubsection{Partial Differential Equations}
\label{sec:PartialDifferentialEquations}

\begin{figure}
    \begin{center}
    \includegraphics[width=1\linewidth]{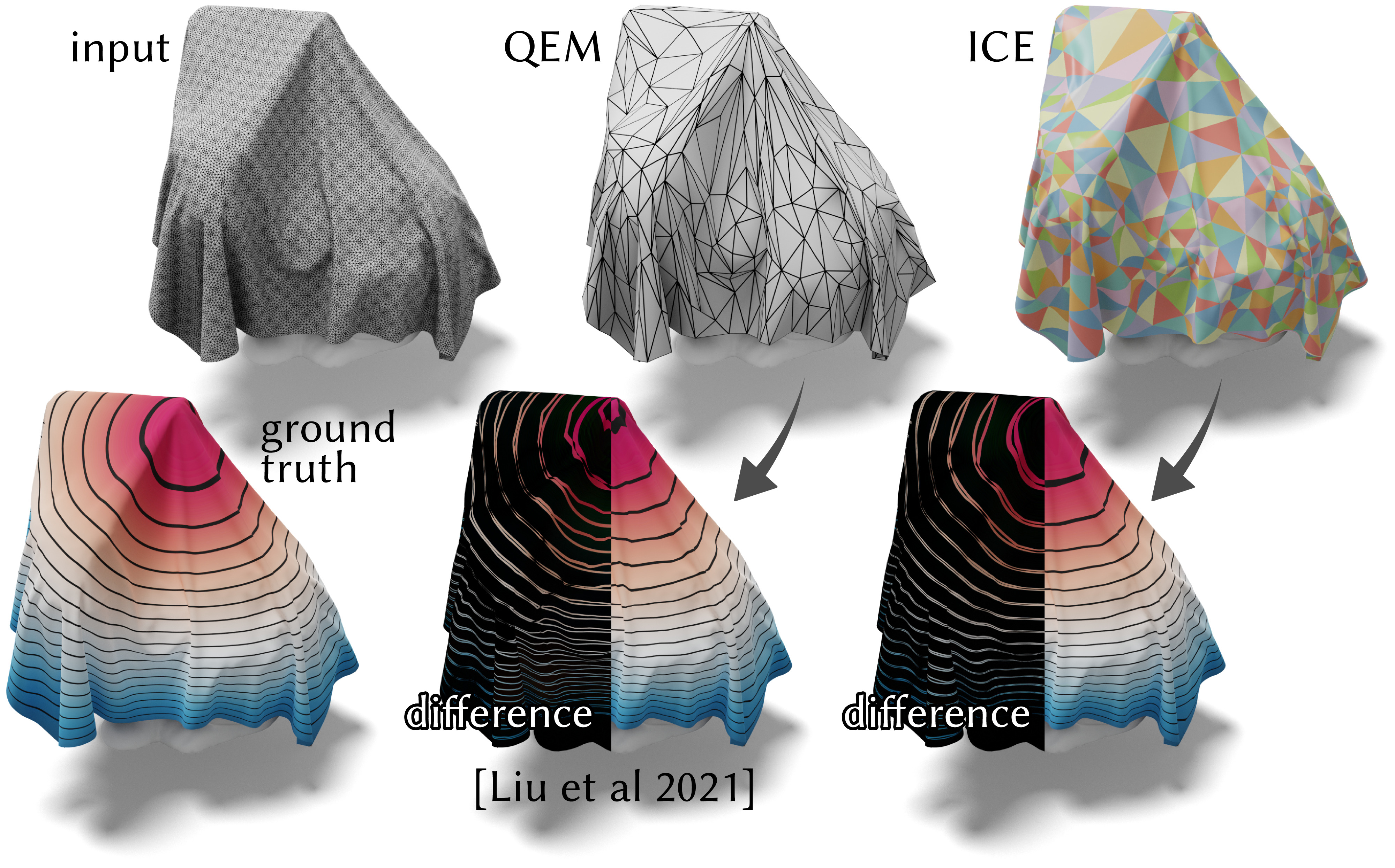}
    \end{center}
    \caption{For the same vertex budget as extrinsic methods like QEM, ICE provides more accurate solutions for basic problems like solving a Poisson equation---seen here via smoother isolines that better approximate the ground truth.}
    \label{fig:cloth_poisson}
\end{figure}

Better domain approximation in turn improves the quality of solutions computed on coarse meshes.  For example, in \figref{cloth_poisson} we coarsen a cloth simulation mesh down to 500 vertices with an extrinsic method (\cite{LiuZBJ21} using QEM simplification) and our intrinsic method. We then solve a Poisson problem on the coarse meshes and apply prolongation, yielding more accurate results in the intrinsic case.

\subsubsection{Single-Source Geodesic Distance}
\label{sec:SingleSourceGeodesicDistance}

\begin{figure}
    \begin{center}
    \includegraphics[width=1\linewidth]{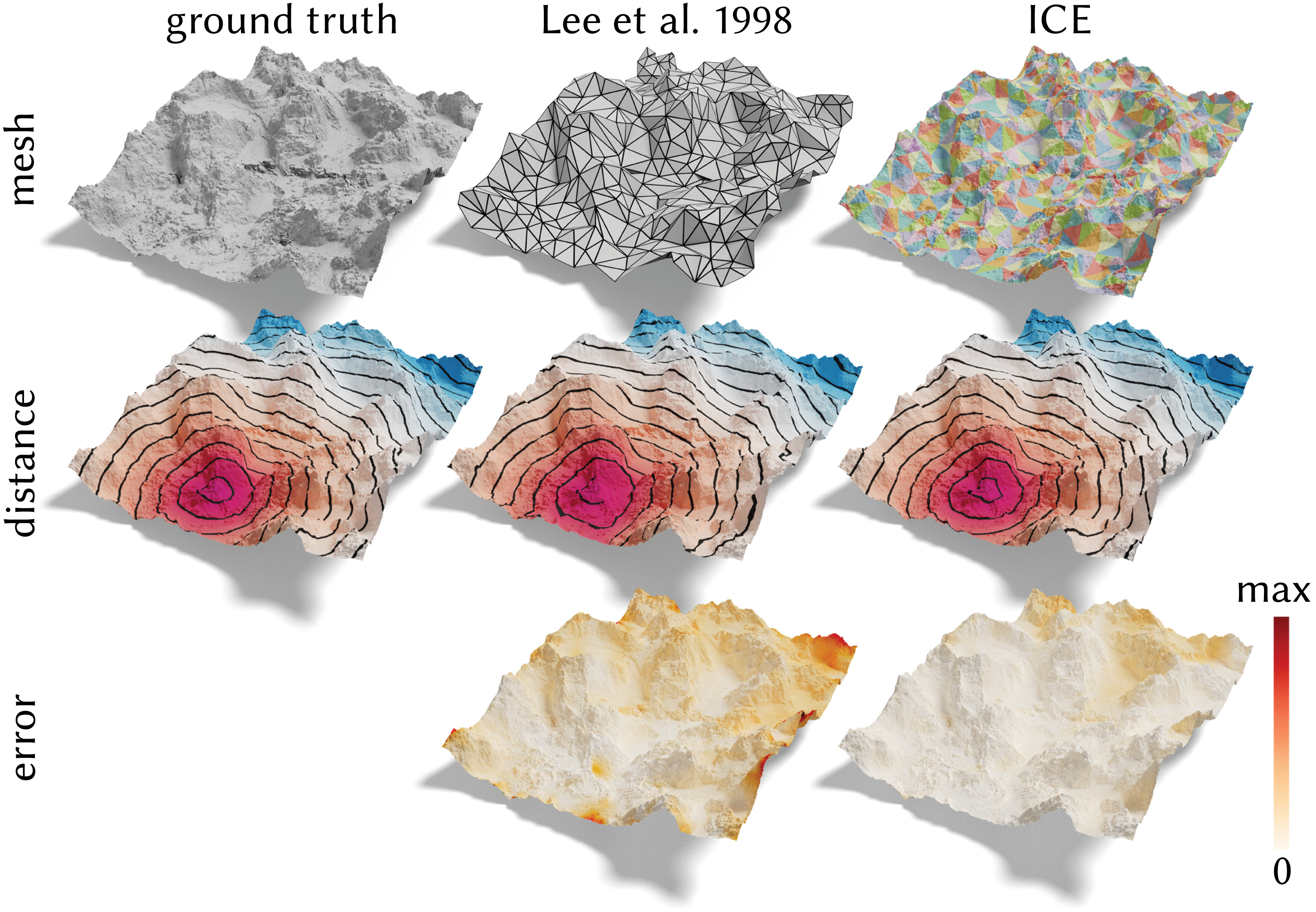}
    \end{center}
    \caption{Since geodesic distance is an intrinsic quantity, it is more accurately approximated via intrinsic coarsening---here providing a 4x reduction in relative error.}
    \label{fig:geodesic_terrain}
\end{figure}

\begin{figure}
    \begin{center}
    \includegraphics[width=1\linewidth]{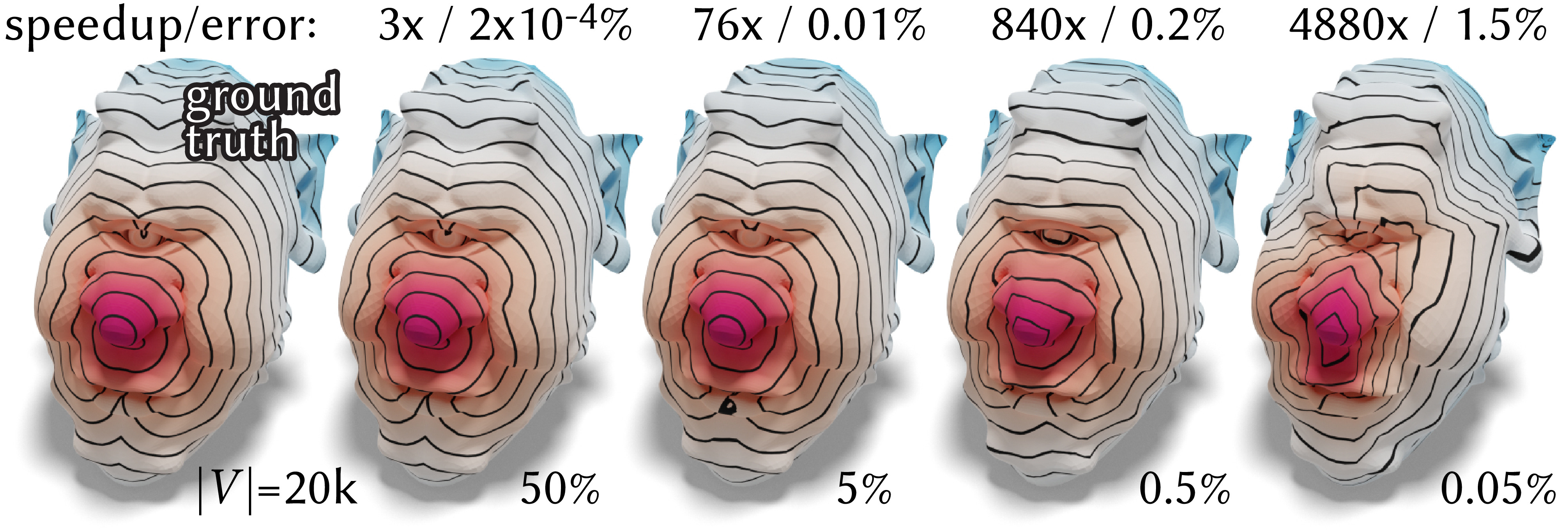}
    \end{center}
    \caption{Intrinsic coarsening offers an attractive approach to approximating single-source geodesic distance, here providing a three orders of magnitude speedup for a fraction of a percent relative error.}
    \label{fig:cost_accuracy}
\end{figure}

Geodesic distance is an intrinsic quantity, making it a natural fit for intrinsic coarsening.  In \figref{geodesic_terrain} we compare ICE to the extrinsic method of \citet{LeeSSCD98} by measuring the difference between the exact distance on the fine input, and prolongated distances from the coarse meshes (both computed via \citep{mitchell1987discrete}); here ICE achieves a roughly 4x reduction in relative error.  \reffig{cost_accuracy} illustrates the speed-accuracy trade off of using ICE, here reducing cost by three orders of magnitude while introducing only \(\sim\!1\%\) relative approximation error.

\subsubsection{All-Pairs Geodesic Distance}
\label{sec:All-PairsGeodesicDistance}

\begin{figure}
    \begin{center}
    \includegraphics[width=1\linewidth]{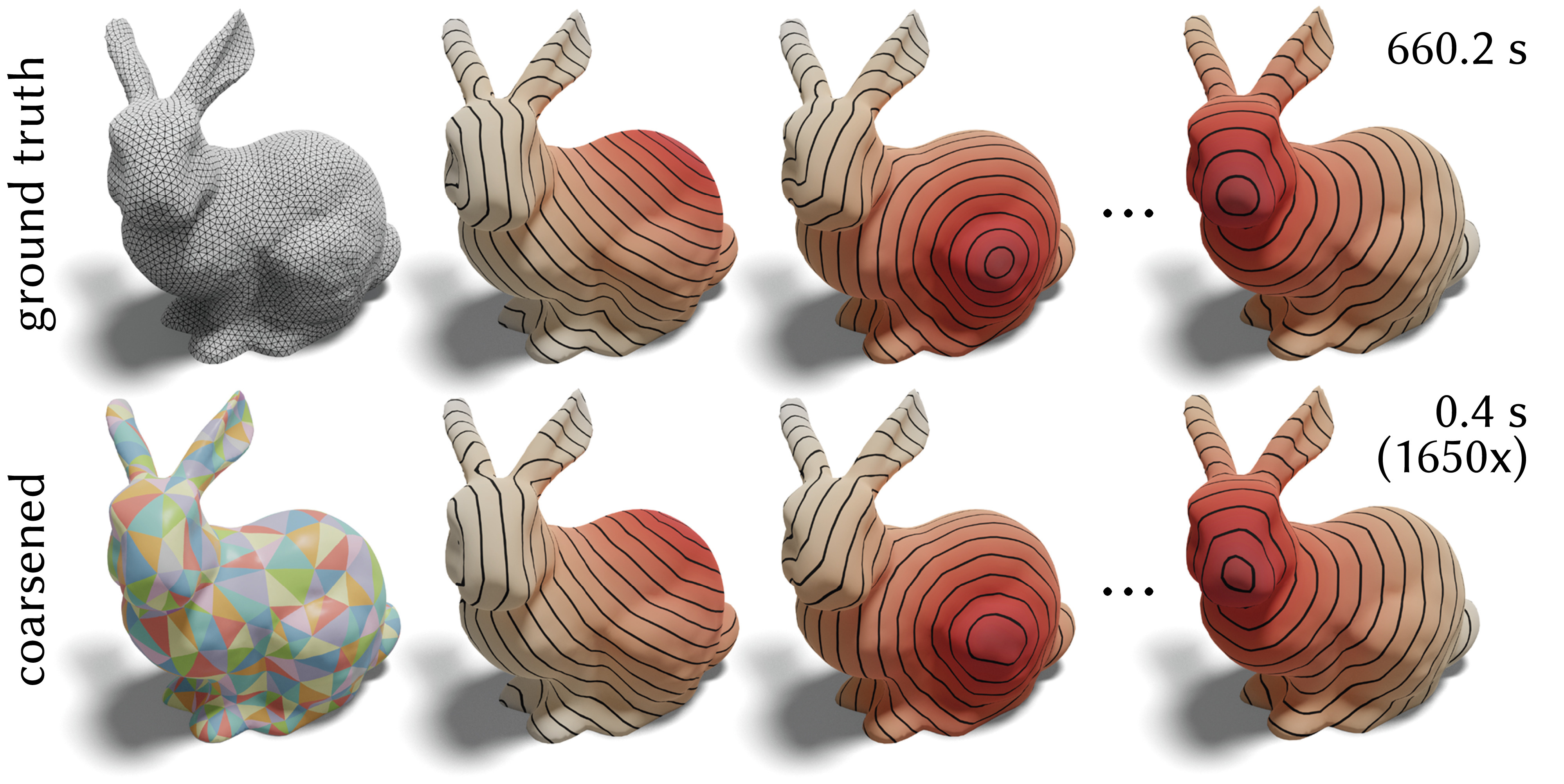}
    \end{center}
    \caption{For a mesh with 6k vertices we obtain an all-pairs geodesic distance matrix 1650x faster, while incurring only 1.4\% relative error.} 
    \label{fig:geodesic_coarse_approx}
\end{figure}

The benefits of an accurate intrinsic approximation become even more pronounced when approximating the dense matrix \(\mD \in \mathbb{R}^{|V| \times |V|}\) of all pairs of geodesic distances---a shape descriptor often used in correspondence and learning methods~\cite{shamai2017geodesic}.  We can compute a low-rank approximation of \(\mathbb{D}\) via
\[
   \widehat{\mD} := \mP \after{\mD} \mP^\top,
\]
where $\after{\mD}$ is the coarse all-pairs matrix (computed again via \citep{mitchell1987discrete}).  See for instance \reffig{geodesic_coarse_approx}---here again we achieve several orders of magnitude speedup, with only 1.4\% relative error.

\subsubsection{Riemannian Computational Geometry}
\label{sec:RiemannianComputationalGeometry}

\begin{wrapfigure}[8]{r}{110pt}
	\includegraphics[width=\linewidth, trim={1mm 4mm 1mm 0mm}]{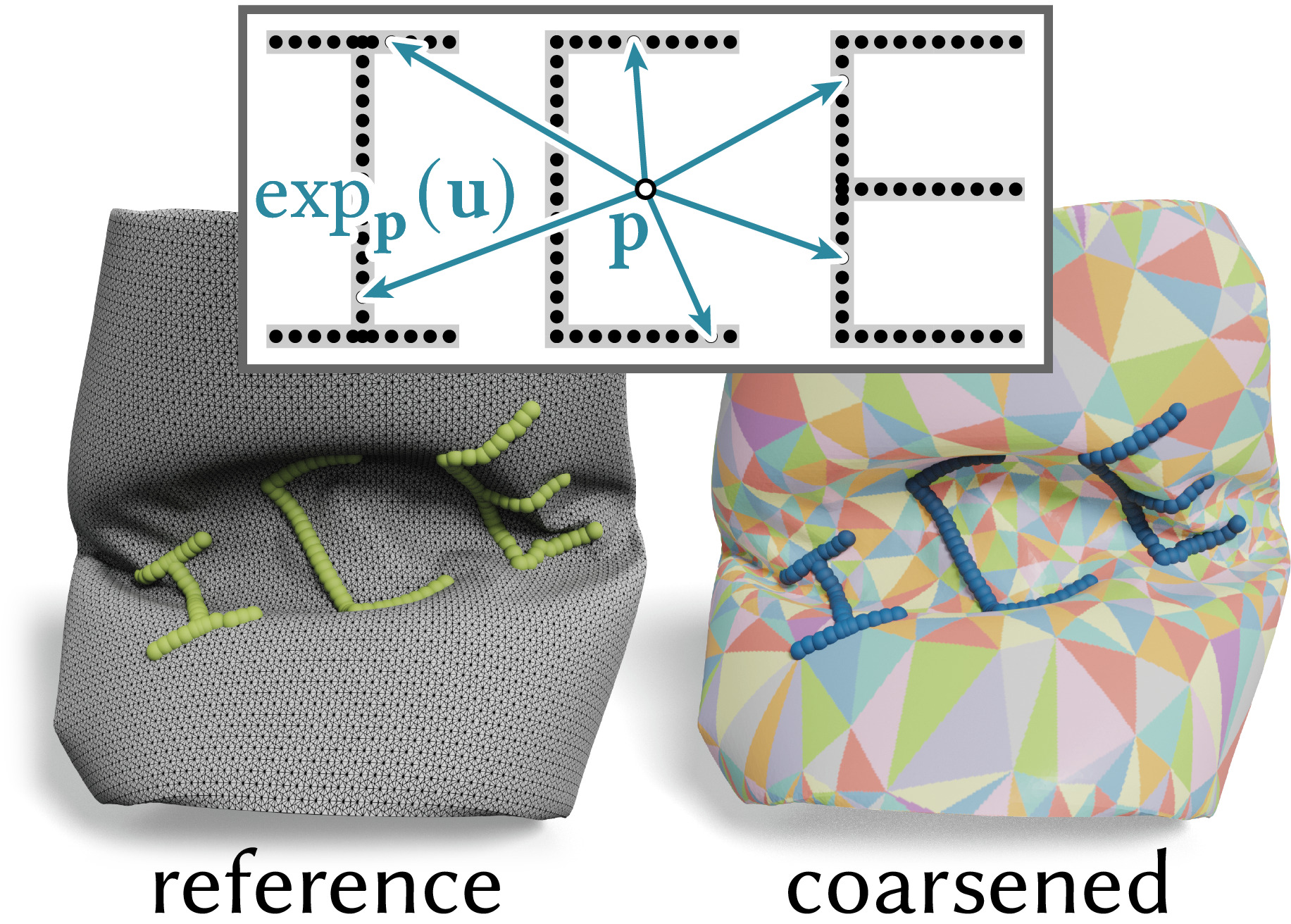}
\end{wrapfigure}
\begin{figure}
    \begin{center}
    \includegraphics[width=1\linewidth]{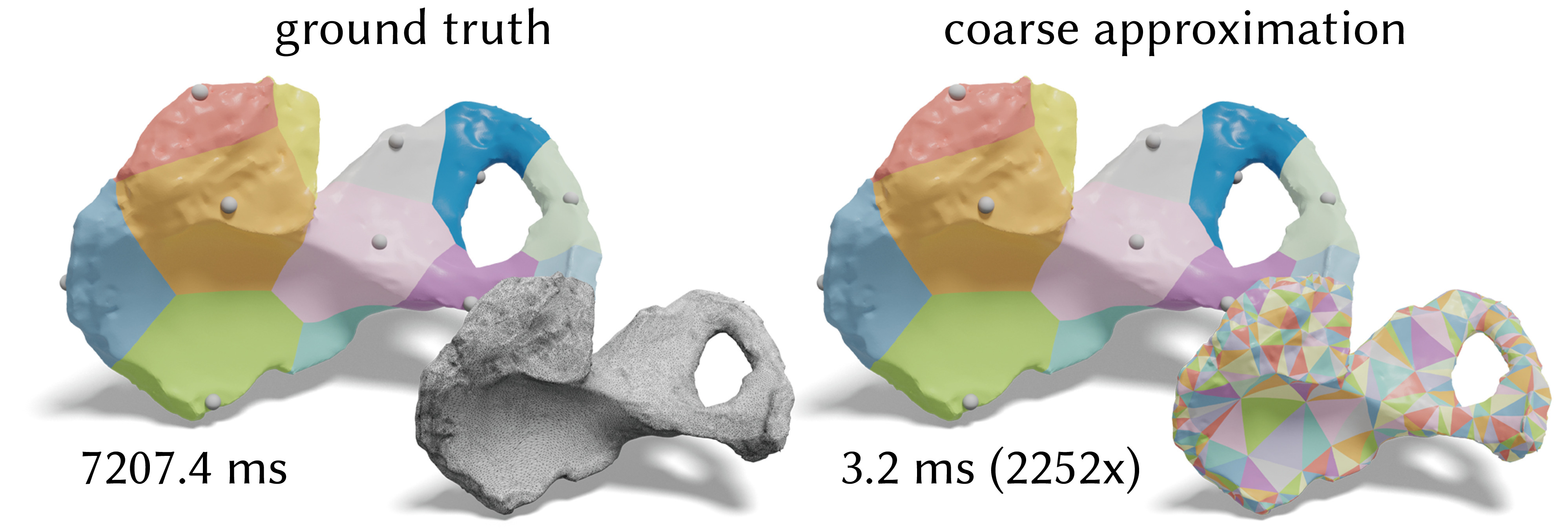}
    \end{center}
    \caption{Fast computation of geodesic distance in turn yields fast computation of other quantities---here we compute geodesic Voronoi diagrams three orders of magnitude faster (2252x) than on the original mesh, while misclassifying only 1\% of fine vertices.}
    \label{fig:voronoi}
\end{figure}

More broadly, standard geometric quantities computed on the coarse mesh provide excellent approximations of the fine solution.  For instance, in \figref{voronoi} we use \citet{mitchell1987discrete} to compute a geodesic Voronoi diagram on the coarse mesh, yielding a near-perfect approximation at a tiny fraction of the cost---such diagrams in turn provide the starting point for remeshing and other applications~\cite{ye2019geodesic}.  Likewise, we can dramatically reduce the cost of evaluating the discrete exponential map over long distances (\ala{} \citep[\S 2.4.2]{SharpGC21}), replacing many small steps through fine triangles with a small handful of ray-edge intersections, while arriving at nearly identical points (see inset).  Other classic algorithms, such as \emph{Steiner tree approximation}~\citep[\S 5.1]{SharpSC19a} could likewise be accelerated by simply swapping out our coarse intrinsic mesh for the fine one.

\subsection{Adaptive Coarsening}
\label{sec:AdaptiveCoarsening}

The local nature of our scheme makes it easy to adaptively coarsen (or preserve) the mesh according to various geometric criteria.  Here we expand on the basic weighting strategy from \secref{AuxiliaryData}.

\subsubsection{Spatial Adaptivity}
\label{sec:SpatialAdaptivity}

\begin{figure}
    \begin{center}
    \includegraphics[width=1\linewidth]{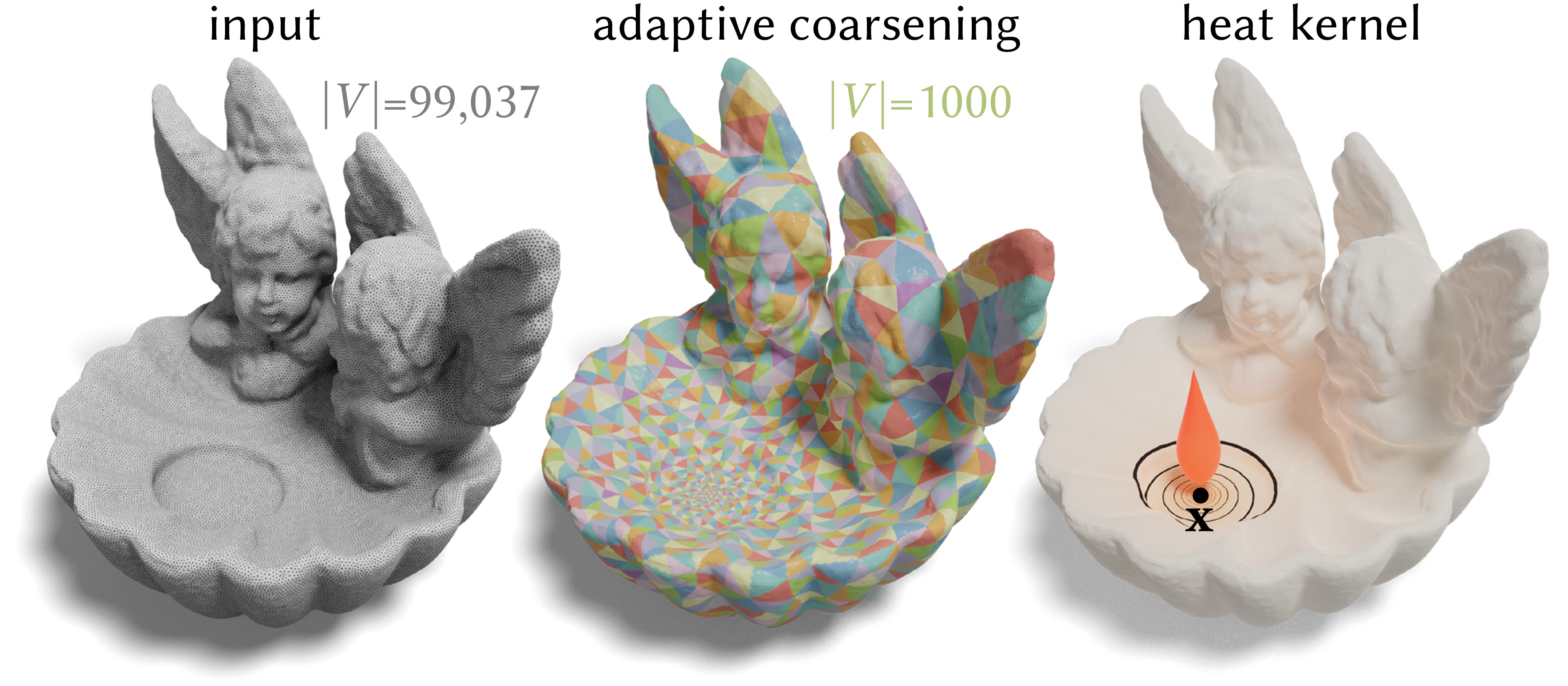}
    \end{center}
    \caption{By coarsening more aggressively away from the point \(\vx\), we better resolve the exponential falloff in a heat kernel centered at \(\vx\).}
    \label{fig:candle_adaptive}
\end{figure}

We can emphasize a region of interest by prescribing spatially-varying masses \(m_i\) on the fine vertices \(i \in V\); regions where \(m_i\) is small are then coarsened less aggressively.  For instance, \figref{candle_adaptive} uses masses \(m_i := 1/d^2_{\vx}(i)\) (where \(d_\vx\) is the distance to \(x\)) to adapt the coarse mesh to the \emph{heat kernel} centered at \(\vx\).

\subsubsection{Anisotropic Coarsening}
\label{sec:AnisotropicCoarsening}

\begin{figure}
    \begin{center}
    \includegraphics[width=1\linewidth]{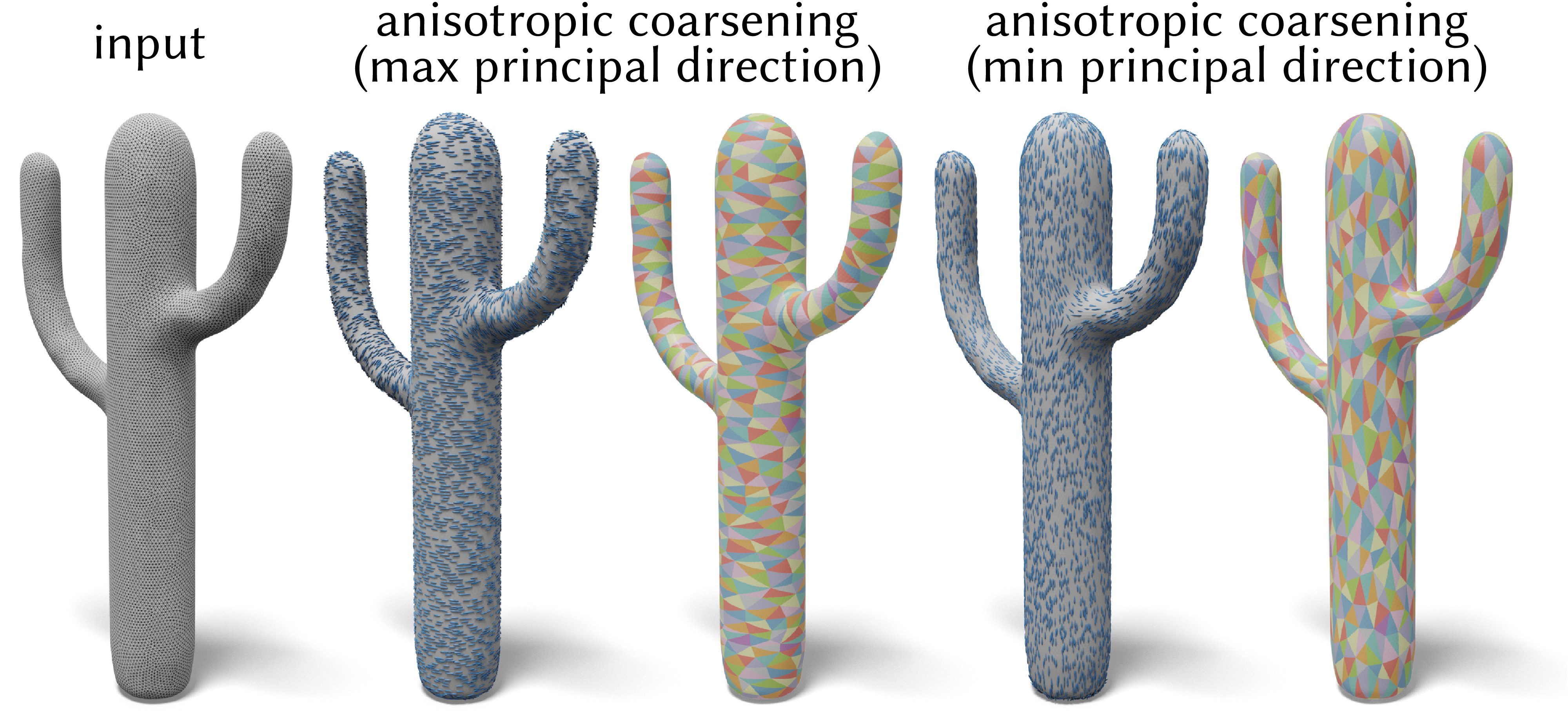}
    \end{center}
    \caption{Anisotropic coarsening can be achieved by applying nonuniform scaling to input edge lengths, here using principal curvature directions.}
    \label{fig:aniso_cactus}
\end{figure} 

Alternatively, we can emphasize important directions by non-uniformly scaling the input edge lengths along directions of interest---such meshes are better suited to, \eg{}, solving PDEs with anisotropic coefficients.  More explicitly, given vectors \(\vu_i\) at vertices, we scale each length \(\ell_{\ij}\) by a factor
\((1-\tau) + \tfrac{\tau}{2}((\vu_i \cdot \widehat{\ve}_{\ij})^2 + (\vu_j  \cdot \widehat{\ve}_{\ji})^2)\), where \(\widehat{\ve}_{\ij} := \ve_{\ij}/\|\ve_{\ij}\| \in \T_i M\) is the unit tangent vector along edge \(\ij\), and the parameter \(\tau \in [0,1]\) controls the strength of anisotropy.  For instance, in \figref{aniso_cactus} the vectors \(\vu_i\) are the (max or min) principal curvature directions, computed via the method of \citet{panozzo2010efficient}.  A challenge here, noted by \citet[Section 4.1]{campen2013practical}, is that a simple rescaling can violate the triangle inequality, which in practice limits the strength of anisotropy.  How to robustly express anisotropic changes to the discrete metric is an interesting question for future work.

\subsubsection{Boundary Preservation}
\label{sec:BoundaryPreservation}

\begin{figure}
    \begin{center}
    \includegraphics[width=1\linewidth]{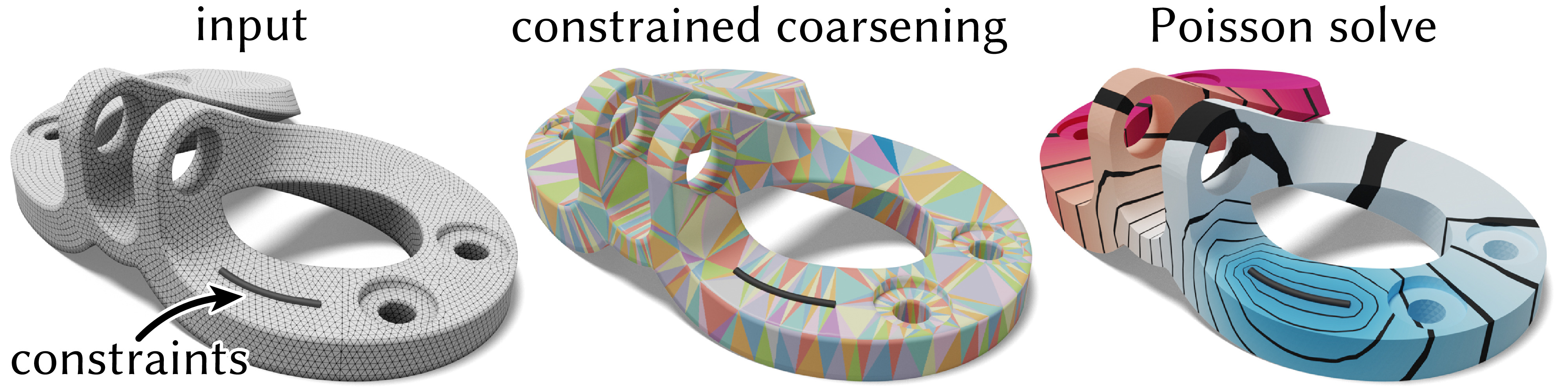}
    \end{center}
    \caption{Fixing vertices during coarsening enables us to exactly preserve boundary conditions and constraint curves when solving PDEs on surfaces.}
    \label{fig:constrained_coarsen}
\end{figure}

Finally, we can fix a user-specified set of vertices in order to, \eg{}, exactly preserve boundary conditions for a PDE.  For instance, in \figref{constrained_coarsen} we fix vertices that encode Dirichlet boundary conditions for a Poisson problem, accurately preserving both the boundary data and the constraint curve.

\subsection{Intrinsic Mesh Hierarchies}
\label{sec:SurfaceMultigrid}

\begin{figure}
    \begin{center}
    \includegraphics[width=1\linewidth]{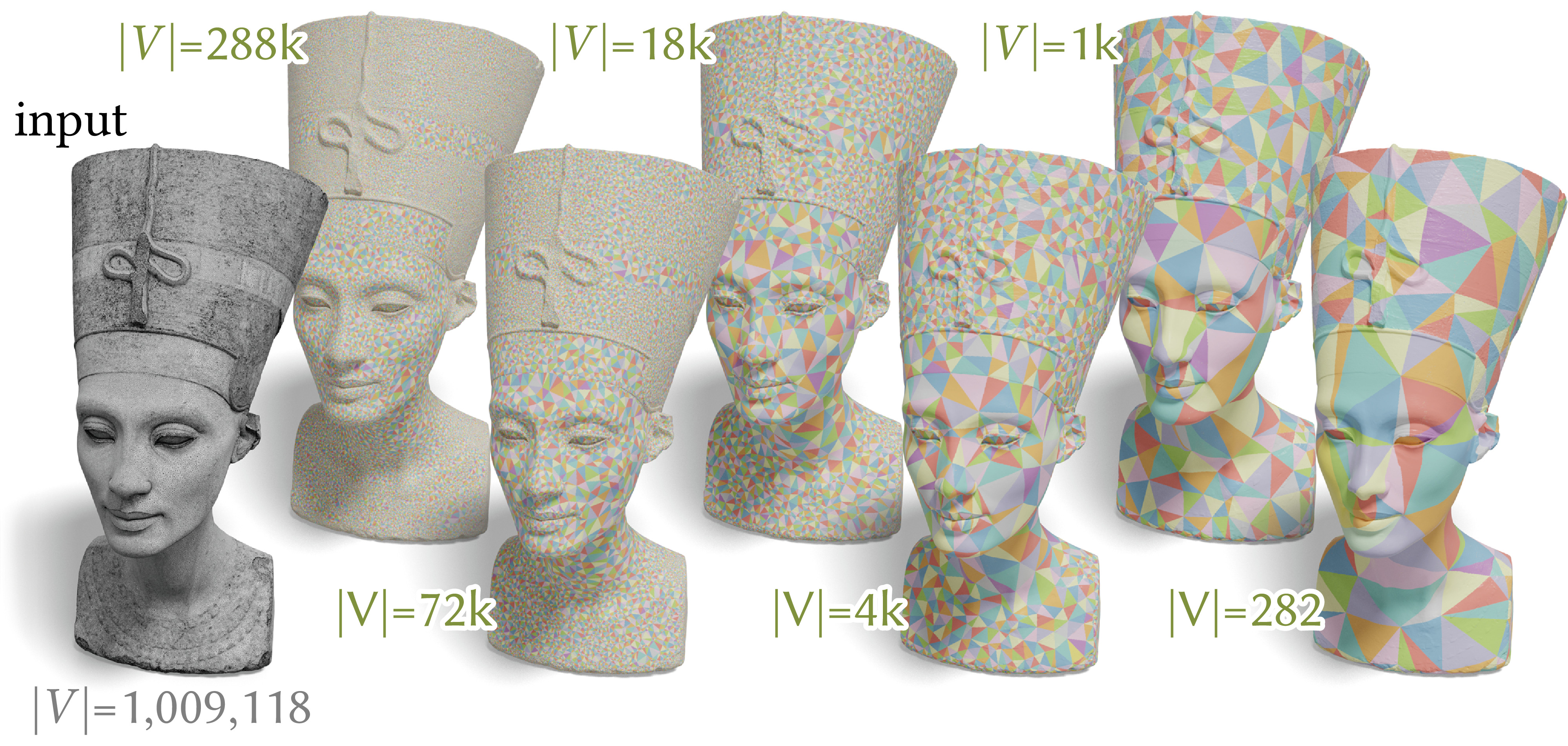}
    \end{center}
    \caption{We can build an intrinsic multigrid mesh hierarchy via repeated coarsening, here reducing the vertex count by a factor 1/4 at each level.}
    \label{fig:mg_hierarchy_highres}
\end{figure}

\begin{figure}
    \begin{center}
    \includegraphics[width=1\linewidth]{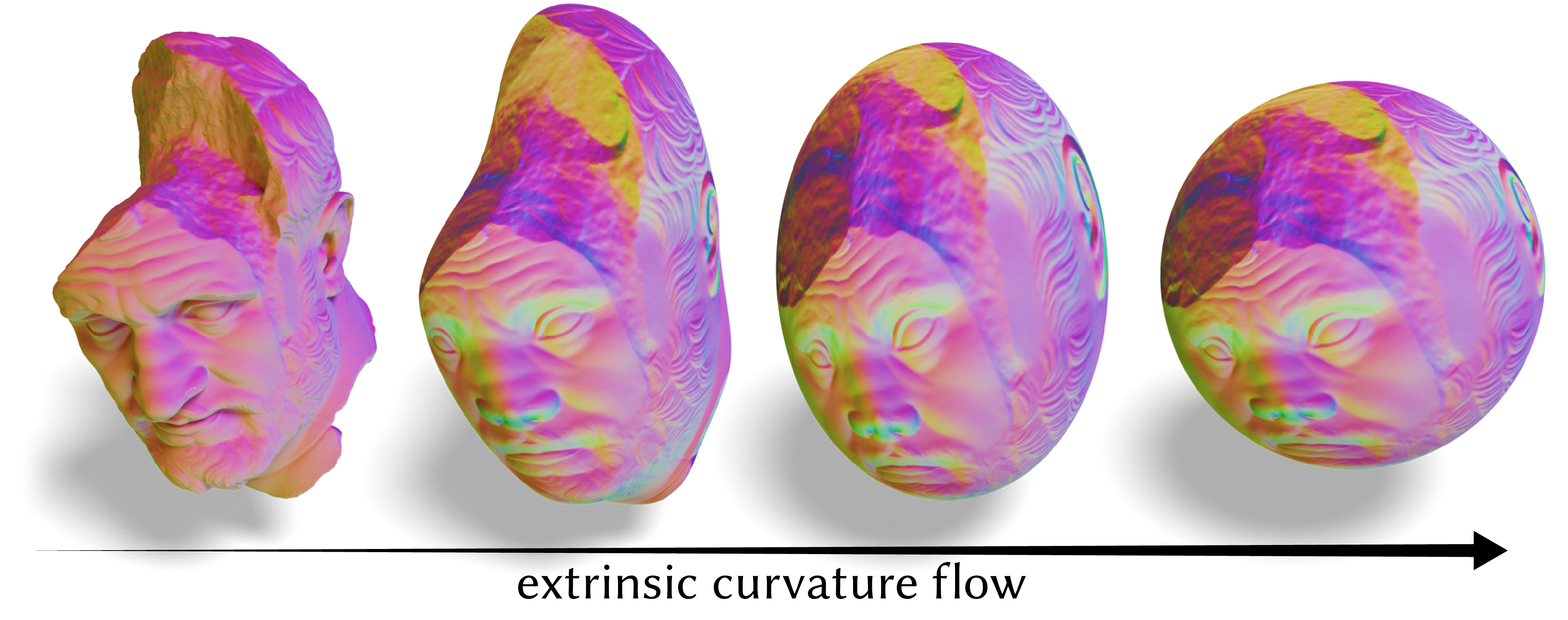}
    \end{center}
    \caption{Here we use our intrinsic multigrid scheme to accelerate the extrinsic curvature flow of \citet{KazhdanSB12}, achieving a 20x speedup.} 
    \label{fig:mcf}
\end{figure}

\begin{figure}
    \includegraphics[width=1\linewidth]{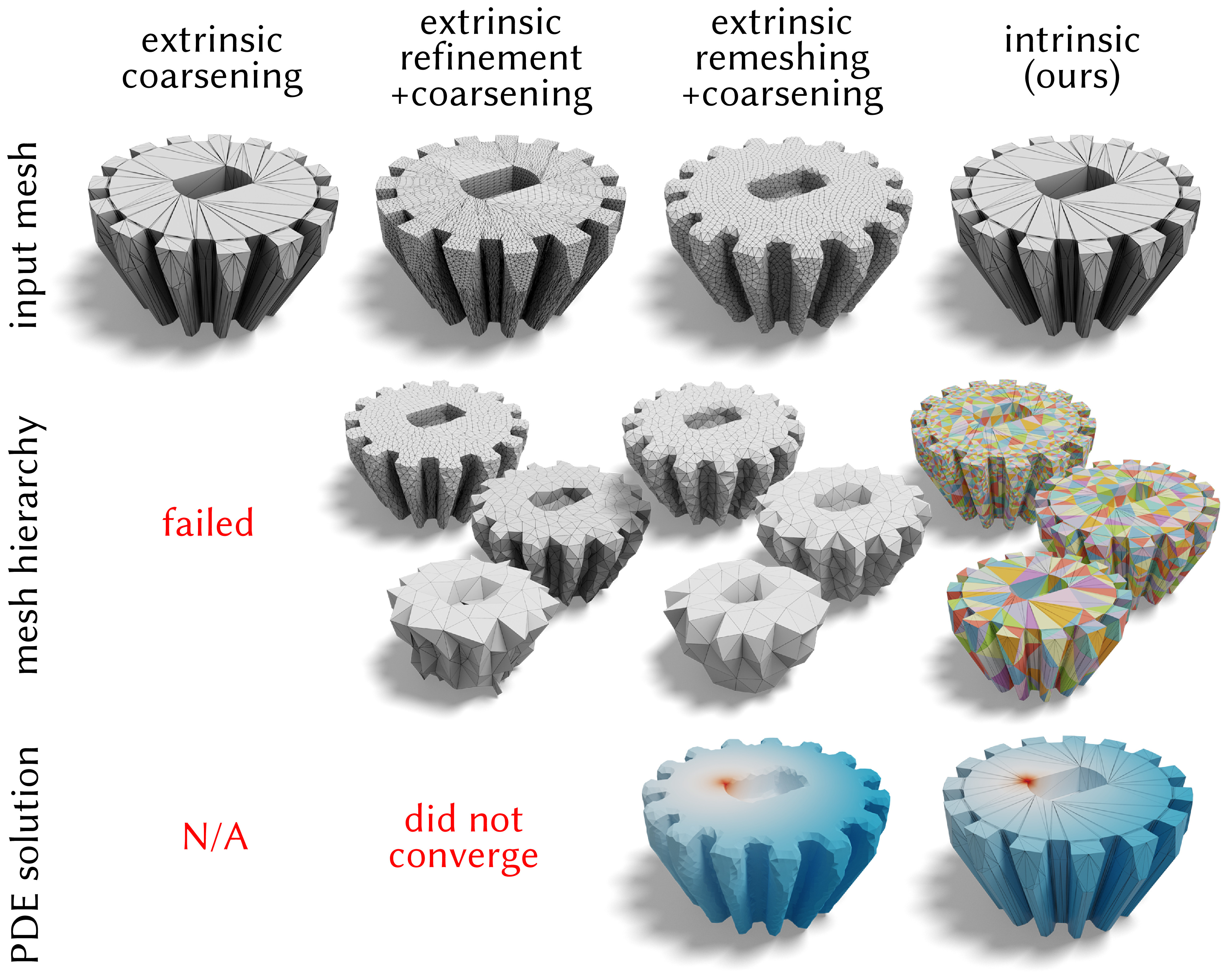}
    \caption{Obtaining a reliable surface mesh hierarchy for geometric multigrid can be surprisingly challenging---often requiring global remeshing of the geometry to obtain acceptable results.  Our intrinsic scheme builds on established guarantees, ensuring success even on extremely low-quality inputs.}
    \label{fig:robust_mg}
\end{figure}

Mesh hierarchies are used throughout visual, geometric, and scientific computing to accelerate solvers via, \eg{}, Cholesky preconditioners \cite{ChenS0D21}, multigrid methods \cite{AksoyluKS05, LiuZBJ21}, or GPU acceleration \citet{MahmoudPO21} further demonstrate the possibility of parallelizing geometry processing with mesh hierarchies.  Despite the fact that many of these applications need only intrinsic operators, these methods do not take full advantage of the intrinsic setting.  Our method can be used to construct an \emph{intrinsic} mesh hierarchy by simplifying the input to a sequence of progressively coarser meshes---\reffig{mg_hierarchy_highres} shows one such example.  Using the prolongation matrices \(\mP\) between consecutive levels, we can then build a surface multigrid method \ala{} \citet{LiuZBJ21}.  This intrinsic multigrid method is well-suited to algorithms expressed in terms of discrete differential operators---even if they involve extrinsic data---such as the modified mean curvature flow of \citet{KazhdanSB12} (\reffig{mcf}).  In general, the additional flexibility of working with intrinsic triangulations yields greater robustness than previous, extrinsic methods, especially on low-quality input.  For instance, in \reffig{robust_mg} the method of \citet{LiuZBJ21} fails to build a valid mesh hierarchy; extrinsically refining the input via \citep{CignoniCCDGR08} preserves the geometry, but the solver now fails due to low-quality triangles; global extrinsic remeshing enables the solver to succeed, but yields a solution on a different domain than the input.  Our intrinsic approach easily succeeds on this example, since (as noted in \figref{teaser}) it does not have to simultaneously juggle mesh quality and element quality, and can rely on hard guarantees about triangulation quality, as discussed in \secref{IntrinsicRetriangulation}.

\section{Limitations \& Future Work}
\label{sec:future_work}

\begin{wrapfigure}{r}{80pt}
   \includegraphics[width=\linewidth]{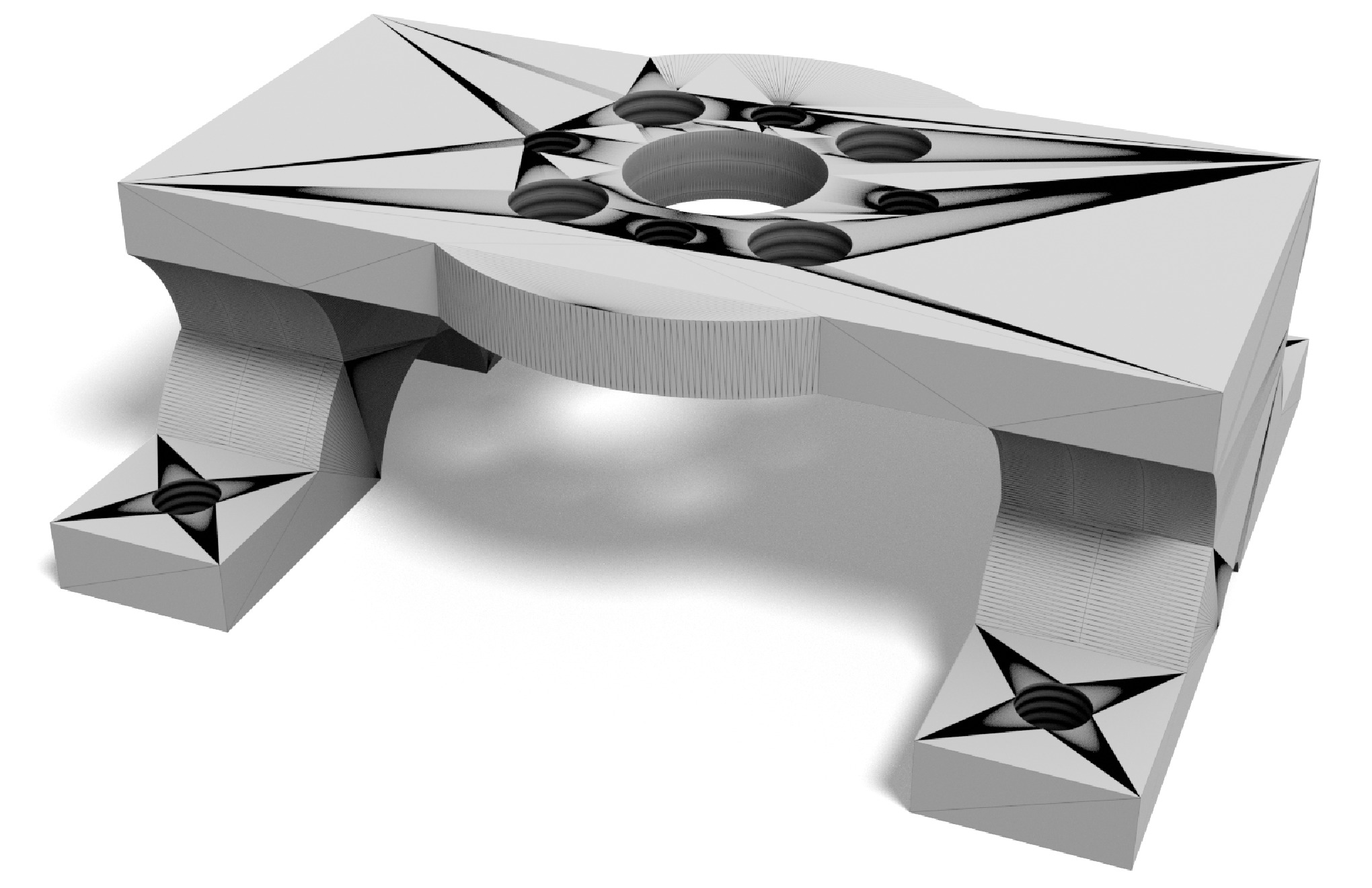}
\end{wrapfigure}

In order to keep computation cheap and local, the ICE metric makes three basic approximations.  First, we approximate the mass distribution of all ``ancestor'' vertices by concentrating their sum at their center of mass (\secref{Memoryless2D}).  This approximation is very much in the spirit of QEM, which approximates the extrinsic distribution of all ancestors via a single quadratic function.  Second, when re-assigning mass to a neighboring vertex, we approximate computation of the logarithmic map via parallel transport along an edge (\secref{memory_error_metric_3D}).  This approximation has no analogue in QEM, and in the future it may be worth considering other approximations---or at least performing an ablation (relative to the exact log map) to better understand the impact of this approximation on overall performance.  Third, we approximate the cost of redistributing curvature from a removed vertex \(i\) to its neighbors \(j\) via the change in curvature at each vertex \(j\) (\refequ{transfer_proportion}).  As noted in \secref{IntrinsicCurvatureErrorMetric} a more principled (but also more expensive) alternative might be to directly compute the transport cost for a signed measure, which still involves only the local vertex neighborhood.

Other aspects of the method could also be improved or generalized.  One significant question, noted in \secref{PointwiseMapping}, is how to more easily compute point correspondences without replaying (or reversing) coarsening operations.  Techniques such as integer coordinates~\citep{GillespieSC21a} might improve floating point robustness on extremely poor-quality inputs (inset).  Likewise, \emph{Ptolemy edge flips}~\citep{GillespieSC21} might further improve robustness to rare violation of the triangle inequality during flattening.  Extending prolongation to \emph{discrete differential forms}~\cite{desbrun2006discrete} could help accelerate a wider variety of geometry processing tasks~\cite{crane2013digital}.  Similarly, it may prove valuable to consider how to best perform simplification in a dynamic context---especially since many natural deformations are near-isometric.  Finally, a more careful treatment of memory management and parallel data structures might help bring performance closer to highly-optimized libraries for extrinsic simplification~\cite{Kapoulkine:2019:MO}.

\begin{acks}
   The authors thank Minchen Li, Silvia Sell\'{a}n, and Jiayi Eris Zhang for providing example data. This work was funded in part by an NSF CAREER Award (IIS 1943123), NSF Award IIS 2212290, a Packard Fellowship, NSERC Discovery (RGPIN–2022–04680), the Ontario Early Research Award program, the Canada Research Chairs Program, a Sloan Research Fellowship, the DSI Catalyst Grant program, the Fields Institute for Mathematics, the Vector Institute for AI, and gifts from Adobe Inc., Facebook Reality Labs, and Google, Inc.
\end{acks}

\bibliographystyle{ACM-Reference-Format}
\bibliography{intrinsic_coarsening}

\appendix

\end{document}